\documentclass[lettersize,journal]{IEEEtran}
\usepackage[noadjust]{cite}
\usepackage{amsmath,amsthm,graphicx,cite,booktabs,amssymb,gensymb, color, xcolor, fancyhdr,float,perpage,tikz, colortbl,enumerate}
\usepackage[linesnumbered,vlined,ruled]{algorithm2e}
\usepackage{algorithmic,array,enumerate,rotating}
\usepackage{subfigure,epsfig,dblfloatfix}
\usepackage{multirow,multicol}
\usepackage[affil-it]{authblk}
\usepackage{hyperref}
\usetikzlibrary{matrix}
\hypersetup{
	colorlinks   = true,
	linkcolor    = blue,
	citecolor    = blue
}

\makeatletter

\renewcommand{\@algocf@capt@plain}{above}

\newcommand{\thickhline}{%
    \noalign {\ifnum 0=`}\fi \hrule height 1pt
    \futurelet \reserved@a \@xhline
}

\definecolor{LightGray}{gray}{0.85}
\definecolor{Gray}{gray}{0.65}

\newcommand\bd[1]{\color{violet} \textbf{#1}}
\newcommand{\sd}{\textcolor{black}}

\makeatother
\def\bsx{{\boldsymbol{x}}}
\def\bsy{{\boldsymbol{y}}}
\def\bsH{{\boldsymbol{H}}}
\def\bsJ{{\boldsymbol{J}}}
\def\bsI{{\boldsymbol{I}}}

\def\bsV{{\boldsymbol{V}}}
\def\bsA{{\boldsymbol{A}}}
\def\bsB{{\boldsymbol{B}}}
\def\bsc{{\boldsymbol{c}}}
\def\bsR{{\boldsymbol{R}}}
\def\bpsi{{\boldsymbol{\psi}}}

\title{A Novel Image Denoising Algorithm Using Concepts of Quantum Many-Body Theory}
\author{Sayantan~Dutta$^{1,2,\ast}$,~Adrian~Basarab$^{3}$,~Bertrand Georgeot$^{2}$,~and~Denis~Kouam\'e$^{1}$}

\affil{
  {\em $^{1}$Institut de Recherche en Informatique de Toulouse, UMR CNRS 5505, Universit\'e de Toulouse, France} \\
 \vspace{.1cm}
  {\em $^{2}$Laboratoire de Physique Th\'eorique, Universit\'e de Toulouse, CNRS, UPS, France} \\
 \vspace{.1cm}
  {\em $^{3}$ Universit\'e de Lyon, INSA-Lyon, Universit\'e Claude Bernard Lyon 1, UJM-Saint Etienne, CNRS, Inserm, CREATIS UMR 5220, U1206, Villeurbanne, France}
  \thanks{$\ast$ Corresponding author: Sayantan Dutta (sayantan.dutta@irit.fr; sayantan.dutta110@gmail.com)}
 }
 
\begin{document}
\onecolumn
%\pagenumbering{gobble}

%
\maketitle

\begin{abstract}

Sparse representation of real-life images is a very effective approach in imaging applications, such as denoising. In recent years, with the growth of computing power, data-driven strategies exploiting the redundancy within patches extracted from one or several images to increase sparsity have become more prominent. This paper presents a novel image denoising algorithm exploiting such an image-dependent basis inspired by the quantum many-body theory. Based on patch analysis, the similarity measures in a local image neighborhood are formalized through a term akin to interaction in quantum mechanics that can efficiently preserve the local structures of real images. The versatile nature of this adaptive basis extends the scope of its application to image-independent or image-dependent noise scenarios without any adjustment. We carry out a rigorous comparison with contemporary methods to demonstrate the denoising capability of the proposed algorithm regardless of the image characteristics, noise statistics and intensity. We illustrate the properties of the hyperparameters and their respective effects on the denoising performance, together with automated rules of selecting their values close to the optimal one in experimental setups with ground truth not available. Finally, we show the ability of our approach to deal with practical images denoising problems such as medical ultrasound image despeckling applications.
\end{abstract}


%%%%%%%%%%%%%%%%%%%%%%%%%%%%%%%%%%%%%%%%%%%%%%%%%%%%%%%%%%%%%%%%%%%%%%%%%%%%%%%%%%%%%%%%%%%%%%%%%%%%%%%%%%%%%%%%%%%%%%%%%%%%%%%%%%%

\section{Introduction}
\label{sec:intro}

Over time, original methods from various branches of science have enriched the literature on digital image processing, and specifically the fundamental question of signal or image denoising, such as statistics \cite{Hamza2001image}, probability theory \cite{Pizurica2006estimating, Lebrun2013anonlocal}, graph theory \cite{Shuman2013the, Sandryhaila2013discrete, Pang2017graph} or differential equations \cite{Kim2006pde, Liu2012remote}. 
For the particular case of image restoration addressed herein, number of methods are based on sparse representations into a given basis, with most of the true image described by the projections on a few basis vectors. This enables to efficiently store and restore the image. Such sparse representations \cite{donoho1994ideal, starck2002curvelet} depend on both the transformation chosen and the nature of the image. Traditionally, all these methods exploit few explicit or underlying hypotheses about the image to restore, for example, piece-wise smoothness, but are not strong enough to capture the complex textures present in the true image.

With the growth of computing power, data-driven strategies to increase the sparsity and overcome the limitations of general transforms have become more prominent in recent decades.
One such approach is to learn overcomplete dictionaries from training image sets \cite{Aharon2006an, Elad2006image}. Another method is based on patch-based schemes, using patch neighborhood as a feature vector. For example, block-matching and 3D filtering known as BM3D creates 3D data arrays by grouping similar image fragments before computing a sparse representation applying 3D transformations \cite{Dabov2007Image, dabov2009bm3d}.

The non-local means (NLM) algorithm brought a different perspective to the image denoising problem, where each estimated image pixel intensity is a weighted average of pixels centered at patches that are similar to the patch centered at the estimated pixel \cite{Buades2005areview}. An alternative patch-based NLM approach projects image patches into a lower dimensional subspace using principal component analysis (PCA) before performing the weighted average for denoising \cite{tasdizen2009principal, deledalle2011image}. Later on, various schemes were proposed in the literature to accelerate or to improve the NLM performance, such as a fast NLM algorithm with a probabilistic early termination \cite{Vignesh2010fast}, quadtree-based NLM with locally adaptive PCA \cite{Zuo2016image}, fast processing using statistical nearest neighbors strategy \cite{Frosio2019statistical}, adaptive neighborhoods \cite{Kervrann2006optimal}, patch-based locally optimal Wiener filtering \cite{Chatterjee2012patch} and others \cite{Mahmoudi2005fast, Van2009sure, Dong2013nonlocally, Li2021patch, Zha2020from, Zha2020image, Zha2021image, Zha2021nonconvex}. These NLM-based schemes are known as a powerful way of denoising exploiting similar patches from the whole image. Hence, the patch neighborhood gives an effective way of preserving the local structures of an image where neighborhood similarity is the key ingredient. 

This paper explores such an approach of exploiting the image neighborhood by borrowing tools from quantum mechanics, precisely, the quantum interactions. Quantum theory is the underlying theory of nature, which governs our world at a fundamental level, and classical mechanics is merely a limiting behavior of quantum mechanics. In classical theory, the position and momentum of a particle are determined precisely, whereas in quantum theory they are given by a probability distribution encoded in the wave function. The wave function in turn can be computed as solution of a wave equation known as the Schr\"odinger equation. Over the past few years, new proposals to use such wave functions as a basis in imaging applications, such as image feature extraction \cite{Aytekin2013Quantum, youssry2015quantum, youssry2019continuous}, denoising \cite{kaisserli2015novel, dutta2021quantum}, deconvolution \cite{dutta2021plug, dutta2021poisson} or others \cite{Eldar2002quantum, dutta2022quantum} have been proposed in the literature. However, all the existing approaches have been built on the theory of single-particle quantum systems.

In this paper, we propose a novel image representation algorithm well adapted for denoising based on the theory of quantum many-body interaction. In the case of a system containing two or more quantum particles, they can influence each other's quantum state through quantum interactions. The main idea of this work is to adapt ideas from this theory to extend the concept of interaction to imaging problems. More precisely, the proposed framework consists in quantum interactions between image patches where interactions reflect patch similarity measures in a local neighborhood. In this way, each patch acts as a single-particle system, and the whole collection, that is the entire image, behaves as a many-body system where interactions describe regional similarities to neighboring patches. Preliminary results on Gaussian denoising were presented in \cite{dutta2021image}. Herein, we show that this method constitutes a robust generalized formalism for image-independent and image-dependent noise models. The extension of \cite{dutta2021image} primarily lies in:
(i) the characterization of the hyperparameters and automated ways to predict their optimal values with limited knowledge about the input image,
(ii) investigation on the denoising possibilities beyond Gaussian statistics without any modification of the algorithms,
(iii) a detailed discussion of denoising performance compared to state-of-the-art methods for both image-independent and image-dependent scenarios,
(iv) application on real medical data for ultrasound image despeckling.

Earlier proposed single-particle based schemes \cite{Aytekin2013Quantum, youssry2015quantum, youssry2019continuous, kaisserli2015novel, dutta2021quantum, dutta2021plug, dutta2021poisson} have proven their good restoration abilities for different noise models, but are too simple to take advantage of the structural properties of the image and are computationally costly at large scale. As we will show, the proposed generalized framework based on the use of quantum many-body physics improves the previous methods on both counts, building a more versatile computationally efficient adaptive basis that considers similarities between neighboring image patches.

In general, it may seem that there is a close architectural resemblance between the NLM and the proposed many-body scheme since similarity measure is the key for both cases. However, the two methods are different from several perspectives. The NLM image denoising algorithm exploits the self-similarities among the image patches to obtain the similarity weights resulting into a non-local weighted average scheme for denoising. The proposed approach brings non-local characteristics within the quantum framework, where interactions between neighboring patches preserve the local structural similarities. For each patch, these interactions convey the structural information into a quantum adaptive basis offering a good sparsifying transformation at a patch level further used for denoising. It turns out that such a theory can be elegantly written using multi-particle quantum theory instead of the single-particle one.

%\textbf{REMOVE THE FOLLOWING BLUE TEXT. WE COULD USE SOME  IDEAS AS PERSPECTIVES OF THIS WORK} \dk{More recently, deep learning-based methods becoming more popular in this image processing domain, especially the convolutional neural network (CNN) based architectures for their very competitive denoising performance. Large modeling capacity and robust training procedure make CNN attractive for image denoising and has been explored with various schemes, such as a fast flexible learning method \cite{Chen2017trainable}, residual learning \cite{Zhang2017beyond}, a fast and flexible denoising with a tunable noise level \cite{Zhang2018ffdnet} and others. Note that our method does not belong to this deep learning paradigm and follows a conventional way. However, embedding this many-body interaction architecture into a CNN is a different algorithm. It provides an intriguing direction for future research and should require a separate study. So keeping aside the deep learning models, in this paper, we mainly concentrate on the development of this many-body interaction framework and will explore its deep learning possibilities in a separate work.}

In the paper, we first present briefly the previously proposed decomposition concept using a quantum adaptive basis \cite{dutta2021quantum} based on single-particle theory with its limitations in Section~\ref{sec:qtsingparti}, and then introduce its generalization using many-body quantum theory for imaging problems in Section~\ref{sec:qtmanybodyth}. Our image denoising algorithm is described in detail in Section~\ref{sec:qmpiimadecom}. We then turn to numerical implementation of the method on several examples in Section~\ref{sec:simuresu}. We first explore ways to propose automated rules for hyperparameters selection, and then display numerical results showing that the ability of the proposed method  in reducing low and high intensity noise regardless of the noise statistics. We also show its good performance in real-life medical ultrasound (US) image despeckling applications.  Finally we end with conclusions and future perspectives in Section~\ref{sec:conclusion}.

% From a macro scale, it may seem that there is a close architectural resemblance of the NLM and proposed many-body scheme.

%\section{Relationship between image representation and quantum many-body model}
%\label{sec:relation}

\begin{figure*}[t!]
\centering
\includegraphics[width=1\textwidth]{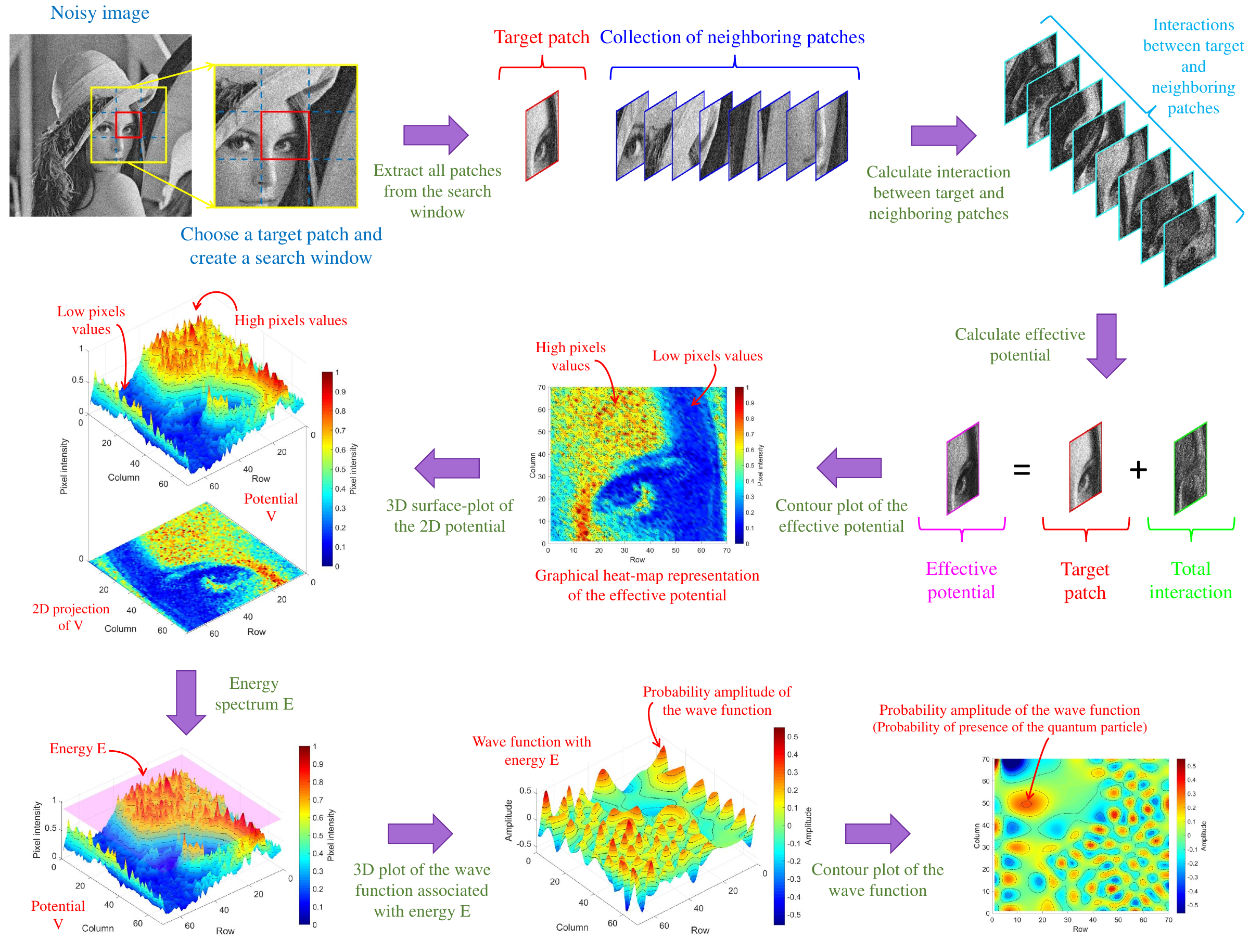}

\caption{A simple example of the construction of adaptive vectors from many-patch interaction.}

\label{fig:flow}
\end{figure*}

\section{Quantum many-body theory for imaging}
\label{sec:quantheo}

\subsection{Quantum theory for a single-particle system}
\label{sec:qtsingparti}

\subsubsection{Quantum theory}
\label{sec:qt_sub1}

Before detailing the proposed method, we briefly review, for self consistency, the quantum mechanical method for denoising built on single-particle theory introduced in \cite{dutta2021quantum}. For more details on quantum theory, one may refer to one of the many textbooks on this subject, e.g. \cite{feynman1977feynman, landau1991quantum, cohen1977quantum}.

In a non-relativistic single-particle quantum system the wave function $\psi(z)$ describes a particle with energy $E$ in a potential $V(z)$ and satisfies the stationary Schr\"odinger equation:
\begin{equation}
 - \frac{\hbar ^2}{2m} \nabla^2 \psi (z) + V(z)  \psi (z) = E \psi (z),
\label{eq:schroedinger}
\end{equation}
\noindent with $m$, $\hbar$, $\nabla$, and $z$ are respectively the mass of the quantum particle, the Planck constant, the gradient operator, and the spatial coordinate. The wave function $\psi(z)$  is an element of the Hilbert space of $L^2$-integrable functions, and its modulus square \textit{i.e.}, $|\psi (z)|^2$, gives the probability of presence of the particle at some point $z$ on the potential $V(z)$. 

The wave function solutions of \eqref{eq:schroedinger} form a complete set of basis vectors of the Hilbert space with the following properties: i) Wave vectors are oscillating functions.
ii) Oscillation frequency increases with increasing energy $E$.
iii) The basis vectors oscillate with a local frequency proportional to $\sqrt{E - V(z)}$, thus for the same wave function the frequency differs locally depending on the local value of $E - V(z)$.
iv) The hyperparameter $\hbar ^2/2m$ controls the dependence of the local frequency on $E - V(z)$.
%\begin{enumerate}[i)]
%\item Wave vectors are oscillating functions.
%\item Oscillation frequency increases with increasing energy $E$.
%\item The basis vectors oscillate with a local frequency proportional to $\sqrt{E - V(z)}$, thus for the same wave function the frequency differs locally depending on the local value of $E - V(z)$.
%\item The hyperparameter $\hbar ^2/2m$ controls the dependence of the local frequency on $E - V(z)$.
%\end{enumerate}
\noindent These properties of the basis vectors are the key features to use them as an adaptive basis for an imaging problem. For a more detailed illustration of these features, we refer readers to \cite{dutta2021quantum}.

\subsubsection{Application to imaging problems}
\label{sec:imaappli}

To adapt these concepts to image processing applications, the wave equation \eqref{eq:schroedinger} is rewritten in operator notation leading to $\bsH \bpsi (z) = E \bpsi (z) $ with Hamiltonian operator $\bsH = -(\hbar ^2/2m) \nabla^2 + \bsV(z)$. The eigenvectors of the Hamiltonian operator are the stationary solutions of \eqref{eq:schroedinger}.

For imaging applications, the space is finite and discretized, and the potential $\bsV$ of the system may be defined as the image pixel values $\bsx$. This leads to a discretized problem, where the Hamiltonian operator becomes a finite matrix and can be used as a tool for constructing an adaptive basis \cite{dutta2021quantum}. This discretized Hamiltonian operator reads:
\begin{eqnarray}
\bsH[i,j]= \left \{
   \begin{array}{c c l}
      \bsx[i]+ 4 \frac{\hbar ^2}{2m} &  & \mbox{for} \; i=j,\\
       -\frac{\hbar ^2}{2m} & & \mbox{for} \; i = j \pm 1,\\
        -\frac{\hbar ^2}{2m} & & \mbox{for} \; i = j \pm n,\\ 
      0 & & \mbox{otherwise}, \vspace*{-1mm}
   \end{array}
   \right.
\label{eq:H}
\end{eqnarray}
\noindent where $\bsx \in \mathbb{R}^{n^2}$ is an image (\textit{i.e.}, $\bsV = \bsx$), and $\bsx[i]$ and $\bsH [i,j]$ represent respectively the $i$-th component of the image $\bsx$, vectorized in lexicographical order and the $(i,j)$-th component of the operator. Note that standard zero padding is used to handle the boundary conditions.
%, corresponding to violations of the rule \eqref{eq:H} at the boundary, more precisely $\bsH[i,j] = \bsx[i]+ 2 \frac{\hbar ^2}{2m}$ for $i=j$ and $i \in \{1,n,n^2-n+1,n^2\}$, $\bsH[i,j] = \bsx[i]+ 3 \frac{\hbar ^2}{2m}$ for $i=j$ and $i \in \{ 2,3,...,n-1,n^2-n+2,n^2-n+3,...,n^2-1\}$, $\bsH[i,j] = \bsx[i]+ 3 \frac{\hbar ^2}{2m}$ for $i=j$ and $i ~\mod~ n\in \{0,1\}$, except for $i \in \{ 1,2,...,n,n^2-n+1,n^2-n+2,...,n^2\}$ in order to respect the boundary conditions, and $\bsH[i,i+1]= \bsH[i+1,i]=0$ for any $i$ multiple of $n$ apart from $n^2$.
A more detailed description of the Hamiltonian construction can be found in \cite{dutta2021quantum, dutta2021plug}. The corresponding set of eigenvectors of the Hamiltonian operator \eqref{eq:H} serves as the quantum adaptive basis on which the image is decomposed before denoising is performed by thresholding the coefficients in energy.

\subsubsection{Shortcomings of the single-particle theory in image processing}
\label{sec:shortcomsingle}

This method of constructing an adaptive basis using quantum principles in a single-particle setting has already been studied in some of our previous works, notably for image denoising \cite{dutta2021quantum} and deconvolution \cite{dutta2021plug, dutta2021poisson}. This adaptive method not only is effective for handling different noise statistics (\textit{e.g.}, Gaussian, Poisson) but also equally efficient for different levels of noise (low as well as high-intensity noise). Nevertheless, there are some technical and intrinsic challenges, such as:
\begin{enumerate}[i)]
\item Structural features are crucial for imaging applications, but this adaptive approach does not take advantage of them.

\item The random noise present in the system leads to the well-known phenomenon of quantum localization \cite{Anderson1958absence} of the wave vectors. The presence of this subtle quantum phenomenon gives additional structures to the adaptive basis and makes it less effective for image denoising. This problem was cured in \cite{dutta2021quantum} by adding an additional step of low-pass filtering, for example, through a Gaussian filter with appropriate standard deviation, of the noisy image. This complicates the method and in particular entails the integration of a new hyperparameter (standard deviation) in the algorithm, which increases the complexity of hyperparameter tuning.

\item The computational burden of such a method can be quite large compared to other sophisticated state-of-the-art methods, thus preventing it from implementation in large-scale images.

\end{enumerate}

In the following, we will show that these drawbacks can be addressed by constructing a new adaptive basis by exploiting quantum many-body theory, more precisely the physics of quantum interactions.

% The single particle theory is more or less well explored for an imaging applications but

\subsection{Quantum many-body theory for image processing}
\label{sec:qtmanybodyth}

\subsubsection{Quantum theory for many particles}
\label{sec:qtmbodyth_sub1}

The quantum theory described above is modified for a system with more than one particle. In particular, particle-to-particle interactions take place inside the quantum system. For a system with $w$ particles the Hamiltonian operator for the many-body system becomes \cite{mahan2013local}:
\begin{equation}
H = - \sum_{a=1}^w \dfrac{\hbar ^2}{2m_a} \nabla^2 + \dfrac{1}{2} \sum_{a=1}^w \sum_{b=1, b\neq a}^w V_{ab},
\label{eq:H_mb}
\end{equation}
where $m_a$ is the mass of the $a$-th particle and the potential $V_{ab}$ is a function of $ z_1,z_2,\cdots,z_w $, the spatial coordinates of the $w$ particles. Thus, for a given energy $E$ the associated wave function $\psi$ depends on $ z_1,z_2,\cdots,z_w $, and satisfies a new Schr\"odinger equation:

\begin{equation}
H \psi( z_1,z_2,\cdots,z_w) = E \psi( z_1,z_2,\cdots,z_w).
\label{eq:schro_mb}
\end{equation}

%The complex nature of $V_{ab}$ makes it difficult to solve the multi-body problems \eqref{eq:H_mb} and \eqref{eq:schro_mb},  which can be dealt with under some prior assumptions from quantum physics.

\subsubsection{Application to image processing}
\label{sec:mabothimapro}

We propose to extend this multi-body theory to build an adaptive basis for imaging applications by assimilating similarities between patches into the quantum framework. Similar to non-local mean filter-based approaches, the proposed algorithm splits the image or a local region into into small patches ranging from $1$ to $w$. Each of these patches acts as a single-particle quantum system, which allows the Hamiltonian operator to be defined for each patch as follows:
\begin{equation}
\bsH_a = \rlap{$\underbrace{\phantom{\sum_{b=1, b\neq a}^z I_{ab} -\dfrac{\hbar ^2}{2m_a}~}}_{ \bsH_{0_a}} $}
	-\dfrac{\hbar ^2}{2m_a} \nabla^2  +  
	\rlap{$\overbrace{\phantom{V(z_a) + \sum_{b=1, b\neq a}^z I_{ab}}}^{\bsV_a^{effective}}$}  \bsV(z_a)
    + \underbrace{ \sum_{b=1, b\neq a}^w \bsI_{ab} }_{\bsH_{I_a}}, ~ a = 1,\cdots,w,
\label{eq:H_hfmf}
\end{equation}
where, $\bsH_{0_a}$ is the Hamiltonian in the patch $\bsA$ for a single particle system (as discretized in \eqref{eq:H}). $\bsI_{ab}$ and $\bsH_{I_a}$ represent respectively the interaction between the $\bsA$ and $\bsB$ patches and the total interaction between the patch $\bsA$ and the other patches in the system. Thus, inside the patch $\bsA$ the effective potential $\bsV_a^{effective}$ is
\begin{equation}
\bsV_a^{effective} = \bsV(z_a) + \sum_{b=1, b\neq a}^w \bsI_{ab} = \bsV(z_a) + \bsH_{I_a}.
\label{eq:V_eff}
\end{equation}
Therefore, we have a different adaptive basis for each patch containing a unique effective potential $\bsV_a^{effective}$ associated with an energy $E_a$. Fig.~\ref{fig:flow} depicts one such simple example of constructing adaptive vectors from the many-patch interaction concept. Thus the problem of finding the adaptive basis is transformed into the solution of the system of $w$ equations, as follows:
\begin{equation}
\bsH_a \bpsi(z_a) = E_a \bpsi(z_a), ~~~ a = 1,2,\cdots,w.
\label{eq:schro_hfmf}
\end{equation}
where similar discretization procedures should be used in each patch as in \eqref{eq:H}.

\subsubsection{Definition of the quantum interaction between two image patches}
\label{sec:defintre}

Interaction between two or more objects is a universal phenomenon that governs the world at a very basic level, fundamentally classified into four groups: gravitational, electromagnetic, strong, and weak interactions. The gravitational and electromagnetic interactions have long-range properties characterized by power laws. We extend this concept to an imaging problem by introducing the interaction between two image patches, as follows:

\begin{itemize}

\item There is an inverse proportionality between the interaction and the square of the Euclidean distance (\textit{i.e.}, physical distance) between the patches, \textit{i.e.,} $\bsI_{ab} \propto \frac{1}{D_{ab}^2}$, where $D_{ab}$ is the Euclidean distance between two patches denoted by $\bsA$ and $\bsB$.

\item There is a linear proportionality between the interaction and the absolute value of the pixel-wise difference between the patches. This process is defined pixel-wise, \textit{i.e.,} $\bsI_{ab}^i \propto |\bsA^i - \bsB^i|$, $i = 1,2,\cdots,P_{dim}$, where superscript $i$ and $P_{dim}$ are associated with the $i$-th pixel and the number of pixels in every image patch respectively.

%indicates the $i$-th element and $P_{dim}$ is the number of pixels in every image patch.

\end{itemize}

Hence, within the proposed image processing framework, the power law for an interacting many-patch system can be defined as
\begin{equation}
\bsI_{ab}^i =  p \frac{|\bsA^i - \bsB^i|}{D_{ab}^2} ,~~ i = 1,2,\cdots,P_{dim},
\label{eq:invsqulaw}
\end{equation}
where the proportionality constant $p$ acts as a hyperparameter for the proposed formalism.

\subsubsection{Interaction and patch similarity in image processing}
\label{sec:interpre}

In our many-patch model the proposed mathematical formalism of the power law interaction can be interpreted in the following way:
i) two patches with similar pixel values have smaller interaction than the ones with very different values,
%ii) far placed patches always show small-scale interaction despite their possible pixel wise dissimilarities.
ii) patches located far from each other have small interaction regardless of their pixel values.
%ii) if two patches are similar but placed far from each other in the image then they present small-scale interaction.
In other words, neighboring patches show high interactions if they are very different from each other based on pixel values, while distant patches are always less interactive despite their possible dissimilarity. Based on these principles, the power law manifests itself in such a way that the effective potential of the patch $\bsA$ is $\bsV_a^{effective}$. This is obtained after the combination of the initial potential (\textit{i.e.}, the target patch itself) with the total interaction between the target patch and its neighboring patches, exploiting the concept of patch similarity in the local neighborhood. This local-similarity is a fundamental building block of real images that preserves structural features \cite{Buades2005areview}. We note that power laws other than the inverse square law could be used, thus modifying the importance of distant patches compared to the nearby ones in the proposed methodology.

%%%%%%%%%%%%%%%% ipr vs snr %%%%%%%%%%%%%%%%

\begin{figure}[t!]
\centering
\includegraphics[width=.6\textwidth]{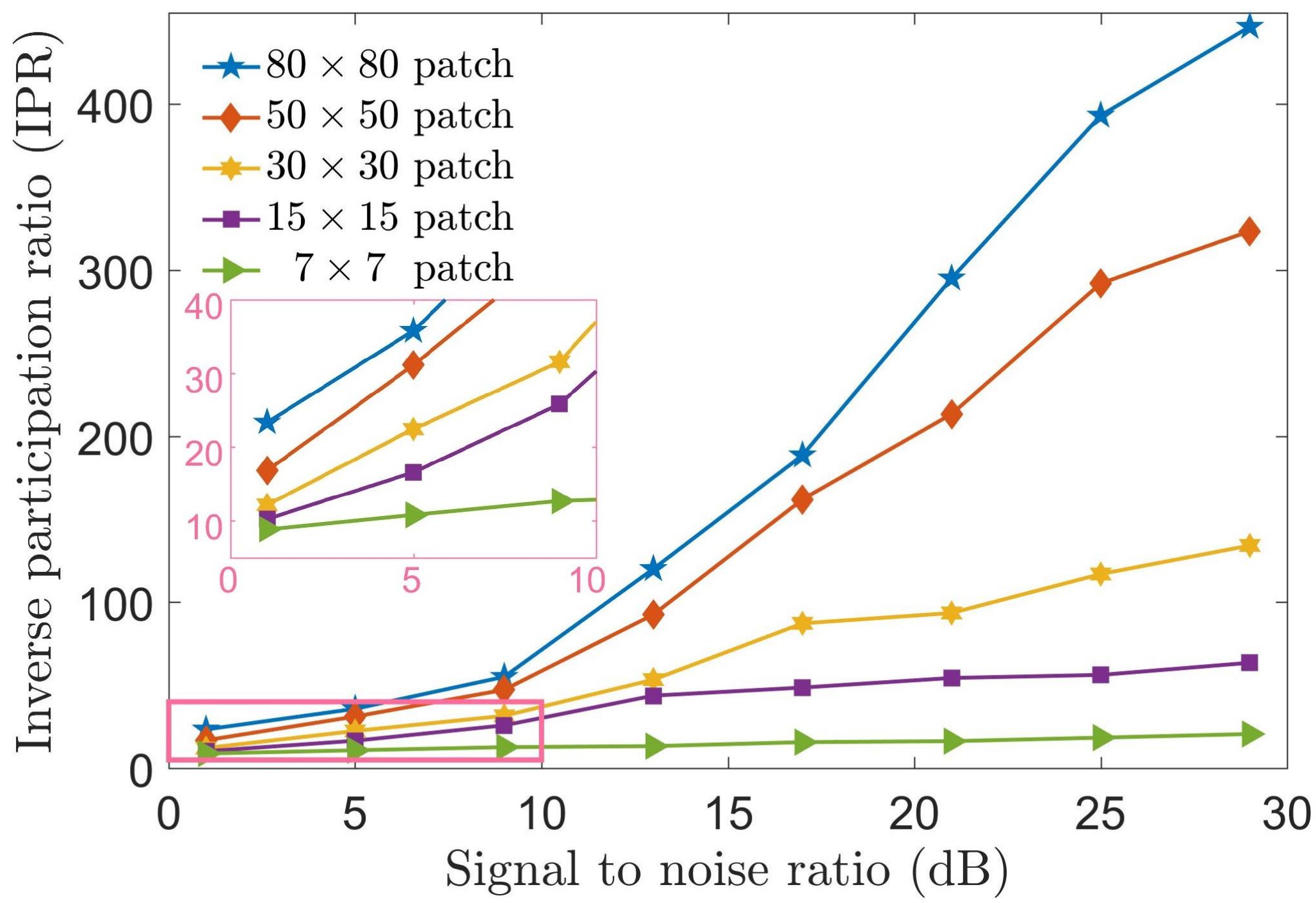}

\caption{Average inverse participation ratio (IPR) of all the adaptive basis vectors as a function of signal to noise ratio for the \textit{Lena} image degraded by AWGN using different sizes of the image patch.}

\label{fig:iprVSsnr}
\end{figure}

\subsubsection{Why the many-patch theory avoids the quantum localization problem}
\label{sec:lolizprobqmpi}

The presence of random fluctuations in the potential of a quantum system leads to the phenomenon of quantum localization, also known as Anderson localization \cite{Anderson1958absence}. This is a property of wave functions in a disordered potential which makes them exponentially localized due to destructive interference. As a consequence, the adaptive basis vectors for various imaging problems are localized at different positions of the potential in presence of random noise, which makes the adaptive basis less suitable for image decomposition tasks. 
In \cite{dutta2021quantum}, this challenge was solved by adding a cumbersome first step of image low-pass filtering, with an additional hyperparameter involved. A more detailed discussion of this phenomenon, in particular for image decomposition and denoising, can be found in our previous work \cite{dutta2021quantum}.

In the framework of the many-patch theory described above, the decomposition is done at the level of the individual patch, much smaller than the full image. The inverse participation ratio (IPR) of the wave functions, defined as $1/\sum_i |\bpsi(i)|^4$ for a wave function $\bpsi$, gives a measure of the localization. For a vector uniformly spread over $L$ indices and zero elsewhere, the IPR is exactly $L$. More generally, the localization length of localized wave functions is proportional to the IPR. It is known from localization theory that this localization length decreases with the intensity of the disorder. Thus unless the noise is  extremely strong, the localization length may be larger than the patch size, making the localization irrelevant for our problem. Fig.~\ref{fig:iprVSsnr} shows the average IPR (measuring the localization length) of all the adaptive basis vectors for the \textit{Lena} image degraded by additive white Gaussian noise (AWGN) with increasing signal to noise ratio (SNR) using different patch sizes. This illustration confirms that the IPR decreases with the SNR, but this effect reduces with patch size. For example, for a $80 \times 80$ patch, the IPR decreases rapidly with decreasing SNR (increasing noise intensity) and becomes less than the patch size for SNR $\leq 12$ dB, making the system extremely localized. However, for smaller patches like $7 \times 7$, almost no such effect is visible for similar noise intensities. In other words, the localization effect becomes negligible in a small patch than in a large one and turns out to be irrelevant beyond a certain level of patch size. We found out that even for fairly strong noise it is always possible to find a patch size smaller than the average IPR that makes irrelevant the localization effect, avoiding the need of the low-pass filtering to create the adaptive basis.

%Our numerical investigations showed that it was always the case even for fairly strong noise, avoiding the need of the low-pass filtering to create the adaptive basis.

\section{Quantum many-patch interaction for imaging applications: The problem of image decomposition}
\label{sec:qmpiimadecom}

\subsection{Key principles of the proposed many-patch model}
\label{sec:genfram}

The objective of this work is to propose a methodology of an explicit construction of an adaptive basis related to the many-body interaction theory under the principles recalled here \cite{dutta2021image}:

\begin{itemize}

% \item In a quantum system, the image (the pixel values) plays the role of the potential.

%\item The wave function $\bpsi (z)$ describes the probability of presence of a quantum particle at some point on the potential (\textit{i.e.}, image) in a single-particle system.

% \item This wave function $\bpsi (z)$ is a stationary solution of \eqref{eq:schroedinger} under the potential $\bsV(z)$ (\textit{i.e.}, image) and an element of the Hilbert space of function.

% \item As an element of the set of oscillatory functions, this wave function uses low oscillation frequencies to probe higher values of the pixels and vice-versa, \textit{i.e.}, local frequencies depend on the image pixels' intensities.

\item Every small patch extracted from an image corresponds to a quantum particle; each of these image-patches or potential surfaces with a quantum particle acts like a single-particle system.

\item These single-particle systems are not isolated from each others, on the contrary, the interaction between them and other patches occurs within the whole image, like a quantum many-body system, where a particle-to-particle interaction takes place in the quantum system.

\item As a consequence of these interactions, the effective potential (see \eqref{eq:V_eff}) of quantum particles changes, thus the local oscillation frequency of the wave function depends on these interactions.

\item These interactions transmit structural features to the wave functions through the effective potential.

\item The effective potentials will be used to construct an adaptive basis for each individual patch, in particular used for the decomposition of that patch.

\item As an element of the set of oscillatory functions, this basis function uses low oscillation frequencies to probe higher values of the effective potential and vice-versa, \textit{i.e.}, local frequencies depend on the effective potential, and thus on the pixel values and inter-patch interactions.

\end{itemize}

\subsection{Denoising algorithm using quantum many-patch interactions}
\label{sec:algorithm}

This subsection illustrates in detail the application of the proposed many-patch scheme to address image denoising. In this application, the construction of an adaptive basis for each individual image-patch is the primary objective, which leads to a three step denoising strategy: decomposition of that patch using the adaptive basis, thresholding of the projection coefficients, and finally recovery of the denoised patch by back-projection. These basis vectors are the eigenvectors of the Hamiltonian matrix \eqref{eq:H}, constructed from the effective potential \eqref{eq:V_eff}.

%These basis vectors for the image-patch, for example, $a$-th patch, are stationary solutions of \eqref{eq:schroedinger} with $\bsV_a^{effective}$ following \eqref{eq:V_eff} as the effective potential. To be specific, under the effective potential $\bsV_a^{effective}$, that incorporate the current patch and its interactions with its neighbours, these basis vectors are the eigenvectors of the Hamiltonian matrix \eqref{eq:H}. Similar to the single-particle system, these adaptive vectors belong to the Hilbert space of oscillatory functions with: i) the frequency of oscillation increases with increasing energy value (\textit{i.e.}, eigenvalue in \eqref{eq:schro_hfmf}), and ii) a given basis vector uses low oscillation frequencies to probe higher values of the effective potential and vice-versa. It is now assumed that the noise primarily rules the high-frequency components of the image, \textit{i.e.}, eigenvectors corresponding to higher energy eigenvalues. Therefore thresholding in energy should be done to eliminate the image components associated with the high energy eigenvectors.

These adaptive vectors belong to the Hilbert space of oscillatory functions with: i) the frequency of oscillation increases with increasing energy value (\textit{i.e.}, eigenvalue in \eqref{eq:schro_hfmf}), and ii) a given basis vector uses low oscillation frequencies to probe higher values of the effective potential and vice-versa.
It is now assumed that the noise primarily rules the high-frequency components of the image, \textit{i.e.}, eigenvectors corresponding to higher energy eigenvalues. Therefore as in the single-particle algorithm, thresholding in energy should be done to eliminate the image components associated with the high energy eigenvectors.

In the proposed interaction framework, the structural similarity between neighboring image patches is assumed to be an innate property of the image. Hence two neighboring patches are assumed to be similar to the extent of random noise. Following the definition \eqref{eq:invsqulaw}, two adjacent patches show high interaction if they are pixel-wise dissimilar (\textit{i.e.}, random noise is present), thus further contributing to the effective potential \eqref{eq:V_eff}. In other words, the interaction term or ultimately the effective potential increases if the noise intensity increases, which eventually shifts the high-frequency noise components of the image to even higher energy eigenvectors. Thus, in order to have a denoised patch, a noisy patch is projected onto a $d$-dimensional subspace that is constructed by the lowest energy solutions of \eqref{eq:schro_hfmf} and rebuild the denoised patch from these projection coefficients. In this way, a lack of similarity between pixels leads to a stronger denoising, since for the same value of the energy these regions will have lower frequencies than the ones with more similarity. Here, $d$ acts as a thresholding hyperparameter. Combining all the denoised patches, following a similar path proposed in the non-local means architecture, one can obtain the final denoised image. Hereafter this proposed adaptive quantum denoiser which integrates the quantum theory of interactions to imaging problems is called Denoising by Quantum Interactive Patches (De-QuIP). The whole denoising process is regrouped in Algorithm~1 in the Supplementary Material file.\footnote{The Matlab code of the proposed denoising algorithm is available at \href{https://github.com/SayantanDutta95/}{github.com/SayantanDutta95/}}. The Computational complexity of the proposed algorithm is discussed in details in the Supp. Mat. file, see Section~B.

%%%%%%%%%%%%%%%%%%% Sample pictures%%%%%%%%%%%%%

\begin{figure}[t!]
\centering
\includegraphics[width=1\textwidth]{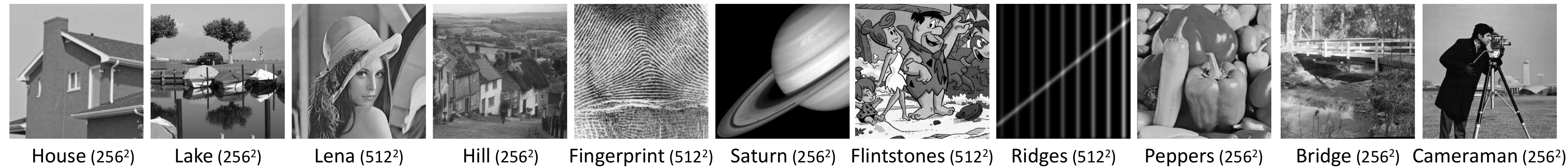}

\caption{Sample images (sizes in parentheses).}
\label{fig:samples}

\end{figure}

\section{Simulation Results}
\label{sec:simuresu}

This section illustrates the interest of the proposed approach in image denoising problems and explores ways to choose the suitable hyperparameters. At the outset, Subsection~\ref{sec:hypara} explains the reliance of the proposed denoising scheme on the optimal choice of the hyperparameters $P_h$, $W_h$, $p$, $\hbar ^2/2m$ and $d$, and explores rules for their possible estimations. For a thorough investigation, we explore cases of four different noise intensities (low to high) with image independent (\textit{e.g.}, Gaussian) and dependent (\textit{e.g.}, Poisson) noise models. The subsequent Subsection~\ref{sec:comparison} provides denoising results and a comparison between the proposed approach and several standard state-of-the-art methods. Finally, the section ends with a real medical application in Subsection~\ref{sec:usdesp}, which highlights the potential of the proposed scheme for the despeckling of real-world ultrasound (US) images.

% Finally, a medical application is discussed in the Supplementary Mat. file, see Table~IX and Fig.~2, 

\subsection{Influence of hyperparameters $P_h$, $W_h$, $p$, $\hbar ^2/2m$ and $d$ and how to select them}
\label{sec:hypara}

%\textbf{REMOVE THIS. The overall text is long enough}\dk{The proposed De-QuIP denoiser is designed using some hyperparameters that control the denoiser's efficiency. In this subsection, we will characterize their influence and investigate the effects on the end result. Finally, we propose ways to predict the optimal values of the hyperparameters with limited knowledge about the original image.}

\subsubsection{Effect of patch size $P_h$}
\label{sec:hy_ph}

The localization of the basis vectors is associated with the length of the image patch, as explained in Subsection~\ref{sec:lolizprobqmpi}. The respective localization length or IPR decreases for increasing noise intensity. To deal with this quantum localization phenomenon, the size of the patch should be always less than or equal to the localization length of the basis vectors for different levels of noise. If the localization length is greater than the size of the patch, the basis vectors probe the entire region of the image patch with different ranges of oscillation frequencies depending on the intensity of the image pixels. On the contrary, a smaller localization length leads to an exponential localization of the basis vectors on a specific part of the image patch. Thus, these localized vectors will not have different frequencies at different pixel values and lose a key asset of this formalism. The drastic effect of this localization phenomenon on image denoising is shown in our previous work \cite{dutta2021quantum}, where an additional Gaussian smoothing was necessary before computing the quantum adaptive basis (QAB), used as a denoiser in that process. On the contrary, the current formalism eliminates this issue without any additional computational requirements.

%%%%%%%%%%% comparison between De-QuIP and QAB  %%%%%%%%%%%%%

\begin{table}[t!]
\begin{scriptsize}

\begin{center}
\caption{Simulation data with different patch sizes for the \textit{Lake} image contaminated by AWGN (SNR = 16dB). For the proposed De-QuIP method hyperparameters $\hbar^2/2m = 1.5$, and $p$ and $d$ are estimated from the equations \eqref{eq:curfit1} and \eqref{eq:curfit2} respectively.}
\label{tab:tab_com_De-QuIPvsQAB}
\begin{tabular}{cc  cccccccc}
\thickhline

\multirow{2}{*}{} & \multirow{2}{*}{Data} & \multicolumn{8}{c}{Patch size}\\ 
			%\cline{3-10}
			& & $1 \times 1$ & $3 \times 3$ & $5 \times 5$ & $7 \times 7$ & $11 \times 11$ & $17 \times 17$ & $27 \times 27$ & $63 \times 63$\\

\thickhline % \midrule

\multirow{3}{*}{\begin{turn}{90}QAB\end{turn}}
& PSNR(dB) 		& 11.36 & 12.78 & 21.56 & 24.40 & 26.54  & 27.12 & 27.33  & 28.09 \\
& SSIM 			& 0.43  & 0.46  & 0.48  & 0.48  & 0.63   & 0.70  & 0.74   & 0.79 \\   \vspace*{2mm}
& Time(sec)		& 30.56 & 17.09 & 41.31 & 70.32 & 161.96 & 328.97 & 881.69 & 5800.72 \\
							
	%	\cline{2-10}

\multirow{3}{*}{\begin{turn}{90}De-QuIP\end{turn}}
& PSNR(dB) 		& 22.12 & 28.16 & 28.73 & 28.84  & 28.58  & 28.23   & 28.16 & 27.77\\
& SSIM 			& 0.37  & 0.78  & 0.83  & 0.83   & 0.82   & 0.81    & 0.80 & 0.79\\   \vspace*{1mm}
& Time(sec)		& 21.93 & 22.75 & 82.61 & 108.01 & 490.52 & 3829.31 & 5644.90 & 22765.18\\

\thickhline

\end{tabular}\end{center} 
\end{scriptsize}

\end{table}

Furthermore, a smaller patch size helps to reduce the computational complexity, as discussed in the section above. As a consequence, De-QuIP denoiser is more computationally efficient than the previously proposed QAB denoiser in \cite{dutta2021quantum}. Table~\ref{tab:tab_com_De-QuIPvsQAB} summarizes the run time using the QAB and De-QuIP denoiser with increasing patch size. The peak signal to noise ratios (PSNR) and the structure similarity
(SSIM), used as denoising quality metrics, are given to have a quantitative analysis concerning the patch size. All the algorithms have been implemented in 
Matlab and tested on a computer with an Intel(R) Core(TM) i7-10510U CPU of 4 cores each with 1.80 GHz, 16 GB memory and using Windows 10 Pro version 20H2 as operating system. From Table~\ref{tab:tab_com_De-QuIPvsQAB}, one can see that the computational time for both denoisers increases as the patch size increases but the denoising performance (\textit{i.e.}, PSNR and SSIM values) for De-QuIP first increases with the patch size and then begins to decrease gradually after size $11 \times 11$. Whereas, QAB requires much larger patches to achieve a similar performance, which essentially imposes a huge computational burden on the process. The gradual decrease in the performance of the De-QuIP denoiser for increasing patch size is expected due to the localization phenomenon, which is discussed above. Therefore, a smaller patch size preserves the fundamental features of these adaptive vectors and reduces the computational complexity and run time. Herein, we will only focus on the patch sizes $5 \times 5$, $7 \times 7$ and $11 \times 11$ for further investigations.

\subsubsection{Effect of the search window size $W_h$}
\label{sec:hy_wh}

The search window is the image region aroung the current patch regrouping all the patches interacting with it. Following the discussion in Subsection~\ref{sec:defintre}, the size of the search window plays an important role in preserving the structural similarities in a local neighborhood. This search window is usually defined as a square window of limited size so that the implementation is restricted to a small neighborhood centered on the target patch (to be denoised) instead of the whole image. In the literature, mostly two types of approaches are used, based on a fixed search window size \cite{tasdizen2009principal, Mahmoudi2005fast, deledalle2011image, Vignesh2010fast} or an adaptive approach \cite{Kervrann2006optimal}. In this work, we concentrate on the fixed size approach for examining the effect of the search window on De-QuIP.

%%%%%%%%%%%%%%%% hyper window size %%%%%%%%%%%%%%%%

\begin{figure}[t!]
\centering
\includegraphics[width=.8\textwidth]{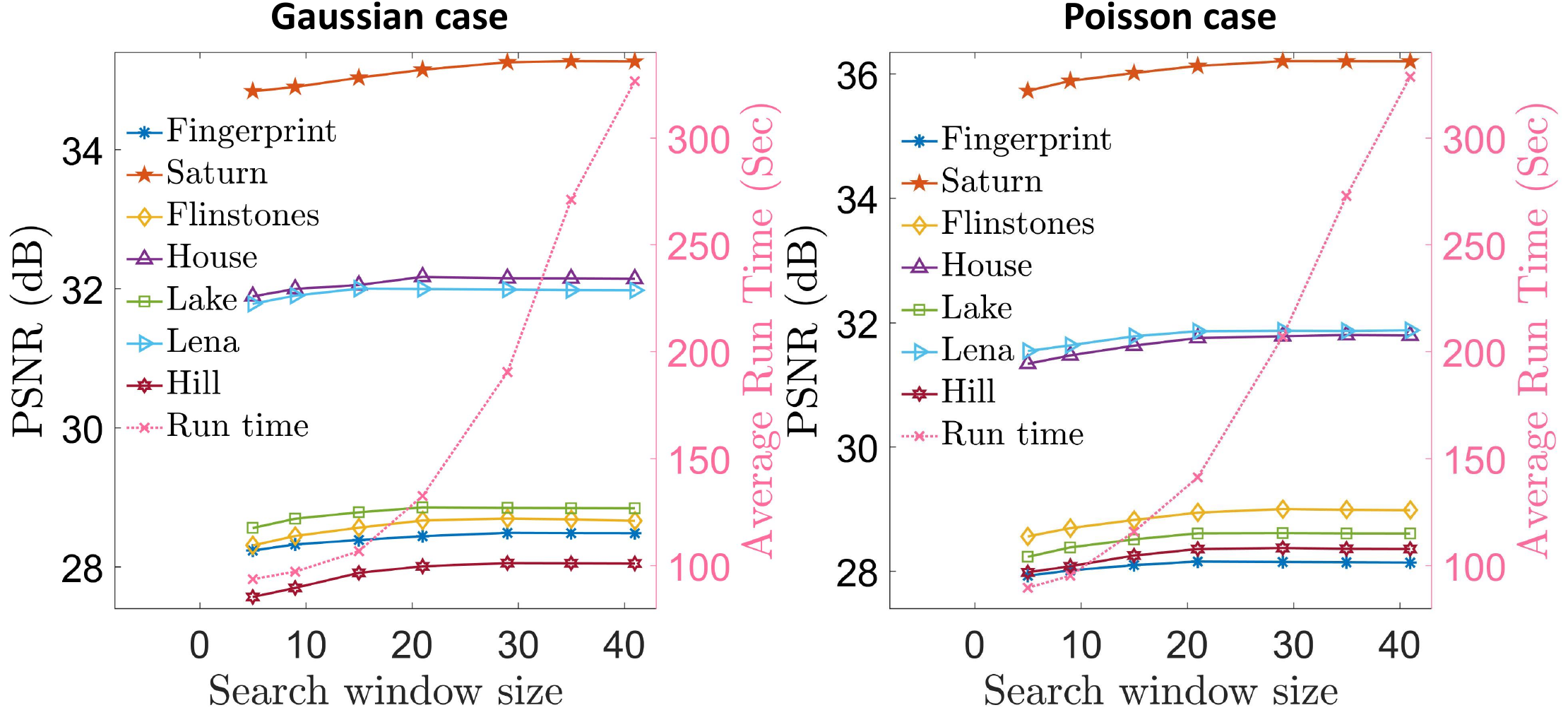}

\caption{Denoising performance in terms of PSNR (left y-axis) and average run time (right y-axis) of De-QuIP as a function of the search window size for the first seven sample images in Fig.~\ref{fig:samples}. The images hyperparameters $P_h = 7$ and others are estimated from the eq.~\eqref{eq:curfit1}-\eqref{eq:curfit4}.}
\label{fig:hywindow}

\end{figure}

Fig.~\ref{fig:hywindow} shows the denoising performance of De-QuIP in terms of PSNR as a function of the search window size for the Gaussian and Poisson noise cases. For these simulations, the patch size is kept fixed at $7 \times 7$ for all images. Note that in these simulations patches overlap, given that consecutive target patches are one pixel away from each other, both in the horizontal and vertical direction. In Fig.~\ref{fig:hywindow}, one can see that in both cases, the denoising ability increases with the size of the search window before roughly stabilizing beyond a size $20 \times 20$ for both noise models. These observations show that the patch neighborhood is important to increase the denoising performance but larger search windows do not bring additional information about the neighborhood due to the inverse square nature of the interaction term. It is also important to notice that the computation time increases with the search window size, as shown in the right y-axis in Fig.~\ref{fig:hywindow}. The use of a relatively moderate size search window is computationally more efficient while preserving the image attributes using the proposed interaction framework. Note that these results are consistent for other patch sizes. Therefore, for simplicity, in this work we choose a search window size of $15 \times 15$, $21 \times 21$ and $33 \times 33$ respectively for the patch sizes of $5 \times 5$, $7 \times 7$ and $11 \times 11$. One interesting observation can be drawn here, that the search window size changes with the patch size and not with the noise model within the proposed algorithm.

% attribute

\subsubsection{Influence of the proportionality constant $p$}
\label{sec:hy_p}

\begin{figure*}[t!]
\centering
\subfigure[PSNR vs $p$]
{\label{subfig:hy_pVSpsnr}
\includegraphics[width=.27\textwidth]{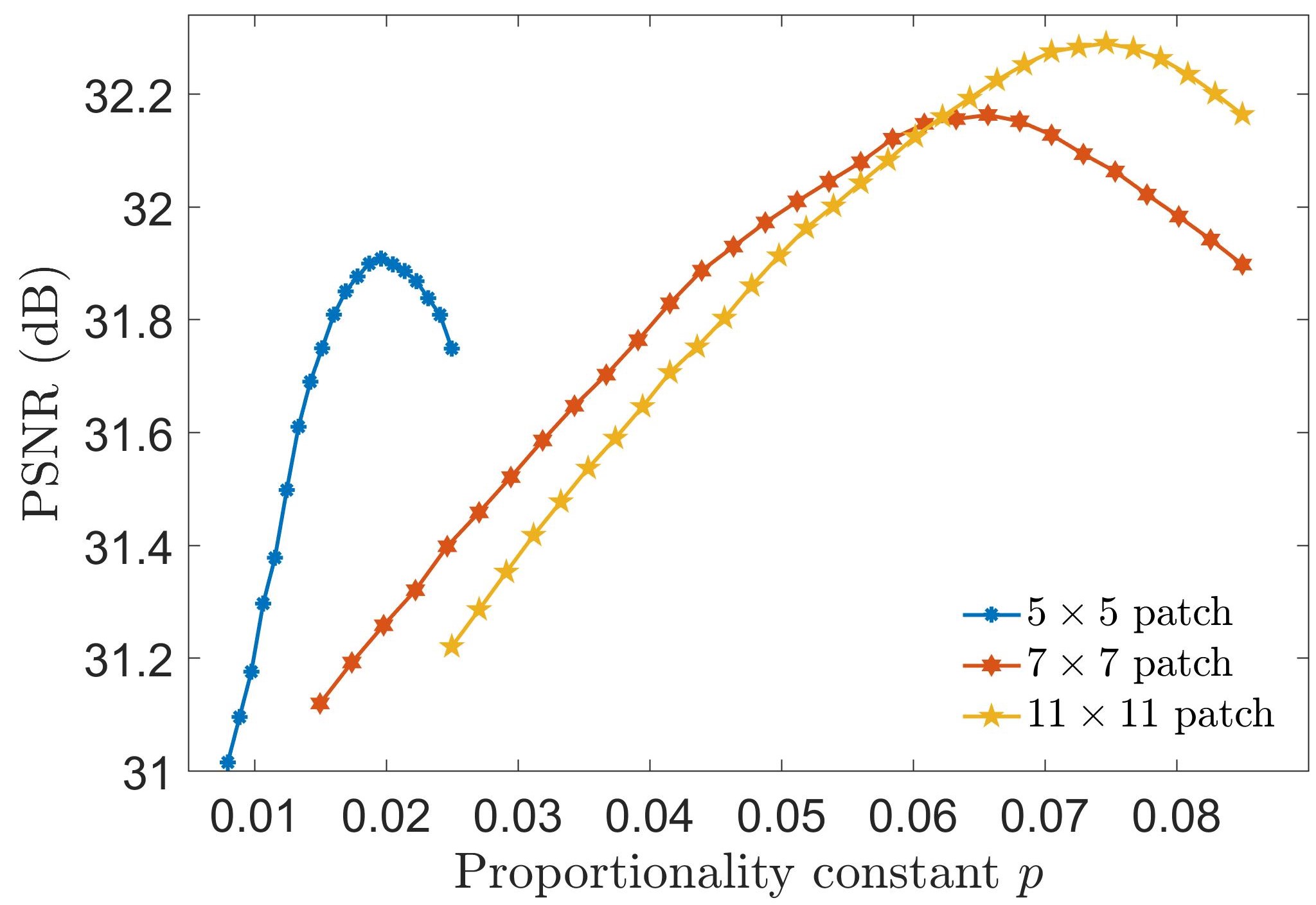}}
\subfigure[PSNR vs $F_{\mbox{factor}}$]
{\label{subfig:hy_factVSpsnr}
\includegraphics[width=.27\textwidth]{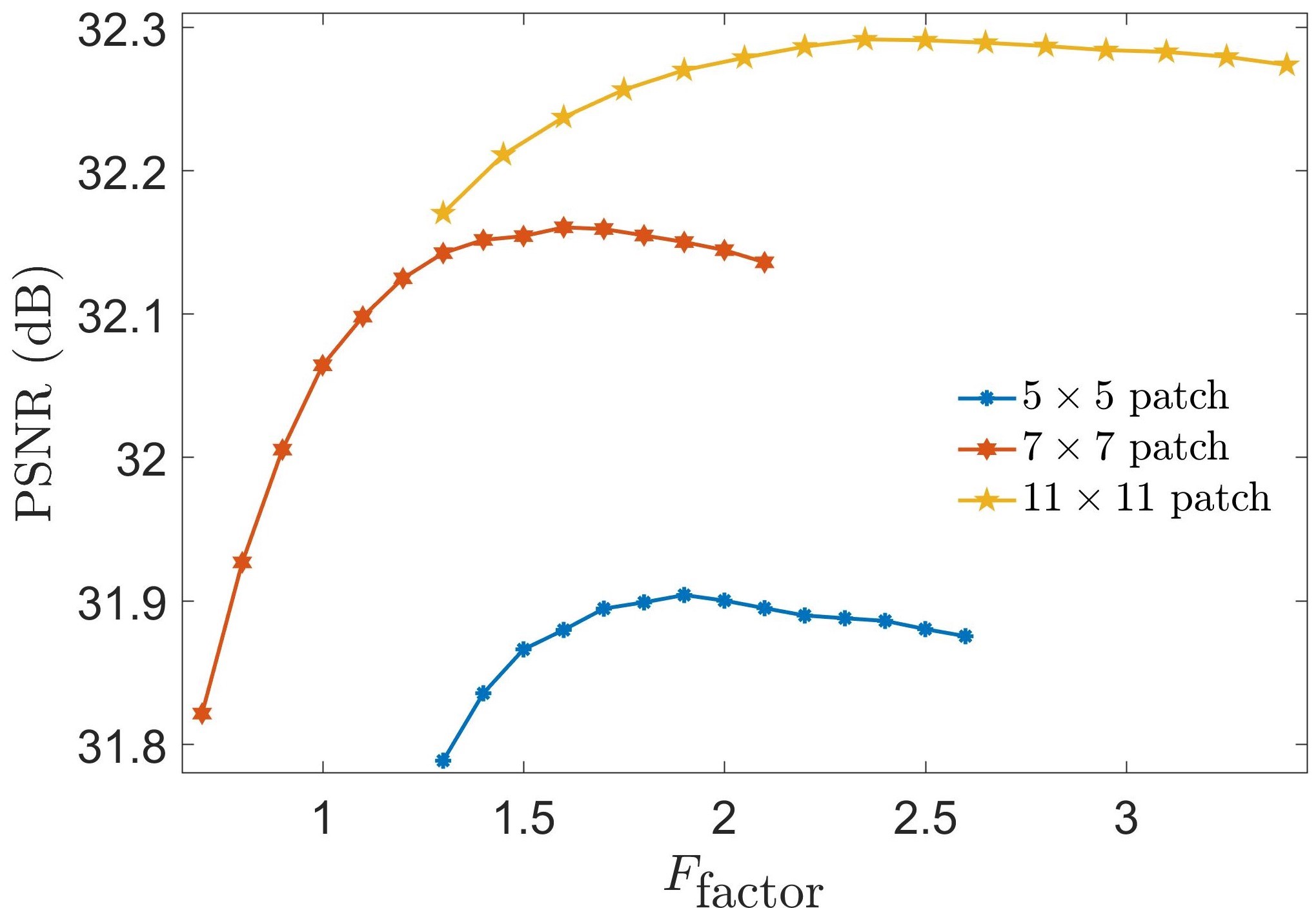}}
\subfigure[PSNR vs $d$]
{\label{subfig:hy_dVSpsnr}
\includegraphics[width=.27\textwidth]{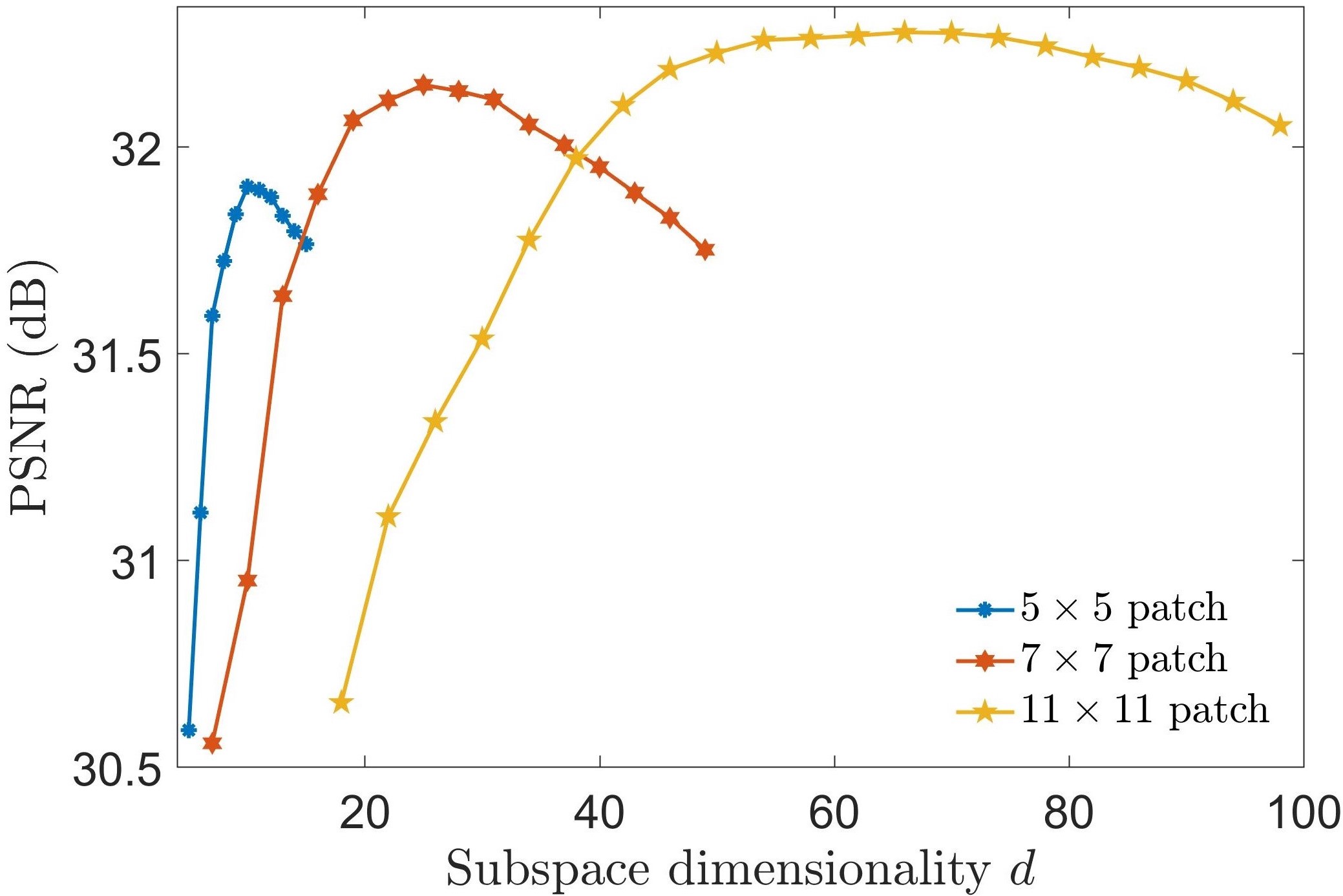}}

\caption{Denoising performance of De-QuIP in terms of PSNR as a function of the hyperparameters for the \textit{house} image corrupted with AWGN (SNR = 16dB) using three different patch sizes.
All hyperparameters are estimated using equations \eqref{eq:curfit1}-\eqref{eq:curfit4}.}
\label{fig:hyVSpsnr}

\end{figure*}

\begin{figure}[t!]
\centering
\includegraphics[width=1\textwidth]{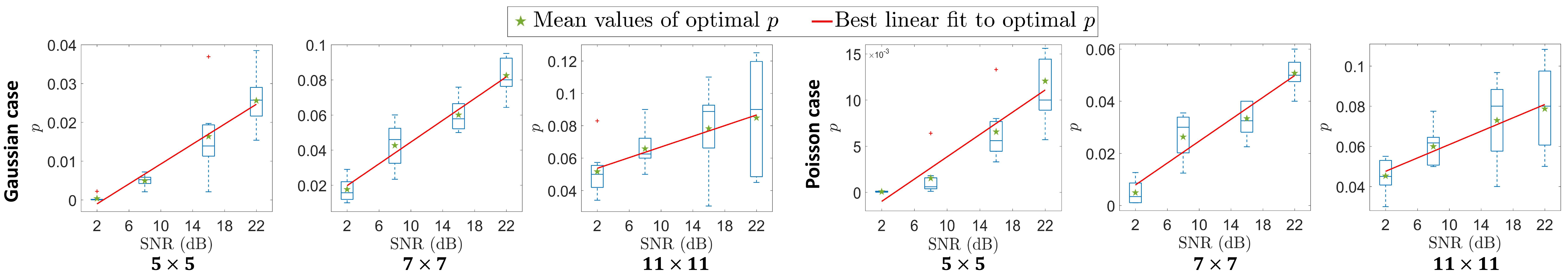}

\caption{Optimal proportionality constant $p$ value as a function of SNR for three patch sizes, where the three left and the three right graphs are associated with the case of Gaussian and Poisson noise models. The bars indicate the minimum and maximum values of the optimal $p$. The bottom and top edges of the blue boxes indicate the $25^{th}$ and $75^{th}$ percentiles and the central mark and green star indicate the median and mean values. The red line is the best linear curve fitted to the data points corresponding to the mean of the optimal $p$ values.}
\label{fig:hy_p_fit}

\end{figure}

As mentioned above, the proportionality constant $p$ regulates the interaction term in the effective potential, and consequently the shape of the basis vectors. Hence, there exists an optimal choice of $p$ depending on the size of the patch for optimal performance of De-QuIP for a given noisy image. Fig.~\ref{subfig:hy_pVSpsnr} presents the denoising performance in terms of PSNR as a function of $p$ for the \textit{house} image corrupted with AWGN (SNR = 16dB), for three different patch sizes. These optimal values also depend on the level of noise present in the image. These $p$ values that maximize the output PSNRs for the first seven sample images corrupted with different noise intensities are highlighted in Fig.~\ref{fig:samples}. These optimal $p$ values are shown as a function of SNR in Fig.~\ref{fig:hy_p_fit} using box-plots for a fixed patch size. The observations confirm that there is a tendency for optimal values to decrease as the noise level increases. For explicit details of these optimal values, we refer readers to the Supp. Mat. file Section~C. A possible explanation for this phenomenon comes from the fact that dissimilarities increase with the noise intensity in a local-neighborhood. Hence, to balance the original potential (patch pixels) and the interactions in the effective potential, the hyperparameter $p$ decreases.

%%%%%%%%%%%%%% hypers VS psnr with diff patchs %%%%%%%%%%%%%%

The data in Fig.~\ref{fig:hy_p_fit} enables rules to fix the $p$ value closer to its optimal values. The distribution of the data gives an intuition about a possible linear relationship between the optimal $p$ and the SNR. Therefore, the proportionality constant $p$ can be chosen from the following rule:
\begin{equation}
p = m_1 \times (\mbox{SNR}) + c_1.
\label{eq:curfit1}
\end{equation}
In Fig.~\ref{fig:hy_p_fit}, the best linear fits to the optimal $p$ as a function of SNR are shown for three different patch sizes as well as for Gaussian and Poisson noise models. These linear fits give a robust way of choosing the suitable $p$ for a given patch size and noise level.

%%%%%%%%%%%%% plot p and curve fitting for p %%%%%%%%%%%%%%

The linear fit parameters are summarized in the Supp. Mat. file Section~C together with the $\ell_2$ error and the resulting average loss in the denoising performance in terms of PSNR and SSIM.
%One can note that the loss in the denoising performance is negligible instead of the optimal choice.
One may notice that the denoising performance loss with rule \eqref{eq:curfit1} rather than the optimal choice is negligible. This is expected due to the smooth nature of the PSNR curve with a broad maxima shown in Fig.~\ref{subfig:hy_pVSpsnr}, which makes the De-QuIP resilient to small sub-optimalities in the adoption of $p$. Hence, it is anticipated that the parameters learned from the sample images to estimate $p$ using \eqref{eq:curfit1}, will be effective for a large set of images. These conclusions are valid for various cases of noise models and patch sizes, as shown in the simulations results. Furthermore, an adaptive approach of tuning $p$ that depends on the image patch gives an alternative to the above rules and opens an interesting perspective for future investigation.

\subsubsection{Influence of the $\hbar ^2/2m$ and subspace dimensionality $d$}
\label{sec:hy_hbar_d}

The last two hyperparameters to be analyzed are $\hbar ^2/2m$ and the subspace dimensionality $d$. Although the utilization of these two hyperparameters seems to be different, the first one being used in the construction of the Hamiltonian operator and the other one acting as a threshold, there is a deep connection between them. In this subsection, we will explain this connection with experimental validation and propose rules for automated estimation of their optimal choices.

As stated above in Subsection~\ref{sec:qt_sub1}, the hyperparameter $\hbar ^2/2m$ controls how the local frequencies of the basis vectors change with the image pixel values. For low values of $\hbar ^2/2m$, the oscillation frequencies are very high, regardless of the low and high pixel values, due to the presence of a very high maximal oscillation in this limit which restricts the wave vectors from properly exploring higher pixel values. On the other side, increasing too much the values of $\hbar ^2/2m$ decreases the ability of the basis vectors to distinguish between high and low values pixels. For more illustrations about the effect of this hyperparameter $\hbar ^2/2m$ on the basis vectors, we refer readers to our previous work \cite{dutta2021quantum}. Therefore, the optimal $\hbar ^2/2m$ value has a strong dependence on the maximum and minimum values of the pixels present in the image patch. Thus, it is more convenient to use an adaptive way to select $\hbar ^2/2m$ that depends on the image patch to have the optimal performance of De-QuIP. Herein, it is possible to write the hyperparameter in terms of the difference between this maximum and minimum pixel values multiplied by a factor $F_{\mbox{factor}}$, for example, for the patch $\bsA$,
\begin{equation}
\hbar ^2/2m = F_{\mbox{factor}} \times ( \bsA_{\mbox{max}} - \bsA_{\mbox{min}} ),
\label{eq:curfit2}
\end{equation}
where $\bsA_{\mbox{max}}$, $\bsA_{\mbox{min}}$ are the maximum and minimum pixel values of the patch $\bsA$. Hence, the optimal choice of $F_{\mbox{factor}}$ is needed to have the best possible output.

\begin{figure}[t!]
\centering
\includegraphics[width=1\textwidth]{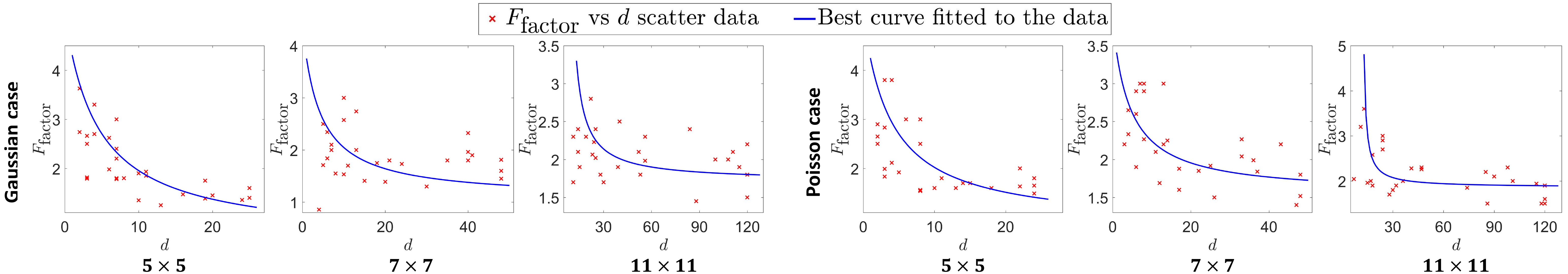}

\caption{$F_{\mbox{factor}}$ vs $d$ scatter plot and the respective best-fitted curve of the form $(F_{\mbox{factor}}-l_1) = l_3/(d-l_2)$.}
\label{fig:hy_hfac_fit}

\end{figure}

In this proposed scheme, the subspace dimensionality $d$ is used as the threshold for truncating high energy wave solutions, which mostly carry noise information. Hence, an optimal choice of $d$ exists for a noisy image that yields the best denoising output depending on the patch size. $\hbar ^2/2m$ or say $F_{\mbox{factor}}$ controls the frequency distribution across the basis vectors since the maximal frequency of a vector with energy $E$ at the local pixel value $V$ is $\sqrt{(E-V)/(\hbar ^2/2m)}$. Hence, the maximal frequency decreases with increasing $F_{\mbox{factor}}$. As a consequence, low-energy basis vectors become more prominent to distinguish low and high pixel regions using different levels of frequency. Thus, the optimal subspace dimensionality $d$ decreases as $F_{\mbox{factor}}$ increases. These optimal choices vary with the image patch size and noise statistics.
%Table~\ref{tab:tab_planckdata} and Table~\ref{tab:tab_subspadim} show these optimal values that give the best output PSNRs for the first seven sample images.
In Fig.~\ref{fig:hy_hfac_fit}, all these optimal values that give the best output PSNRs for the first seven sample images are shown as a scatter-plot of $F_{\mbox{factor}}$ vs $d$, which clearly shows their inverse relationship, \textit{i.e.}, $d$ decreases with $F_{\mbox{factor}}$'s growth or vice-versa and validates our above arguments. More details of these optimal values can be found in the Supp. Mat. file Section~D.

%%%%%%%%%%%%% plot d vs \hbar_factor and curve fitting %%%%%%%%%%%%%%

%%%%%%%%%%%% plot d and curve fitting for d %%%%%%%%%%%%%

\begin{figure}[t!]
\centering
\includegraphics[width=1\textwidth]{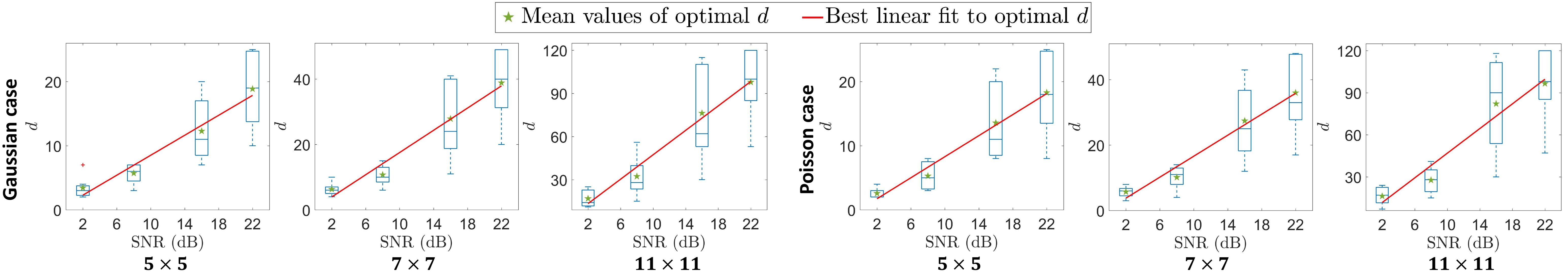}

\caption{Optimal subspace dimensionality $d$ value as a function of SNR for three patch sizes are shown for the Gaussian and Poisson noise models. Similar to Fig.~\ref{fig:hy_p_fit}, a box-plot diagram is used for the optimal $d$. The red line is the best linear curve fitted to the data points corresponding to the mean of the optimal $d$.}
\label{fig:hy_d_fit}

\end{figure}

% these optimal $p$ values are shown as a function of SNR in Fig.~\ref{fig:hy_p_fit} using box-plots for a fixed patch size. The observations confirm that there is a tendency for optimal values to decrease as the noise level increases.

These experimental data enable an automated way of selecting the values of $d$ and $F_{\mbox{factor}}$ close to their optimalities. To do this, the optimal $d$ values are shown in Fig.~\ref{fig:hy_d_fit} as a function of SNR using box-plots for a fixed patch size, for the Gaussian and Poisson cases. The observation shows a very predictable behaviour of this optimal $d$ as a function of SNR which is expected as it needs to be further thresholded as the noise increases.
%A similar observation can be made for the Poisson model, whose data are available in the Supp. Mat. file Fig.~1.
For a specific patch size, the optimal $d$ and SNR follow a linear relationship. Therefore, the subspace dimensionality $d$ and $F_{\mbox{factor}}$ can be inferred from the following two rules,
\begin{equation}
d = m_2 \times (\mbox{SNR}) + c_2 ,
\label{eq:curfit3}
\end{equation}
\begin{equation}
F_{\mbox{factor}} - l_1 = l_3 / ( d - l_2 ).
\label{eq:curfit4}
\end{equation}
Figs.~\ref{fig:hy_hfac_fit} and Fig.~\ref{fig:hy_d_fit} show the best-fitted curves to the optimal $F_{\mbox{factor}}$ and $d$, and the respective fit parameters are regrouped in the Supp. Mat. file Section~D. These rules give an efficient way of selecting the hyperparameters close to their optimality depending on the size of the given patch and the intensity of the noise. Our data show that the respective costs in terms of performance loss are minimal, since the output PSNR curves are smooth and have broad maxima, shown in Fig.~\ref{subfig:hy_factVSpsnr}-\ref{subfig:hy_dVSpsnr} for the choice of $F_{\mbox{factor}}$ and $d$, as discussed in Subsection~\ref{sec:hy_p} for the hyperparameter $p$. Hence, the rules for automated selecting hyperparameters are expected to be valid for other images as well.

% Moreover, the respective costs in terms of performance loss are minimal as shown in Table~\ref{tab:tab_hbar_d_fit}. This is because of the broad maxima of the output PSNR curves with respect to the optimal $\hbar_{\mbox{factor}}$ and $d$, as discussed in Subsection~\ref{sec:hy_p} for the hyperparameter $p$.

% For low values of $\hbar ^2/2m$, the oscillation frequencies are very high irrespective of the low and high pixel values due to the presence of a maximal oscillation period in this limit that restricts the wave vectors to prob higher pixel values properly.

\subsection{Denoising efficiency of the proposed scheme in comparison with standard methods}
\label{sec:comparison}

This subsection presents the denoising performance of the De-QuIP algorithm depending on the noise statistics and intensity, and also how this performance varies with patch size for the sample images. The denoising outputs using three patch sizes are summarized in the Supp. Mat. file, see Section~E. The numerical simulations show that $11 \times 11$ is the suitable patch size for most of the cases, but for low-level noise, smaller sizes give a small advantage. It is expected to have a better result with a large patch for a strong noise scenario since high noise intensity refers to an extreme random system and a large patch is more efficient to capture the similarity measures from this strong randomness. Obviously, the size should not be so large because it is affected by the phenomenon of localization, as illustrated in Subsection~\ref{sec:lolizprobqmpi}.

As explained earlier, the De-QuIP follows a similar principle to the NLM approach. Comparisons with NLM-based state-of-the-art methods are thus provided in order to prove the efficiency of the proposed algorithm. However, for a comprehensive survey of the denoising ability of De-QuIP, rigorous comparisons with contemporary noise removal methods from the literature are also presented. For the recovery of Gaussian corrupted images, the following methods were used for comparison: NLM method using PCA called PND in \cite{tasdizen2009principal}, two patch-based PCA for NLM denoising methods referred to as PGPCA (global approach) and PLPCA (local approach) in \cite{deledalle2011image}, BM3D \cite{Dabov2007Image}, dictionary learning (DL) method \cite{Elad2006image}, graph signal processing (GSP) method \cite{Pang2017graph}, and finally, our earlier implementations of quantum adaptive basis (QAB) for image denoising based on the single particle theory \cite{dutta2021quantum}.

For the recovery of Poisson corrupted images, comparisons have been carried out with recent algorithms dedicated to the Poissonian model such as Poisson non-local PCA (PNLPCA) \cite{salmon2014poisson}, BM3D consolidated with the Anscombe transform \cite{Makitalo2011optimal} leveled as ATBM3D, and finally the QAB \cite{dutta2021quantum} method.

\begin{figure*}[t!]
\centering
\includegraphics[width=1\textwidth]{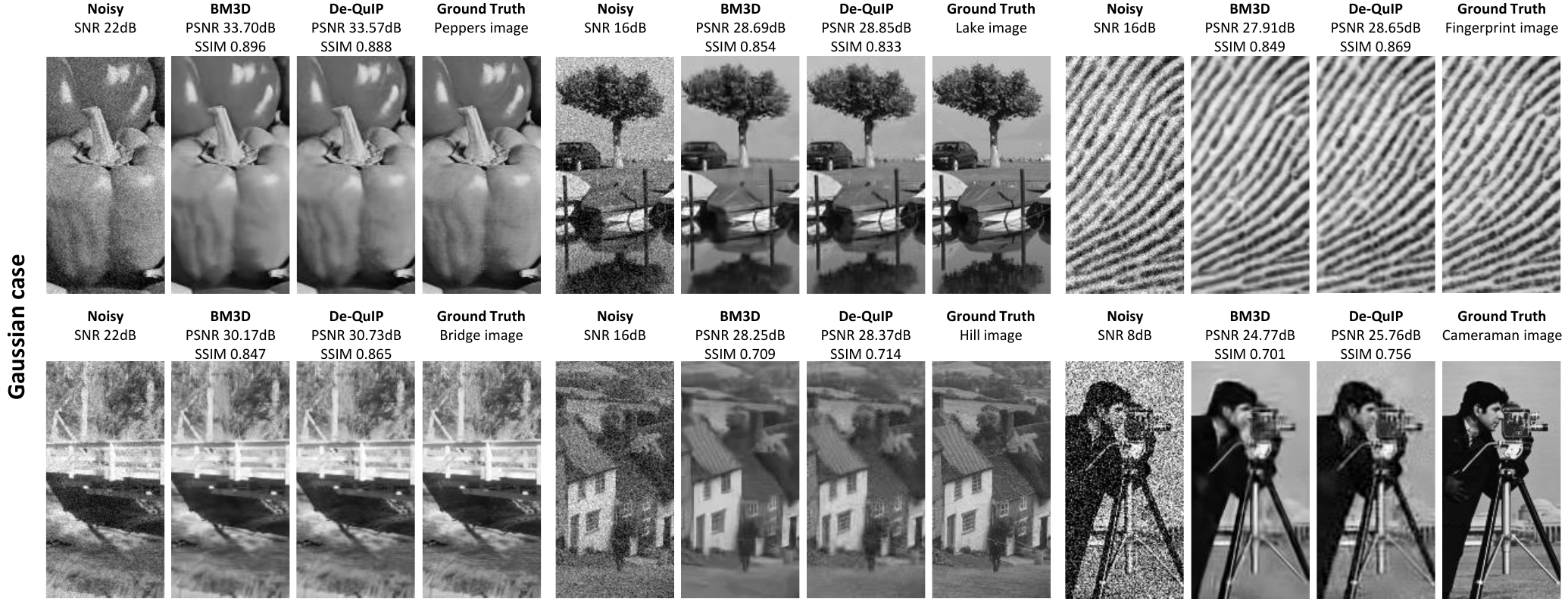}
\vspace{2mm}

\includegraphics[width=1\textwidth]{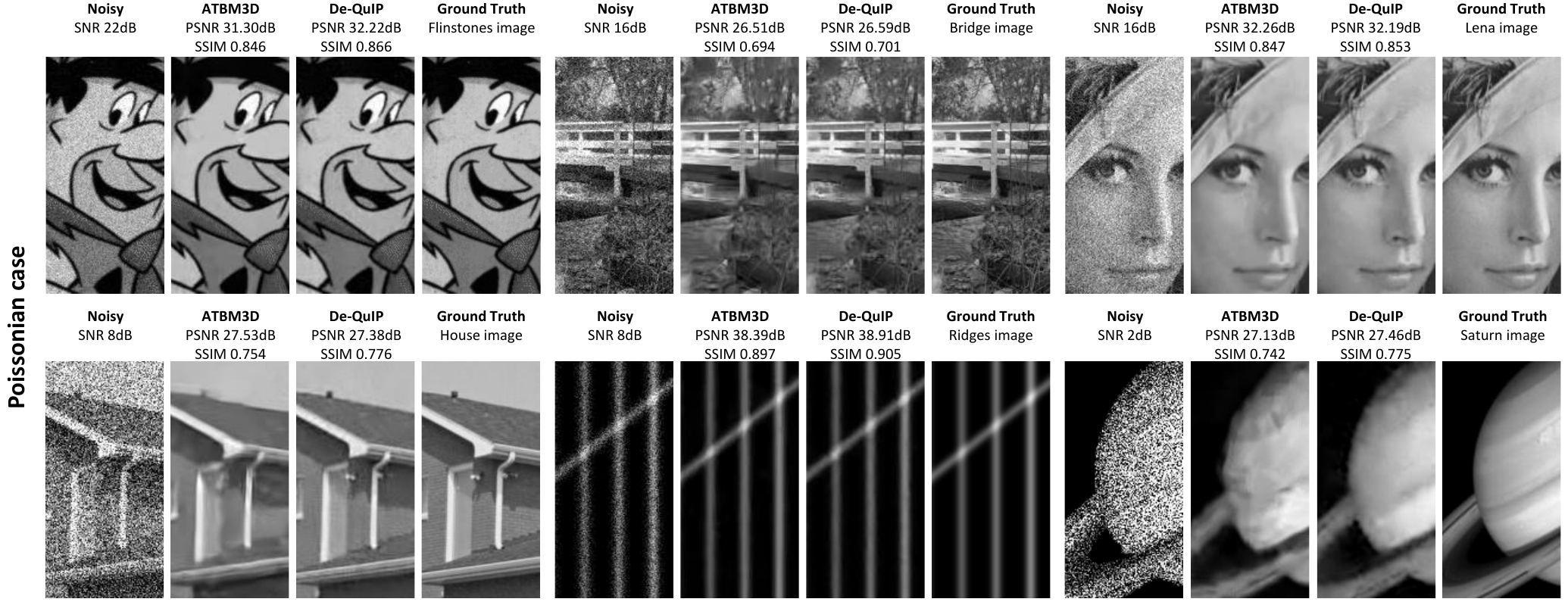}

\caption{The top two and bottom two rows depict the Gaussian and Poissonian denoising results, respectively. The noisy, BM3D-based results, De-QuIP results, and ground-truth images are presented here accordingly. The BM3D-based and De-QuIP schemes are listed as these are always among the two best-performing methods.}
\label{fig:result_1GP}

\end{figure*}

%%%%%%%%%%%% denoised comparisons %%%%%%%%%%%%%

\begin{figure*}[t!]
\centering
\includegraphics[width=1\textwidth]{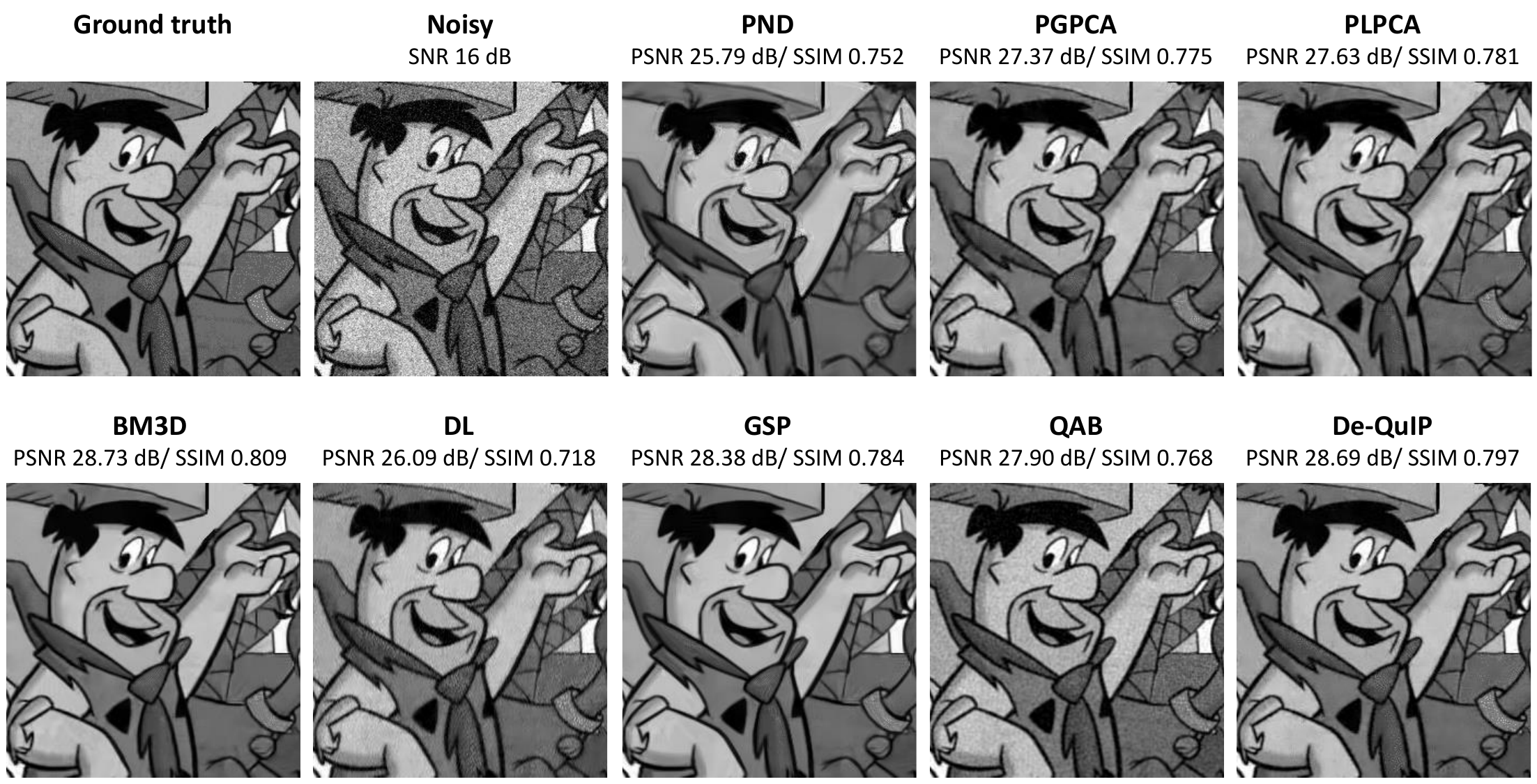}

\caption{Zoomed segments of the denoised estimations of the \textit{Flintstones} image while corrupted with AWGN (SNR $16$dB) using different methods.}
\label{fig:result_1G}

\end{figure*}

\begin{figure*}[t!]
\centering
\includegraphics[width=1\textwidth]{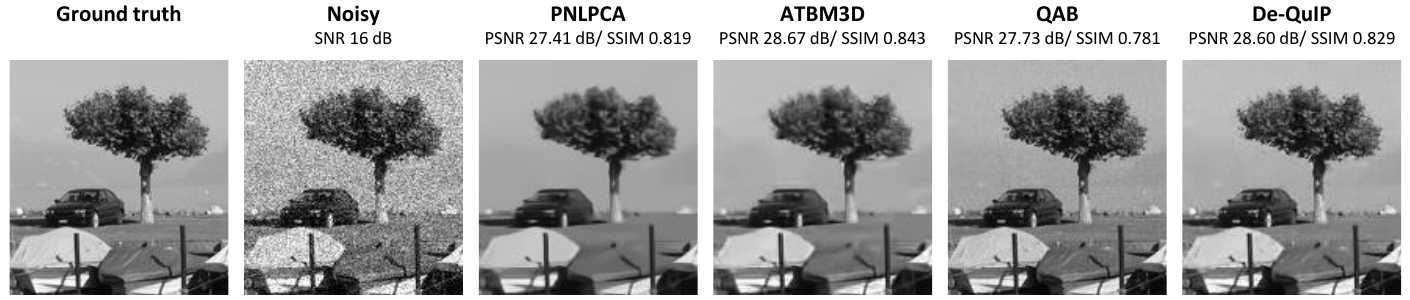}

\caption{Zoomed segments of the denoised estimations of the \textit{Lake} image while corrupted with Poisson noise (SNR $16$dB) using different methods.}
\label{fig:result_1P}

\end{figure*}

%%%%%%%%%%%% denoising results box-plot %%%%%%%%%%%%%

\begin{figure*}[t!]
\centering
%\textbf{Gaussian corrupted images}
\subfigure[Recovery of Gaussian corrupted images]
{\label{subfig:rescomp_boxplotG}
\includegraphics[width=.48\textwidth]{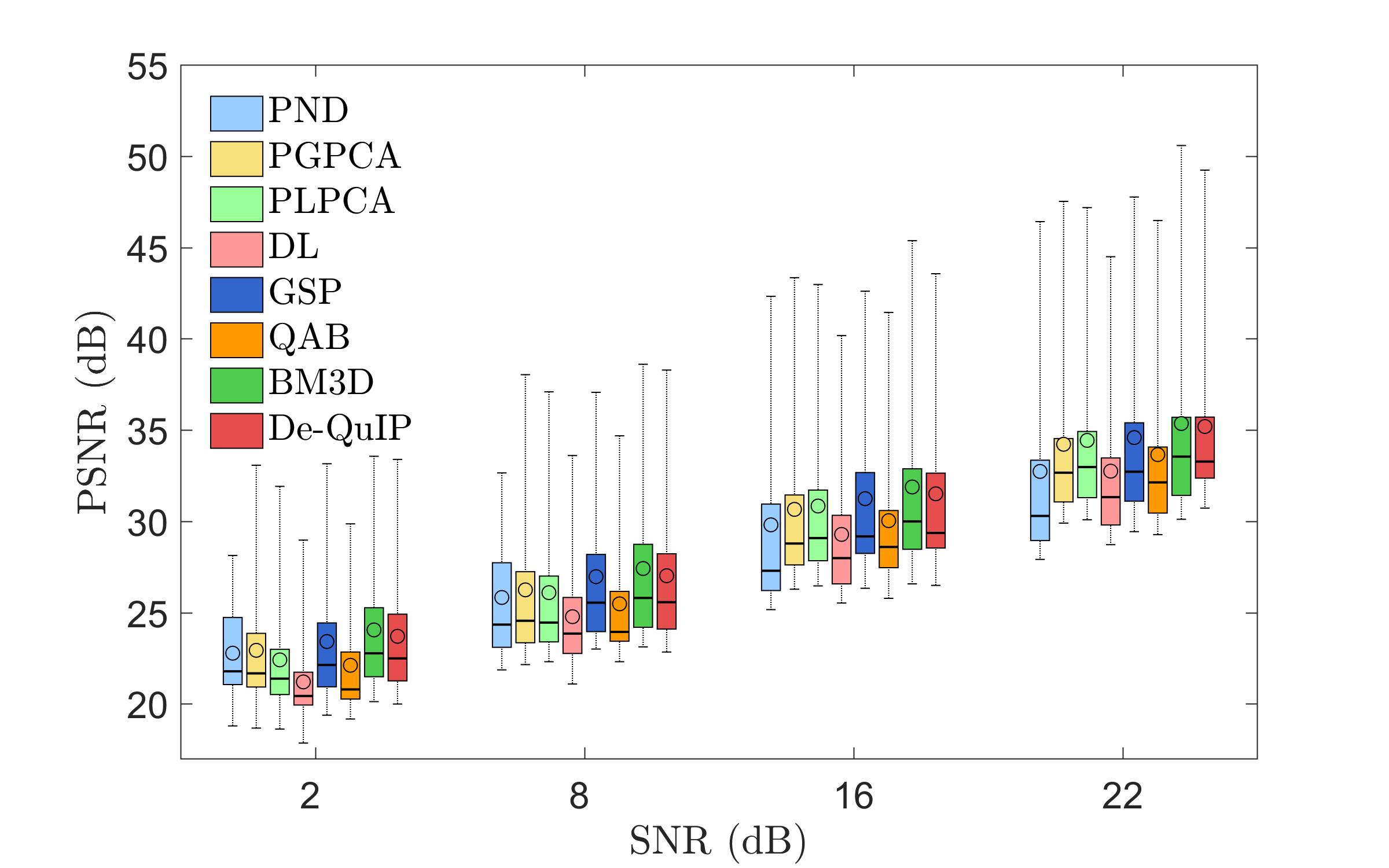}
\includegraphics[width=.48\textwidth]{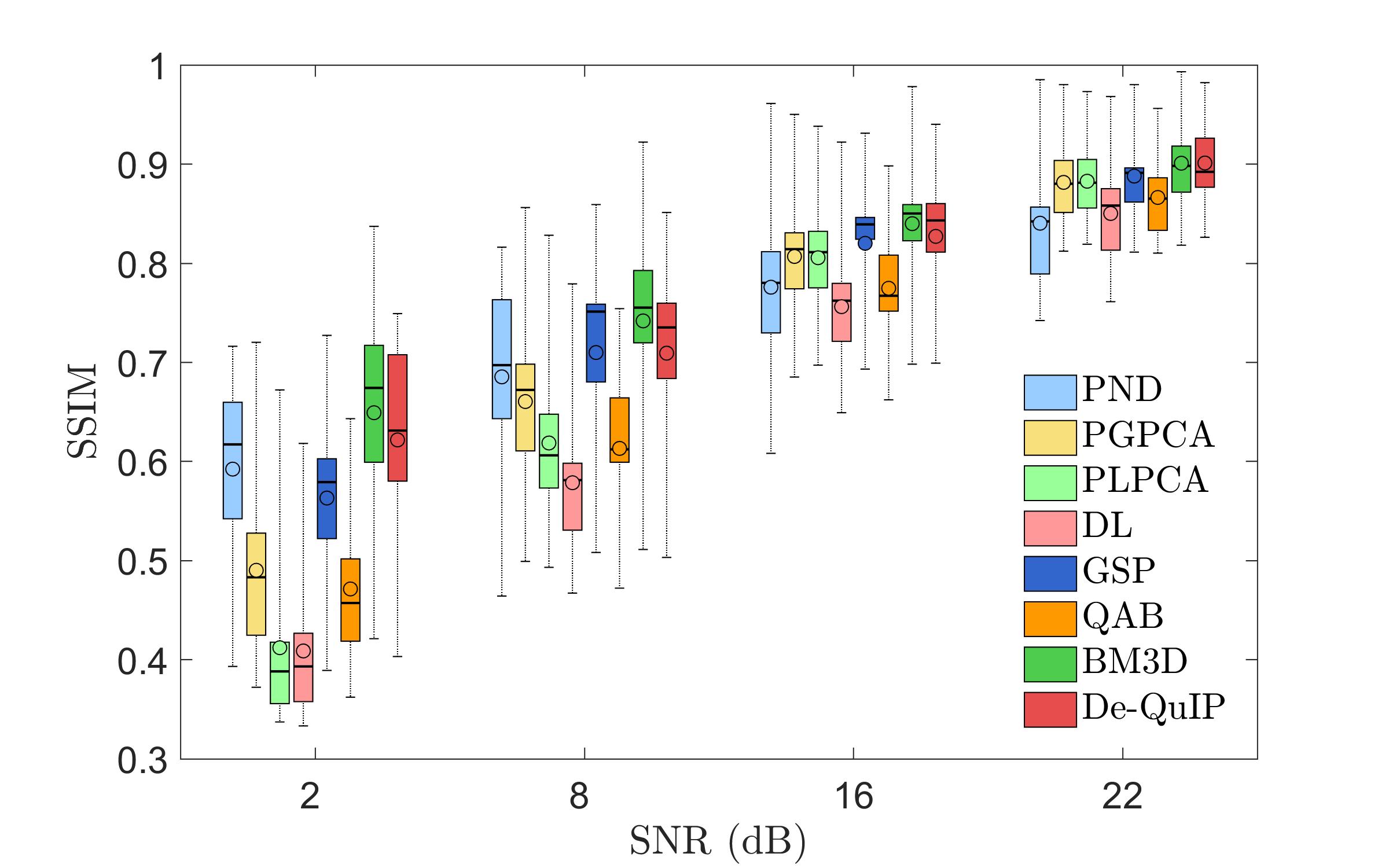}
}

%\textbf{Poisson corrupted images}
\subfigure[Recovery of Poisson corrupted images]
{\label{subfig:rescomp_boxplotP}
\includegraphics[width=.48\textwidth]{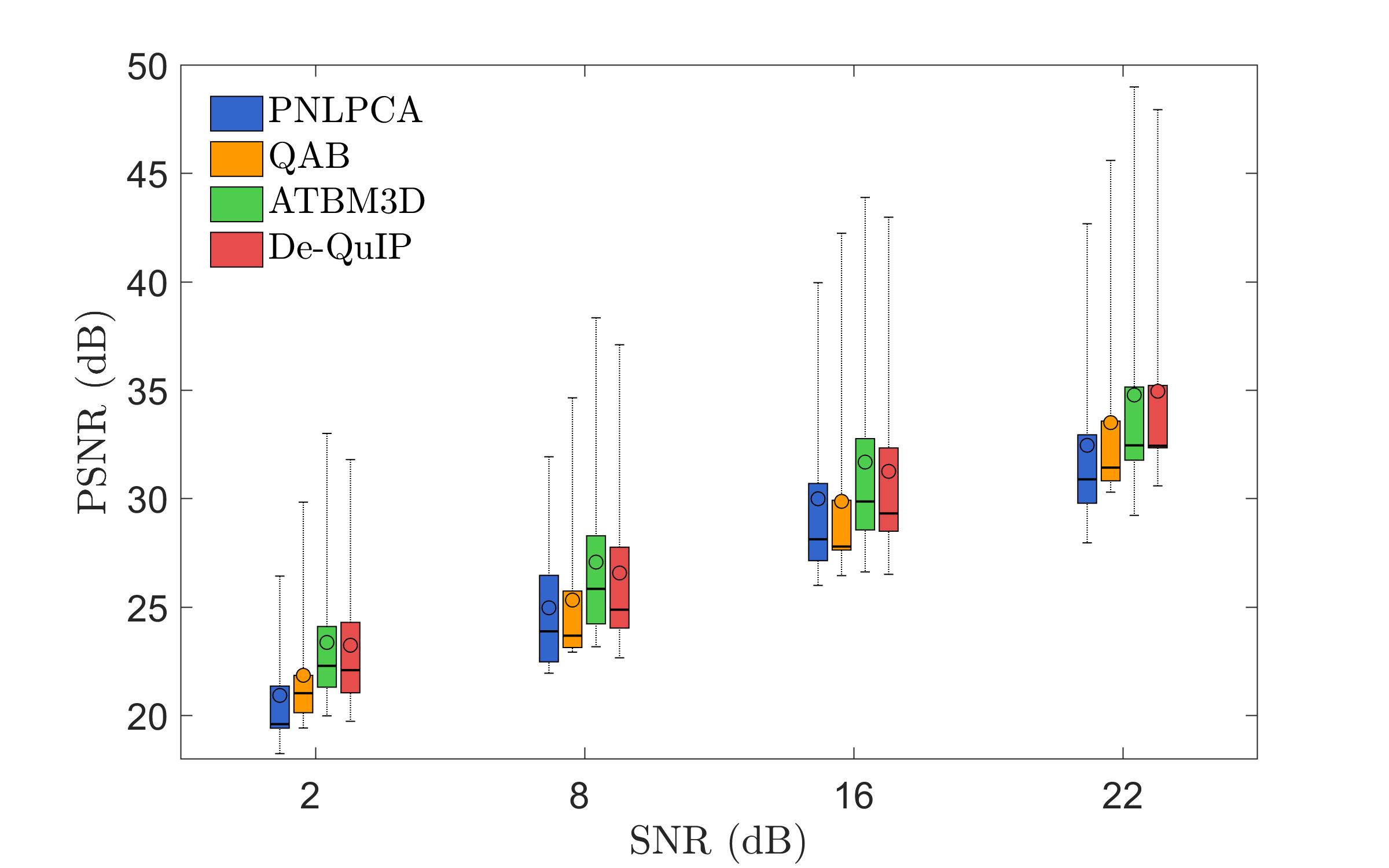}
\includegraphics[width=.48\textwidth]{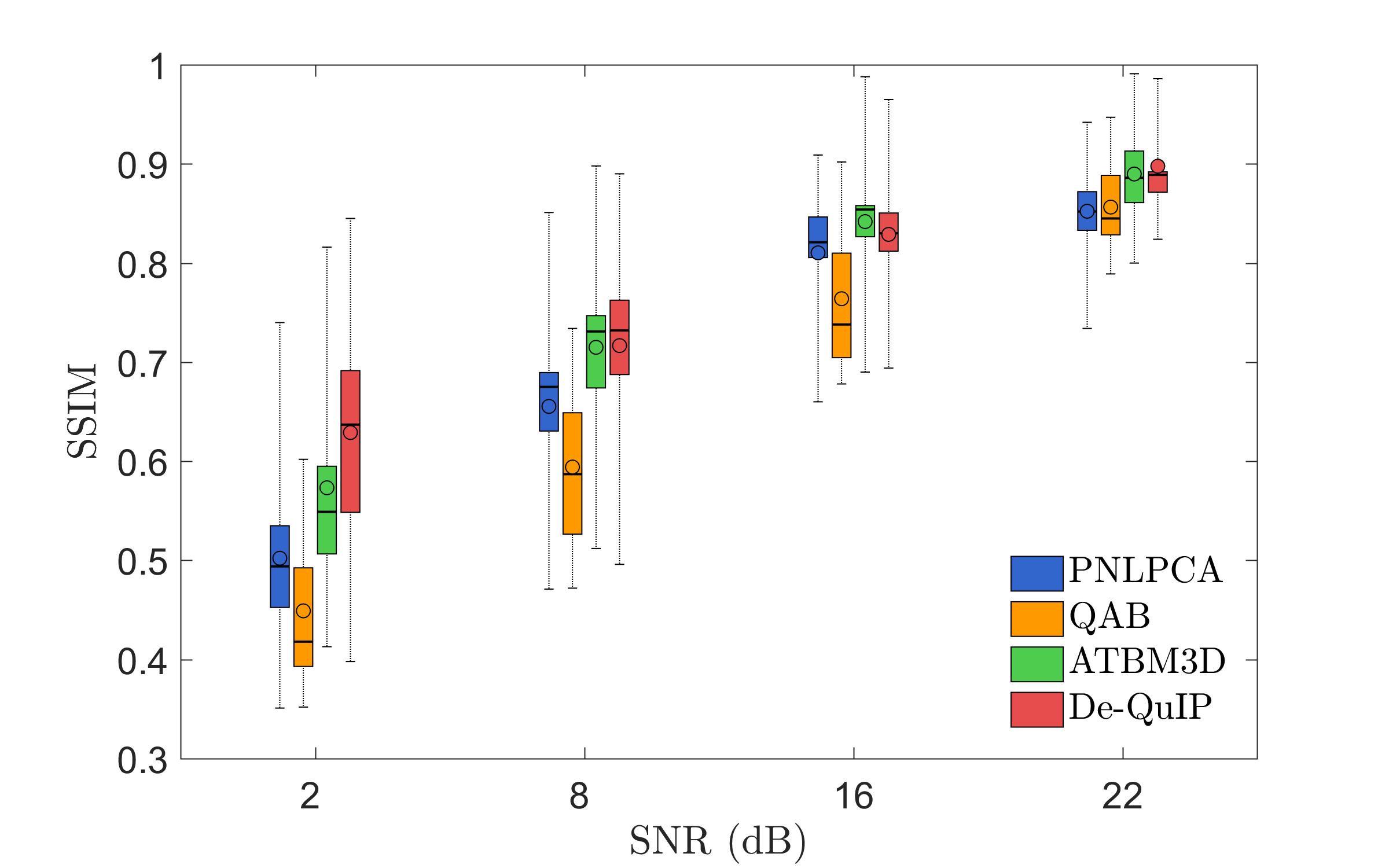}
}

\caption{Quantitative denoising results using different methods for Gaussian and Poisson corrupted images with four different noise levels. The bottom and top edges of the boxes indicate the $25^{th}$ and $75^{th}$ percentiles, and the central black line and circle indicate the median and mean relative to the data points.} 
% Here the bars demonstrate the minimum and maximum values.
\label{fig:rescomp_boxplot}

\end{figure*}

% Extensive comparisons have been conducted to demonstrate the potential of our proposed denoising algorithm De-QuIP against other standard methods.

The proposed denoising algorithm has been extensively compared to other standard methods to demonstrate the accuracy of De-QuIP. Detailed quantitative evaluations in terms of PSNR and SSIM for both noise models are available in the Supp. Mat. file, see Section~E. From these results, on can observe that De-QuIP scheme significantly outperforms all NLM-based methods with an average gain of $1.1$ to $2.6$ dB in PSNR and $3$ to $20$\% in SSIM. One can also observe that De-QuIP and BM3D-based methods stand out as the two best-performing algorithms for both Gaussian and Poissonian cases. The denoising performance of De-QuIP and BM3D-based methods are presented in Fig.~\ref{fig:result_1GP} for visual inspection, where ground truth, noisy, and corresponding denoised images are shown. These results confirm the good performance of De-QuIP regardless of the noise model and intensity. In the denoised images, image features and details, for example, patterns (in \textit{Fingerprint} and \textit{Ridges}), sharp edges (in \textit{Lake}, \textit{Bridge}, \textit{Cameraman} and \textit{House}), smooth areas (in \textit{Peppers}, \textit{Flintstones}, and \textit{Lena}), are well-preserved. Although BM3D and respectively ATBM3D are slightly more accurate in some of the experiments, a smoothing effect is present in their corresponding denoised images and becomes more prominent as the noise level increases. This effect is clearly visible around the windows and roof of the \textit{Hill}, on the patterns of the \textit{Fingerprint}, near the eye of the \textit{Lena}, on the face of the \textit{Flintstones}, and around the sharp edges of the \textit{House} images, while De-QuIP preserves all these image features in a better way and consequently provides a denoised image closer to the original one. This is due to the interaction term that allows De-QuIP to better extract the image information. Although, for the increasing noise intensity, some artefacts can be observed in the denoised images (for example in \textit{House}, \textit{Saturn}, \textit{Cameraman} images) due to the presence of strong noise, they are very few and negligible for low-level noises.

For further inspection in Fig.~\ref{fig:result_1G}, we show zoomed segments of the denoised results of the \textit{Flintstones} image while corrupted with AWGN (SNR $16$dB). Similarly, for Poisson corrupted (SNR $16$dB) \textit{Lake} image the zoomed segments of the denoised estimations are shown in Fig.~\ref{fig:result_1P}. Quantitative performance in terms of PSNR and SSIM adopting different methods for Gaussian and Poisson contaminated images are presented  in Fig.~\ref{fig:rescomp_boxplot} using box-plots as a function of SNR. Readers may refere to the Supp. Mat. file to have more details related to these experiments.

%Table~\ref{tab:tab_psnrssimG} and Table~\ref{tab:tab_psnrssimP} compare quantitative performance in terms of PSNR and SSIM in the case of Gaussian and Poisson, respectively.

Through visual and quantitative inspections of Figs.~\ref{fig:result_1G}-\ref{fig:result_1P}, it is clear that the proposed De-QuIP uniformly outperforms all the NLM-based approaches with a significant increase in terms of PSNR and SSIM. For Gaussian corrupted images, BM3D is still the best method in most cases, but De-QuIP allows competitive comparisons in all scenarios. Furthermore, for both noise models, the positive effects of local similarity considerations are clearly visible in the Fig.~\ref{fig:rescomp_boxplot} data of QAB and De-QuIP, as it gives much better PSNR and SSIM values with significantly fewer computations. The Figs.~\ref{fig:result_1G}-\ref{fig:result_1P} show a pronounced gain in the qualitative performance of the proposed DeQuIP model against the QAB. Therefore, exploiting the structural details through interaction terms notably contributes to the preservation of image details, as conveyed by quantitative and visual assessments. Additionally, regardless of the noise intensity, De-QuIP always provides good PSNR and SSIM for the recovery of Gaussian corrupted images as shown in Fig~\ref{subfig:rescomp_boxplotG}, which is not the case with most algorithms, as highlighted by SSIM values. Although in Fig.~\ref{subfig:rescomp_boxplotG}, our results look comparable with BM3D, one should note that a beautification happens in the BM3D outputs due to the smoothing effect as illustrated in Fig.~\ref{fig:result_1GP}, which is not present in our outcomes and makes our resultant image texture closer to the original one.

For Poisson corrupted images, De-QuIP provides better outcomes compared to the other methods. ATBM3D generates comparable PSNR and SSIM data in some scenarios, but the visual assessment clearly shows an extra smoothing effect present on the denoised image, which causes lower SSIM values for low SNR images as shown in Fig.~\ref{subfig:rescomp_boxplotP}.
This is due to the process of data Gaussianization through the Anscombe transformation. In addition, for increasing noise intensity, this Anscombe transformation loses its accuracy \cite{dutta2021plug}, which is clearly observable in the Fig.~\ref{subfig:rescomp_boxplotP} in the cases of low SNR. On contrary. De-QuIP is a straightforward method without having any such transformation and efficiently shows good denoising performance in all situations. Similar to the Gaussian case, De-QuIP outperforms PNLPCA, a NLM based method, by a large margin. This proves its adaptability for high as well as for low SNR images regardless of their noise statistics which can be viewed as a strong point in several practical applications. %This indicates the reliable applicability of De-QuIP for denoising in a variety of situations.

%This indicates that the De-QuIP is reliably applicable for denoising context in a variety of situations.

%%%%%%%%%%%% US despeckling %%%%%%%%%%%%%

\begin{figure}[t!]
\centering
\includegraphics[width=.65\textwidth]{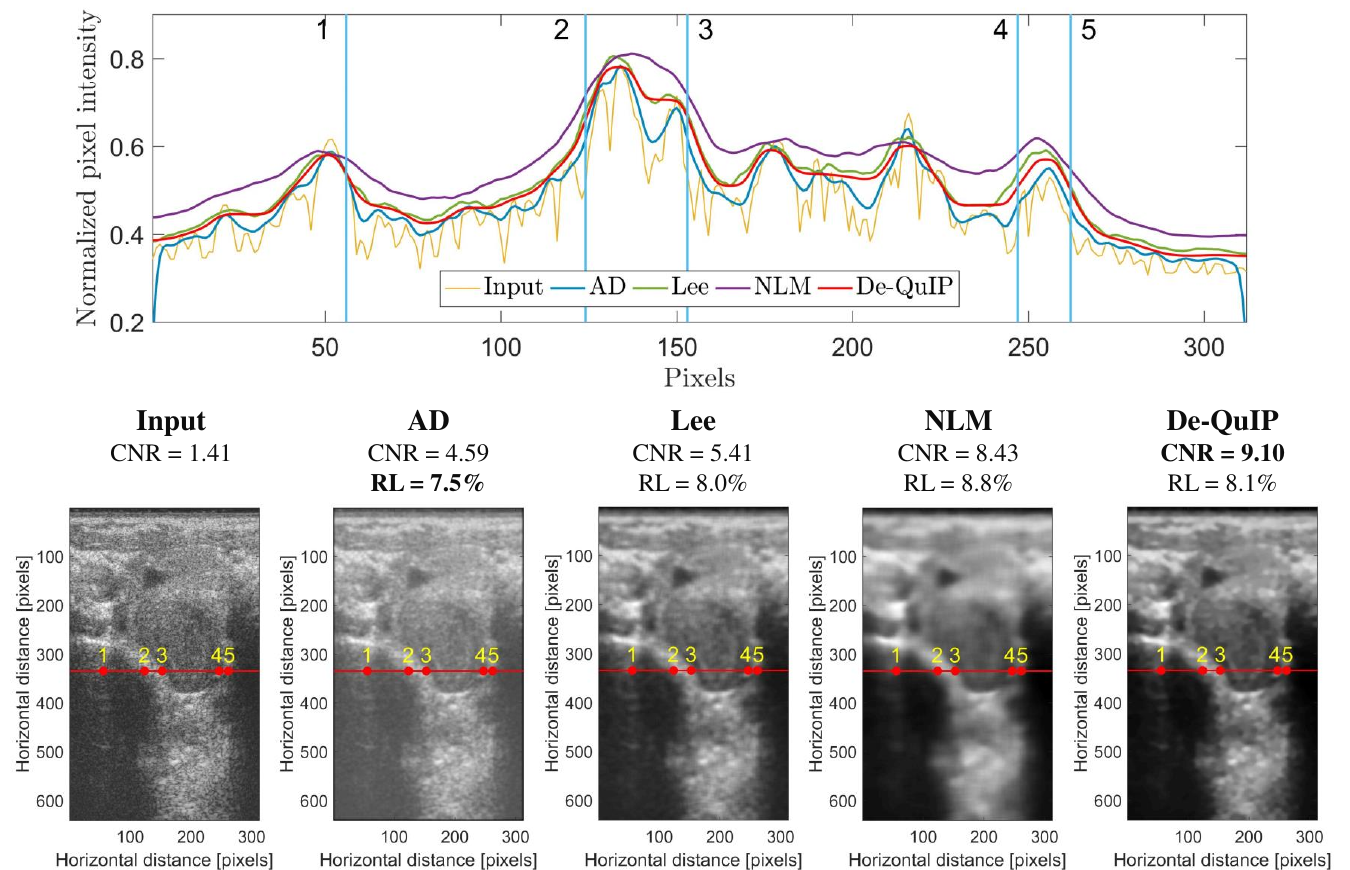}

\caption{US image despeckling results using different methods for cancer thyroid image. The normalized pixel intensities of the extracted red lines from speckled and despeckled US images are shown.}

\label{fig:result_US}
\end{figure}

\subsection{Application to ultrasound (US) image despeckling}
\label{sec:usdesp}

In this subsection, for further illustration of the potential of De-QuIP, we investigated its ability for real medical US image despeckling. US imaging is an integral part of modern medical science as it gives harmless, non-invasive, real-time images in an affordable way. The main artefact affecting US images is a random granular pattern, the speckle, which is generated by random constructive and destructive interference between US waves. This phenomenon related to the acquisition system is used as a source of information about the tissues in several applications, but can also affect the interpretability of the images by diminishing their readability. Indeed, the speckle does not follow an additive rule and has a complex noise distribution. Therefore, the important task of estimating speckle-free US images, known as despeckling \cite{Liu2021SAR} in the relevant literature, has been extensively explored using various schemes \cite{Yu2002speckle, Santos2017ultrasound, Achim2001novel, Lee1980digital} to enhance the readability of the US images.

Despeckling performance of De-QuIP is investigated through a phantom as well as four real cancer and two non-cancer thyroid US images acquired with a 7.5 MHz linear probe. As an extension of our preliminary results in \cite{dutta2021despeckling}, we are proposing a comprehensive study of this problem here. The estimated despeckled outcomes are compared with three existing despeckling algorithms, the anisotropic diffusion (AD) \cite{Yu2002speckle}, Lee \cite{Lee1980digital} and NLM \cite{tasdizen2009principal} filters. For the quantitative analysis, the contrast-to-noise-ratio (CNR) and resolution loss (RL) are regrouped for a cancer image with visual demonstrations in Fig.~\ref{fig:result_US}. Observation shows that De-QuIP offers a better image contrast (higher CNR than AD, Lee and slightly lower than NLM, which over-smooths the images and yields poor resolution) while having less spatial resolution loss (De-QuIP has less spatial resolution loss compared to the native US image). Note that these images are chosen arbitrarily, that is, the quality of the results should not depend on the data tested. More data related to these experiments can be found in the Supp. Mat. file, see Section~F.

\section{Conclusions and Perspectives}
\label{sec:conclusion}

A novel image denoising  algorithm inspired by the quantum many-body theory has been developed in this paper. This gives a way to adapt the concept of interaction from the many-body physics to an imaging problem. More precisely, the interactions between image patches are nothing more than a reflection of the similarity-measures in a local image neighborhood and provide an efficient way to capture the local structures of real images. Through these interactions, structural details are transmitted on a patch-based adaptive basis created by the solutions of the Schr\"odinger equation of quantum mechanics, which can be exploited as filters for denoising the patches. The versatile nature of the adaptive basis that conveys the structural similarities of image neighborhood, extends its scope of applications beyond AWGN without modification.

A rigorous comparison with contemporary methods exemplifies the denoising ability of our De-QuIP algorithm regardless of the image nature, noise statistics and intensity. Simulation results demonstrate that the proposed method clearly outperforms other schemes and gives a good comparison with the best outcome for both image independent and dependent noise models. Additionally, De-QuIP achieves much better results at a significantly less computational cost in comparison with the earlier single-particle based quantum scheme of e.g. \cite{dutta2021quantum}. To make De-QuIP more robust, automated rules are discussed in the paper to efficiently select the values of the hyperparameters close to the optimal ones when less information is available. Furthermore, its good performance in real-life medical US image despeckling applications further shows its ability in handling multiplicative noise efficiently.

Adaptation of this new quantum many-body idea opens up a new domain of future explorations. Since De-QuIP primarily has a non-local nature and significantly outperforms contemporary NLM-based methods, the first obvious perspective comes from the extension of this idea of interactions for collaborative patch denoising, as originally proposed in \cite{Dabov2007Image}.

A second interesting point would be to embed this interaction architecture into a convolutional neural network, as explored with various schemes, such as a fast flexible learning method \cite{Chen2017trainable, Zhang2018ffdnet}, residual learning \cite{Zhang2017beyond} and others, and study imaging problems through this many-body network where each node shows interaction with others.

Another possibility is to use a more advanced theory of physics, such as the time-dependent Schr\"odinger equation, a fascinating perspective for further research.

Further expansion of the framework in three dimensional data or RGB color images can be easily done by simply bypassing data across different processing channels. Finally, other imaging applications, such as deconvolution or super-resolution, could also take benefit of this interaction model.
%\dk{The end of this sentence is not usefull. It is removed }
% through, for instance, Plug-and-Play \cite{dutta2021plug} or Regularization by Denoising schemes.

%\sd{In recent years deep learning, especially convolutional neural network (CNN) based architectures becoming more attractive in this image denoising domain for their large modeling capacity and robust training procedure and have been explored with various schemes, such as a fast flexible learning method \cite{Chen2017trainable}, residual learning \cite{Zhang2017beyond}, a fast and flexible denoising with a tunable noise level \cite{Zhang2018ffdnet} and others. Note that our method does not belong to this deep learning paradigm and follows a conventional way. However, embedding this many-body interaction architecture into a CNN provides an intriguing direction for future research where each node shows interaction with others.}

% To make De-QuIP more robust, automated rules to efficiently select the hyperparameters values close to the optimal when less informations are available, are discussed in the paper

\section*{Acknowledgments}
\label{sec:acknowledge}

%\noindent 
This work was supported in part by Centre National de la Recherche Scientifique (CNRS) through the 80 Prime Program.

\bibliographystyle{elsarticle-num}

\bibliography{DeQuIP2022}

\begin{thebibliography}{10}
\expandafter\ifx\csname url\endcsname\relax
  \def\url#1{\texttt{#1}}\fi
\expandafter\ifx\csname urlprefix\endcsname\relax\def\urlprefix{URL }\fi
\expandafter\ifx\csname href\endcsname\relax
  \def\href#1#2{#2} \def\path#1{#1}\fi

\bibitem{Hamza2001image}
A.~Hamza, H.~Krim, Image denoising: a nonlinear robust statistical approach,
  IEEE Trans. Signal Process. 49~(12) (2001) 3045--3054.
\newblock \href {https://doi.org/10.1109/78.969512}
  {\path{doi:10.1109/78.969512}}.

\bibitem{Pizurica2006estimating}
A.~Pizurica, W.~Philips, Estimating the probability of the presence of a signal
  of interest in multiresolution single- and multiband image denoising, IEEE
  Trans. Image Process. 15~(3) (2006) 654--665.
\newblock \href {https://doi.org/10.1109/TIP.2005.863698}
  {\path{doi:10.1109/TIP.2005.863698}}.

\bibitem{Lebrun2013anonlocal}
M.~Lebrun, A.~Buades, J.~Morel, A nonlocal bayesian image denoising algorithm,
  SIAM J. Imag. Sci. 6~(3) (2013) 1665--1688.
\newblock \href {https://doi.org/10.1137/120874989}
  {\path{doi:10.1137/120874989}}.

\bibitem{Shuman2013the}
D.~Shuman, S.~Narang, P.~Frossard, A.~Ortega, P.~Vandergheynst, The emerging
  field of signal processing on graphs: Extending high-dimensional data
  analysis to networks and other irregular domains, IEEE Signal Process. Mag.
  30~(3) (2013) 83--98.
\newblock \href {https://doi.org/10.1109/MSP.2012.2235192}
  {\path{doi:10.1109/MSP.2012.2235192}}.

\bibitem{Sandryhaila2013discrete}
A.~Sandryhaila, J.~Moura, Discrete signal processing on graphs, IEEE Trans.
  Signal Process. 61~(7) (2013) 1644--1656.
\newblock \href {https://doi.org/10.1109/TSP.2013.2238935}
  {\path{doi:10.1109/TSP.2013.2238935}}.

\bibitem{Pang2017graph}
J.~Pang, G.~Cheung, Graph laplacian regularization for image denoising:
  Analysis in the continuous domain, IEEE Trans. Image Process. 26~(4) (2017)
  1770--1785.
\newblock \href {https://doi.org/10.1109/TIP.2017.2651400}
  {\path{doi:10.1109/TIP.2017.2651400}}.

\bibitem{Kim2006pde}
S.~Kim, Pde-based image restoration: a hybrid model and color image denoising,
  IEEE Trans. Image Process. 15~(5) (2006) 1163--1170.
\newblock \href {https://doi.org/10.1109/TIP.2005.864184}
  {\path{doi:10.1109/TIP.2005.864184}}.

\bibitem{Liu2012remote}
P.~Liu, F.~Huang, G.~Li, Z.~Liu, Remote-sensing image denoising using partial
  differential equations and auxiliary images as priors, IEEE Geosci. Remote
  Sens. Lett. 9~(3) (2012) 358--362.
\newblock \href {https://doi.org/10.1109/LGRS.2011.2168598}
  {\path{doi:10.1109/LGRS.2011.2168598}}.

\bibitem{donoho1994ideal}
D.~Donoho, I.~Johnstone, Ideal spatial adaptation by wavelet shrinkage,
  Biometrika 81~(3) (1994) 425--455.
\newblock \href {https://doi.org/10.1093/biomet/81.3.425}
  {\path{doi:10.1093/biomet/81.3.425}}.

\bibitem{starck2002curvelet}
J.-L. Starck, E.~Candes, D.~Donoho, The curvelet transform for image denoising,
  IEEE Trans. Image Process. 11~(6) (2002) 670--684.
\newblock \href {https://doi.org/10.1109/TIP.2002.1014998}
  {\path{doi:10.1109/TIP.2002.1014998}}.

\bibitem{Aharon2006an}
M.~Aharon, M.~Elad, A.~Bruckstein, K-svd: An algorithm for designing
  overcomplete dictionaries for sparse representation, IEEE Trans. Signal
  Process. 54~(11) (2006) 4311--4322.
\newblock \href {https://doi.org/10.1109/TSP.2006.881199}
  {\path{doi:10.1109/TSP.2006.881199}}.

\bibitem{Elad2006image}
M.~Elad, M.~Aharon, Image denoising via sparse and redundant representations
  over learned dictionaries, IEEE Trans. Image Process. 15~(12) (2006)
  3736--3745.
\newblock \href {https://doi.org/10.1109/TIP.2006.881969}
  {\path{doi:10.1109/TIP.2006.881969}}.

\bibitem{Dabov2007Image}
K.~Dabov, A.~Foi, V.~Katkovnik, K.~Egiazarian, Image denoising by sparse 3-d
  transform-domain collaborative filtering, IEEE Trans. Image Process. 16~(8)
  (2007) 2080--2095.
\newblock \href {https://doi.org/10.1109/TIP.2007.901238}
  {\path{doi:10.1109/TIP.2007.901238}}.

\bibitem{dabov2009bm3d}
K.~Dabov, A.~Foi, V.~Katkovnik, K.~Egiazarian, Bm3d image denoising with
  shape-adaptive principal component analysis, in: Proc. Signal Process. Adapt.
  Sparse Struct. Represent. (SPARS), 2009, pp. 1--7.

\bibitem{Buades2005areview}
A.~Buades, B.~Coll, J.-M. Morel, A review of image denoising algorithms, with a
  new one, SIAM Multiscale Model. Simul. 4~(2) (2005) 490--530.
\newblock \href {https://doi.org/10.1137/040616024}
  {\path{doi:10.1137/040616024}}.

\bibitem{tasdizen2009principal}
T.~Tasdizen, Principal neighborhood dictionaries for nonlocal means image
  denoising, IEEE Trans. Image Process. 18~(12) (2009) 2649--2660.
\newblock \href {https://doi.org/10.1109/TIP.2009.2028259}
  {\path{doi:10.1109/TIP.2009.2028259}}.

\bibitem{deledalle2011image}
C.-A. Deledalle, J.~Salmon, A.~Dalalyan, Image denoising with patch based pca:
  local versus global, in: Proc. BMVC, Vol.~81, 2011, pp. 425--455.

\bibitem{Vignesh2010fast}
R.~Vignesh, B.~Oh, C.-C. Kuo, Fast non-local means (nlm) computation with
  probabilistic early termination, IEEE Signal Process. Lett. 17~(3) (2010)
  277--280.
\newblock \href {https://doi.org/10.1109/LSP.2009.2038956}
  {\path{doi:10.1109/LSP.2009.2038956}}.

\bibitem{Zuo2016image}
C.~Zuo, L.~Jovanov, B.~Goossens, H.~Luong, W.~Philips, Y.~Liu, M.~Zhang, Image
  denoising using quadtree-based nonlocal means with locally adaptive principal
  component analysis, IEEE Signal Process. Lett. 23~(4) (2016) 434--438.
\newblock \href {https://doi.org/10.1109/LSP.2016.2530406}
  {\path{doi:10.1109/LSP.2016.2530406}}.

\bibitem{Frosio2019statistical}
I.~Frosio, J.~Kautz, Statistical nearest neighbors for image denoising, IEEE
  Trans. Image Process. 28~(2) (2019) 723--738.
\newblock \href {https://doi.org/10.1109/TIP.2018.2869685}
  {\path{doi:10.1109/TIP.2018.2869685}}.

\bibitem{Kervrann2006optimal}
C.~Kervrann, J.~Boulanger, Optimal spatial adaptation for patch-based image
  denoising, IEEE Trans. Image Process. 15~(10) (2006) 2866--2878.
\newblock \href {https://doi.org/10.1109/TIP.2006.877529}
  {\path{doi:10.1109/TIP.2006.877529}}.

\bibitem{Chatterjee2012patch}
P.~Chatterjee, P.~Milanfar, Patch-based near-optimal image denoising, IEEE
  Trans. Image Process. 21~(4) (2012) 1635--1649.
\newblock \href {https://doi.org/10.1109/TIP.2011.2172799}
  {\path{doi:10.1109/TIP.2011.2172799}}.

\bibitem{Mahmoudi2005fast}
M.~Mahmoudi, G.~Sapiro, Fast image and video denoising via nonlocal means of
  similar neighborhoods, IEEE Signal Process. Lett. 12~(12) (2005) 839--842.
\newblock \href {https://doi.org/10.1109/LSP.2005.859509}
  {\path{doi:10.1109/LSP.2005.859509}}.

\bibitem{Van2009sure}
D.~Van De~Ville, M.~Kocher, Sure-based non-local means, IEEE Signal Process.
  Lett. 16~(11) (2009) 973--976.
\newblock \href {https://doi.org/10.1109/LSP.2009.2027669}
  {\path{doi:10.1109/LSP.2009.2027669}}.

\bibitem{Dong2013nonlocally}
W.~Dong, L.~Zhang, G.~Shi, X.~Li, Nonlocally centralized sparse representation
  for image restoration, IEEE Trans. Image Process. 22~(4) (2013) 1620--1630.
\newblock \href {https://doi.org/10.1109/TIP.2012.2235847}
  {\path{doi:10.1109/TIP.2012.2235847}}.

\bibitem{Li2021patch}
F.~Li, Y.~Ru, X.~Lv,
  \href{https://doi.org/10.1007/s10915-021-01688-5}{Patch-based weighted scad
  prior for rician noise removal}, J. Sci. Comput. 90~(26) (2021) 1573--7691.
\newblock \href {https://doi.org/10.1007/s10915-021-01688-5}
  {\path{doi:10.1007/s10915-021-01688-5}}.
\newline\urlprefix\url{https://doi.org/10.1007/s10915-021-01688-5}

\bibitem{Zha2020from}
Z.~Zha, X.~Yuan, B.~Wen, J.~Zhou, J.~Zhang, C.~Zhu, From rank estimation to
  rank approximation: Rank residual constraint for image restoration, IEEE
  Trans. Image Process. 29 (2020) 3254--3269.
\newblock \href {https://doi.org/10.1109/TIP.2019.2958309}
  {\path{doi:10.1109/TIP.2019.2958309}}.

\bibitem{Zha2020image}
Z.~Zha, X.~Yuan, J.~Zhou, C.~Zhu, B.~Wen, Image restoration via simultaneous
  nonlocal self-similarity priors, IEEE Trans. Image Process. 29 (2020)
  8561--8576.
\newblock \href {https://doi.org/10.1109/TIP.2020.3015545}
  {\path{doi:10.1109/TIP.2020.3015545}}.

\bibitem{Zha2021image}
Z.~Zha, B.~Wen, X.~Yuan, J.~Zhou, C.~Zhu, Image restoration via reconciliation
  of group sparsity and low-rank models, IEEE Trans. Image Process. 30 (2021)
  5223--5238.
\newblock \href {https://doi.org/10.1109/TIP.2021.3078329}
  {\path{doi:10.1109/TIP.2021.3078329}}.

\bibitem{Zha2021nonconvex}
Z.~Zha, X.~Yuan, B.~Wen, J.~Zhang, C.~Zhu, Nonconvex structural sparsity
  residual constraint for image restoration, IEEE Trans. Cyber. (2021)
  1--14\href {https://doi.org/10.1109/TCYB.2021.3084931}
  {\path{doi:10.1109/TCYB.2021.3084931}}.

\bibitem{Aytekin2013Quantum}
{\c{C}}.~Aytekin, S.~Kiranyaz, M.~Gabbouj, Quantum mechanics in computer
  vision: Automatic object extraction, in: Proc. IEEE Int. Conf. Image
  Process., 2013, pp. 2489--2493.
\newblock \href {https://doi.org/10.1109/ICIP.2013.6738513}
  {\path{doi:10.1109/ICIP.2013.6738513}}.

\bibitem{youssry2015quantum}
A.~Youssry, A.~El-Rafei, S.~Elramly, A quantum mechanics-based framework for
  image processing and its application to image segmentation, Quantum Inf.
  Process. 14~(10) (2015) 3613--3638.
\newblock \href {https://doi.org/https://doi.org/10.1007/s11128-015-1072-3}
  {\path{doi:https://doi.org/10.1007/s11128-015-1072-3}}.

\bibitem{youssry2019continuous}
A.~Youssry, A.~El-Rafei, R.-G. Zhou, A continuous-variable quantum-inspired
  algorithm for classical image segmentation, Quantum Mach. Intell. 1~(3-4)
  (2019) 97--111.
\newblock \href {https://doi.org/https://doi.org/10.1007/s42484-019-00009-2}
  {\path{doi:https://doi.org/10.1007/s42484-019-00009-2}}.

\bibitem{kaisserli2015novel}
Z.~Kaisserli, T.-M. Laleg-Kirati, A.~Lahmar-Benbernou, A novel algorithm for
  image representation using discrete spectrum of the schr{\"o}dinger operator,
  Digit. Signal Process. 40 (2015) 80--87.
\newblock \href {https://doi.org/https://doi.org/10.1016/j.dsp.2015.01.005}
  {\path{doi:https://doi.org/10.1016/j.dsp.2015.01.005}}.

\bibitem{dutta2021quantum}
S.~Dutta, A.~Basarab, B.~Georgeot, D.~Kouam\'e, Quantum mechanics-based signal
  and image representation: Application to denoising, IEEE Open J. Signal
  Process. 2 (2021) 190--206.
\newblock \href {https://doi.org/10.1109/OJSP.2021.3067507}
  {\path{doi:10.1109/OJSP.2021.3067507}}.

\bibitem{dutta2021plug}
S.~Dutta, A.~Basarab, B.~Georgeot, D.~Kouam\'e, Plug-and-play quantum adaptive
  denoiser for deconvolving poisson noisy images, IEEE Access 9 (2021)
  139771--139791.
\newblock \href {https://doi.org/10.1109/ACCESS.2021.3118608}
  {\path{doi:10.1109/ACCESS.2021.3118608}}.

\bibitem{dutta2021poisson}
S.~Dutta, A.~Basarab, B.~Georgeot, D.~Kouam\'e, Poisson image deconvolution by
  a plug-and-play quantum denoising scheme, in: Proc. 29th Eur. Signal Process.
  Conf. (EUSIPCO), 2021, pp. 646--650.
\newblock \href {https://doi.org/10.23919/EUSIPCO54536.2021.9616253}
  {\path{doi:10.23919/EUSIPCO54536.2021.9616253}}.

\bibitem{Eldar2002quantum}
Y.~Eldar, A.~Oppenheim, Quantum signal processing, IEEE Signal Process. Mag.
  19~(6) (2002) 12--32.
\newblock \href {https://doi.org/10.1109/MSP.2002.1043298}
  {\path{doi:10.1109/MSP.2002.1043298}}.

\bibitem{dutta2022quantum}
S.~Dutta, N.~K. Tuador, J.~Michetti, B.~Georgeot, D.~H. Pham, A.~Basarab,
  D.~Kouam{\'e}, Quantum denoising-based super-resolution algorithm applied to
  dental tomography images, in: Proc. IEEE 19th Int. Symp. Biomed. Imag.
  (ISBI), 2022, pp. 1--4.
\newblock \href {https://doi.org/10.1109/ISBI52829.2022.9761623}
  {\path{doi:10.1109/ISBI52829.2022.9761623}}.

\bibitem{dutta2021image}
S.~Dutta, A.~Basarab, B.~Georgeot, D.~Kouam\'e, Image denoising inspired by
  quantum many-body physics, in: Proc. 28th IEEE Int. Conf. Image Process.
  (ICIP), 2021, pp. 1619--1623.
\newblock \href {https://doi.org/10.1109/ICIP42928.2021.9506794}
  {\path{doi:10.1109/ICIP42928.2021.9506794}}.

\bibitem{feynman1977feynman}
R.~Feynman, R.~Leighton, M.~Sands, \href{New Millennium ed. New York, NY, USA,
  2010.}{The Feynman Lectures on Physics}, Addison-Wesley world student series,
  Addison-Wesley, 1977.
\newline\urlprefix\url{New Millennium ed. New York, NY, USA, 2010.}

\bibitem{landau1991quantum}
L.~Landau, E.~Lifshitz,
  \href{https://books.google.fr/books?id=J9ui6KwC4mMC}{Quantum Mechanics:
  Non-Relativistic Theory}, Course of theoretical physics, Elsevier Science,
  1991.
\newline\urlprefix\url{https://books.google.fr/books?id=J9ui6KwC4mMC}

\bibitem{cohen1977quantum}
C.~Cohen-Tannoudji, B.~Diu, F.~Lalo{\"e},
  \href{https://books.google.fr/books?id=tVI\_EAAAQBAJ}{Quantum Mechanics}, 1st
  ed, New York, NY, USA: Wiley, 1977.
\newline\urlprefix\url{https://books.google.fr/books?id=tVI\_EAAAQBAJ}

\bibitem{Anderson1958absence}
P.~Anderson, \href{https://link.aps.org/doi/10.1103/PhysRev.109.1492}{Absence
  of diffusion in certain random lattices}, Phys. Rev. 109 (1958) 1492--1505.
\newblock \href {https://doi.org/10.1103/PhysRev.109.1492}
  {\path{doi:10.1103/PhysRev.109.1492}}.
\newline\urlprefix\url{https://link.aps.org/doi/10.1103/PhysRev.109.1492}

\bibitem{mahan2013local}
G.~Mahan, K.~Subbaswamy,
  \href{https://books.google.fr/books?id=7v0ICAAAQBAJ}{Local Density Theory of
  Polarizability}, Physics of Solids and Liquids, Springer US, 2013.
\newline\urlprefix\url{https://books.google.fr/books?id=7v0ICAAAQBAJ}

\bibitem{salmon2014poisson}
J.~Salmon, Z.~Harmany, C.-A. Deledalle, R.~Willett, Poisson noise reduction
  with non-local pca, J. math. imag. vis. 48~(2) (2014) 279--294.
\newblock \href {https://doi.org/https://doi.org/10.1007/s10851-013-0435-6}
  {\path{doi:https://doi.org/10.1007/s10851-013-0435-6}}.

\bibitem{Makitalo2011optimal}
M.~Makitalo, A.~Foi, Optimal inversion of the anscombe transformation in
  low-count poisson image denoising, IEEE Trans. Image Process. 20~(1) (2011)
  99--109.
\newblock \href {https://doi.org/10.1109/TIP.2010.2056693}
  {\path{doi:10.1109/TIP.2010.2056693}}.

\bibitem{Liu2021SAR}
S.~Liu, L.~Gao, Y.~Lei, M.~Wang, Q.~Hu, X.~Ma, Y.~Zhang, Sar speckle removal
  using hybrid frequency modulations, IEEE Trans. Geosci. Remote Sens. 59~(5)
  (2021) 3956--3966.
\newblock \href {https://doi.org/10.1109/TGRS.2020.3014130}
  {\path{doi:10.1109/TGRS.2020.3014130}}.

\bibitem{Yu2002speckle}
Y.~Yu, S.~Acton, Speckle reducing anisotropic diffusion, IEEE Trans. Image
  Process. 11~(11) (2002) 1260--1270.
\newblock \href {https://doi.org/10.1109/TIP.2002.804276}
  {\path{doi:10.1109/TIP.2002.804276}}.

\bibitem{Santos2017ultrasound}
C.~Santos, D.~Martins, N.~Mascarenhas, Ultrasound image despeckling using
  stochastic distance-based bm3d, IEEE Trans. Image Process. 26~(6) (2017)
  2632--2643.
\newblock \href {https://doi.org/10.1109/TIP.2017.2685339}
  {\path{doi:10.1109/TIP.2017.2685339}}.

\bibitem{Achim2001novel}
A.~Achim, A.~Bezerianos, P.~Tsakalides, Novel bayesian multiscale method for
  speckle removal in medical ultrasound images, IEEE Trans. Med. Imag. 20~(8)
  (2001) 772--783.
\newblock \href {https://doi.org/10.1109/42.938245}
  {\path{doi:10.1109/42.938245}}.

\bibitem{Lee1980digital}
J.-S. Lee, Digital image enhancement and noise filtering by use of local
  statistics, IEEE Trans. Pattern Anal. Mach. Intell. PAMI-2~(2) (1980)
  165--168.
\newblock \href {https://doi.org/10.1109/TPAMI.1980.4766994}
  {\path{doi:10.1109/TPAMI.1980.4766994}}.

\bibitem{dutta2021despeckling}
S.~Dutta, A.~Basarab, B.~Georgeot, D.~Kouam\'e, Despeckling ultrasound images
  using quantum many-body physics, in: Proc. IEEE Int. Ultrason. Symp. (IUS),
  2021, pp. 1--4.
\newblock \href {https://doi.org/10.1109/IUS52206.2021.9593778}
  {\path{doi:10.1109/IUS52206.2021.9593778}}.

\bibitem{Chen2017trainable}
Y.~Chen, T.~Pock, Trainable nonlinear reaction diffusion: A flexible framework
  for fast and effective image restoration, IEEE Trans. Pattern Anal. Mach.
  Intell. 39~(6) (2017) 1256--1272.
\newblock \href {https://doi.org/10.1109/TPAMI.2016.2596743}
  {\path{doi:10.1109/TPAMI.2016.2596743}}.

\bibitem{Zhang2018ffdnet}
K.~Zhang, W.~Zuo, L.~Zhang, Ffdnet: Toward a fast and flexible solution for
  cnn-based image denoising, IEEE Trans. Image Process. 27~(9) (2018)
  4608--4622.
\newblock \href {https://doi.org/10.1109/TIP.2018.2839891}
  {\path{doi:10.1109/TIP.2018.2839891}}.

\bibitem{Zhang2017beyond}
K.~Zhang, W.~Zuo, Y.~Chen, D.~Meng, L.~Zhang, Beyond a gaussian denoiser:
  Residual learning of deep cnn for image denoising, IEEE Trans. Image Process.
  26~(7) (2017) 3142--3155.
\newblock \href {https://doi.org/10.1109/TIP.2017.2662206}
  {\path{doi:10.1109/TIP.2017.2662206}}.

\end{thebibliography}

%%%%%%%%%%%%%%%%%%%%%%%%%%%%%%%%%%%%%%%%%%%%%%%%%%%%%%%%%%%%%%%%%%%%%%%%%%%%%%%%%%%%%%%%%%%%%%%%%%%%%%%%%%%%%%%%%%%%%%%%%%%%%%%%%%%

\onecolumn
\section*{\Huge{Supplementary Material:}\\ \Large{for}\\ \LARGE{A Novel Image Denoising Algorithm Using Concepts of Quantum Many-Body Theory}}
\label{sec:supplement}
\vspace*{5mm}

\begin{center}

{Sayantan~Dutta$^{1,2}$,~Adrian~Basarab$^{3}$,~Bertrand Georgeot$^{2}$,~and~Denis~Kouam\'e$^{1}$}

\vspace{.2cm}
$^{1}$Institut de Recherche en Informatique de Toulouse, UMR CNRS 5505, Universit\'e de Toulouse, France

\vspace{.1cm}
$^{2}$Laboratoire de Physique Th\'eorique, Universit\'e de Toulouse, CNRS, UPS, France

\vspace{.1cm}
$^{3}$Universit\'e de Lyon, INSA-Lyon, Universit\'e Claude Bernard Lyon 1, UJM-Saint Etienne, CNRS, Inserm, CREATIS UMR 5220, U1206, Villeurbanne, France.

\vspace*{5mm}
\end{center}

In this paper, a novel image restoration and denoising algorithm inspired by the quantum many-body theory has been discussed, precisely, using interactions between image patches. To illustrate the interest of the proposed approach in image denoising problems and explore ways to choose the suitable hyperparameters, we study in this supplementary file the optimal hyperparameter values for four different levels of noise intensities with image independent (Gaussian) and dependent (Poisson) noise models. We also present additional information on comparisons of our procedure to other contemporary methods, as well as on a real-life medical application on ultrasound image despeckling.

\subsection{Proposed De-QuIP algorithm}
\label{sec:algo}

\sd{The complete denoising process is summarized in the following Algorithm~\ref{Algo:QMBI}.}

%%%%%%%%%%%%%% algorithm %%%%%%%%%%%%%%%%%%%%%
%\vspace*{-3mm}
%\newpage

\begin{algorithm}[h!]
\begin{small}

\label{Algo:QMBI}

\KwIn{ $\bsy$ , $P_h$, $W_h$, $d$, $p$, $\hbar ^2/2m$}
%\BlankLine

 {Divide the noisy image $\bsy$ into small patches of size $P_h$; say total number is $T_{patch}$. So, the patch dimension $P_{dim} = {P_h}^2$}\\

  \For{ $w = 1 : T_{patch}$}{
  
 {Choose one small image patch $\bsJ_w$}\\
 
 {Create a search window of size $W_h$ centering at $\bsJ_w$ and using cyclic boundary conditions}\\
 
 {Collect all the small image patches inside this search window; say the total number is $S_{patch}$}\\

     \For{ $l = 1 : S_{patch}$}{
     
     {Calculate Euclidean distance $D_{wl}$ between the $\bsJ_w$ and $\bsJ_l$ patch inside the search window}\\
     
     {Calculate interaction $\bsI_{wl}$ between the $\bsJ_w$ and $\bsJ_l$ patch inside the search window as, $\bsI_{wl}^k = p \dfrac{ | \bsJ_w^k - \bsJ_l^k | }{ D_{wl}^2}, ~~k = 1,\cdots, P_{dim}$}\\
         
     } 
 
 {Calculate total interaction $\bsI^{~total}_w$ between the patch $\bsJ_w$ and the patches inside the search window by taking sum over all $l$; \textit{i.e.}, $\bsI^{{~total}^k}_w = \sum_{l = 1}^{S_{patch}}  \bsI_{wl}^k , ~~ k = 1,\cdots,P_{dim}$}\\
  
 {Effective potential for the $\bsJ_w$ patch is $ \bsV_w^{{~effective}^k} = \bsJ_w^k + \bsI^{{~total}^k}_w , ~~ k = 1,\cdots,P_{dim}$}\\
 
 {Construct the Hamiltonian matrix $\bsH_w$ using the effective potential $ \bsV_w^{~effective}$}\\
 
 {Calculate the eigenvalues and eigenvectors of $\bsH_w$}\\
 
 {Construct adaptive basis $\bsB_w^{adaptive}$ using the eigenvectors $\bpsi_w^k , ~~ k = 1,\cdots,P_{dim}$}\\

     {Project the noisy patche $\bsJ_w$ onto this adaptive basis $\bsB_w^{adaptive}$}\\
     
     {Calculate projection coefficients $\bsc_w$ in the $P_{dim}$-dimensional space. Note that, $P_{dim} > d$}\\
     
     {Redefine the projection coefficients in the $d$-dim subspace as $\bsc^{{new}^k}_{w} = \bsc^k_{w} , k = 1,\cdots,d$}\\
     
     {Reconstruct the patch by $\bsR_w = \sum_{k = 1}^d \bsc^{{new}^k}_{w} \bpsi_w^k$}\\
          
 }
 
 {Combining all $T_{patch}$ number of small denoised image patches $\bsR_w$ restores the full denoised image $\hat{\bsx}$}\\
 
% \BlankLine
 \KwOut{$\hat{\bsx}$}

\caption{De-QuIP algorithm}

\DecMargin{1em}

\end{small}

\end{algorithm}
%\vspace{-4mm}

\subsection{Computational complexity}
\label{sec:compucomplx}
\vspace*{-2mm}
In the precedingly developed algorithm based on single particle quantum physics \cite{dutta2021quantum}, the computational complexity of the algorithm was essentially controlled by the diagonalization of a large Hamiltonian matrix and the identification of its eigenvectors. For an image of size $n\times n$, this matrix is $n^2\times n^2$. In general, for an arbitrary matrix, the diagonalization process would require $O(n^6)$ operations and $O(n^4)$ storage space. However, for a highly sparse matrix (like the Hamiltonian matrix), efficient iterative methods such as the Lanczos method reduce the computational complexity to $O(n^4)$ operations with $O(n^4)$ space complexity required for the diagonalization.

In the case of the many patch algorithm, the denoising is done patch-wisely (of size $P_h \times P_h$), the time and space complexity become $O(P_h^4)$ for each denoise region, much smaller than the previous one for $P_h \ll n$. Yet, the best time complexity one can achieve is $O(dP_h^2)$ if one computes only the $d$ eigenvectors used for the restoration task (with $d \leq P_h^2$), with a space complexity also in $O(dP_h^2)$.

The second major contribution comes from the computations of the transform coefficients using an iterative scheme that would require $O(dP_h)$ operations for each denoise region.

The interaction count for each denoise region gives a complexity in total of $O((S_{patch}+1) P_h)$ if there are $S_{patch}$ patches inside the $W_h \times W_h$ size search window.

Therefore, if the image consists of $T_{patch}$ regions (patches), then the dominant computational cost of the proposed denoising algorithm is $ O( T_{patch}d P_h^2 )$.
Additionally, parallel computation can be used to boost up the process even further.

\subsection{Influence of the proportionality constant $p$}
%\label{sec:hy_p}

In Table~\ref{tab:tab_proporconst}, we observed the proportionality constant $p$ values that maximize the output PSNR for the first seven sample images in Fig.~\ref{fig:samples} (in the primary manuscript) corrupted with four different noise intensities. For the numerical experiment, we consider three different patch sizes when the images are corrupted with image independent (\textit{e.g.}, Gaussian) and dependent (\textit{e.g.}, Poisson) noise models. The observations confirm that there is a tendency for optimal values to decrease as the noise level increases.

%%%%%%%%%%% Proportionality constant value %%%%%%%%%%%%%%%%%

\begin{table}[h!]
\begin{scriptsize}

\begin{center}
\caption{Optimal proportionality constant $p$ for De-QuIP}
%\resizebox{.8\hsize}{!}{
\label{tab:tab_proporconst}
\begin{tabular}{cc  ccc c ccc}
\thickhline

\multirow{3}{*}{Sample} & \multirow{3}{*}{Input} & \multicolumn{3}{c}{Gaussian case} && \multicolumn{3}{c}{Poisson case}\\
% \cline{3-5} \cline{7-9}
 & & \multicolumn{3}{c}{Patch size} && \multicolumn{3}{c}{Patch size}\\
		%	\cline{3-5}
			  & SNR(dB) & $5 \times 5$ & $7 \times 7$ & $11 \times 11$ && $5 \times 5$ & $7 \times 7$ & $11 \times 11$\\
\thickhline

\multirow{4}{*}{house}	&22& 0.0385 & 0.0750 & 0.0550 && 0.0109  & 0.0467 & 0.0800\\
						&16& 0.0197 & 0.0667 & 0.0700 && 0.0046  & 0.0325 & 0.0900\\
						&8& 0.0053 & 0.0600  & 0.0625 && 0.0064  & 0.0305 & 0.0500\\  \vspace*{1mm}
						&2& 0.0001 & 0.0225  & 0.0475 && 0.00013 & 0.0034 & 0.0400\\

\multirow{4}{*}{lake}	&22& 0.0220 & 0.0900 & 0.0900 && 0.0057  & 0.0555 & 0.0600\\
						&16& 0.0130 & 0.0522 & 0.0936 && 0.0033  & 0.0269 & 0.0400\\
						&8& 0.0060 & 0.0460  & 0.0755 && 0.0018  & 0.0355 & 0.0499\\  \vspace*{1mm}
						&2& 0.0001 & 0.0290  & 0.0573 && 0.00013 & 0.0096 & 0.0300\\
						
	%	\cline{2-5}

\multirow{4}{*}{lena}	&22& 0.0215 & 0.0950 & 0.1250 && 0.0156  & 0.0533 & 0.0625\\
						&16& 0.0107 & 0.0758 & 0.1100 && 0.0067  & 0.0317 & 0.0550\\
						&8& 0.0046 & 0.0467  & 0.0600 && 0.00031 & 0.0207 & 0.0775\\  \vspace*{1mm}
						&2& 0.00001 & 0.0100 & 0.0400 && 0.00001 & 0.0010 & 0.0550\\
						
	%	\cline{2-5}

\multirow{4}{*}{hill}	&22& 0.0154 & 0.0643 & 0.1229 && 0.0089  & 0.0500 & 0.0900\\
						&16& 0.0139 & 0.0521 & 0.0888 && 0.0056  & 0.0400 & 0.0800\\
						&8& 0.0072 & 0.0375  & 0.0625 && 0.00088 & 0.0300 & 0.0629\\  \vspace*{1mm}
						&2& 0.00001 & 0.0146 & 0.0340 && 0.00001 & 0.0055 & 0.0429\\
						
	%	\cline{2-5}

\multirow{4}{*}{fingerprint}	&22& 0.0500 & 0.0700 & 0.0450 && 0.0244  & 0.0600 & 0.1000\\
							&16& 0.0369 & 0.0657 & 0.0650 && 0.0133  & 0.0400 & 0.0833\\
							&8& 0.0041 & 0.0543  & 0.0900 && 0.00011 & 0.0350 & 0.0650\\  \vspace*{1mm}
							&2& 0.0022 & 0.0110  & 0.0830 && 0.00005 & 0.0011 & 0.0500\\

%		\cline{2-5}

\multirow{4}{*}{saturn}	&22& 0.0257 & 0.0700 & 0.1100 && 0.0100  & 0.0500 & 0.1083\\
						&16& 0.0021 & 0.0578 & 0.0900 && 0.0080  & 0.0400 & 0.0967\\
						&8& 0.0031 & 0.0234  & 0.0600 && 0.0006  & 0.0124 & 0.0617\\  \vspace*{1mm}
						&2& 0.00001 & 0.0157  & 0.0500 && 0.00001 & 0.0010 & 0.0540\\
						
%		\cline{2-5}

\multirow{4}{*}{flintstones}	&22& 0.0260 & 0.0931 & 0.0463 && 0.0089 & 0.0400 & 0.0500\\
							&16& 0.0183 & 0.0500 & 0.0304 && 0.0044& 0.0225 & 0.0655\\
							&8& 0.0052 & 0.0308  & 0.0500 && 0.0006  & 0.0200 & 0.0525\\
							&2& 0.0016 & 0.0205  & 0.0500 && 0.00017 & 0.0126 & 0.0450\\
						
\thickhline

\end{tabular}
%}
\end{center}

\end{scriptsize}
\end{table}

The linear fit parameters are summarized in Table~\ref{tab:tab_proporconstfit} together with the $\ell_2$ error and the resulting average loss in the denoising performance in terms of PSNR and SSIM. One may notice that the denoising performance loss with rule (9) (in the primary manuscript) rather than the optimal choice is negligible.

%%%%%%%%%%%%%%%% curve fitting for p %%%%%%%%%%%%%%

\begin{table}[h!]
\begin{scriptsize}

\begin{center}
\caption{Slope and intercept used in determining proportionality constant $p$ for various patch sizes for Gaussian and Poisson noise models. Also, the associative $\ell_2$ error, PSNR (dB) loss and SSIM loss in linear curve fitting to the optimal $p$.}
%\vspace{-2mm}
%\resizebox{.8\hsize}{!}
%{
\label{tab:tab_proporconstfit}
\begin{tabular}{c l  r r r}
\thickhline

\multirow{2}{*}{~} & \multirow{2}{*}{~} & \multicolumn{3}{c}{Size of the patches}\\

			 && $5 \times 5$ & $7 \times 7$ & $11 \times 11$ \\
\thickhline

\multirow{5}{*}{\begin{turn}{90}Gaussian\end{turn}}

& Slope ($m_1$)		 & $12.84\times 10^{-4}$ & $30.96\times 10^{-4}$ & $16.46\times 10^{-4}$ \\

& Intercept ($c_1$) 	& $-35.96\times 10^{-4}$ & $13.56\times 10^{-3}$ & $50.40\times 10^{-3}$\\

& $\ell_2$ error for $p$ fit & 0.0327 & 0.0528 & 0.1196\\

& PSNR(dB) loss	& 0.278 & 0.306 & 0.179\\ \vspace*{1mm}

& SSIM loss	& 0.0139 & 0.0172 & 0.0106\\

\multirow{5}{*}{\begin{turn}{90}Poisson\end{turn}}

& Slope ($m_1$)		 & $60.33\times 10^{-5}$ & $21.00\times 10^{-4}$ & $16.64\times 10^{-4}$ \\

& Intercept ($c_1$) 	& $-21.85\times 10^{-4}$ & $36.31\times 10^{-4}$ & $44.23\times 10^{-3}$\\

& $\ell_2$ error for $p$ fit & 0.0189 & 0.0392 & 0.0811\\

& PSNR(dB) loss	& 0.380 & 0.422 & 0.485\\

& SSIM loss	& 0.0197 & 0.0185 & 0.0150\\

\thickhline

\end{tabular}
%}
\end{center}

\end{scriptsize}
%\vspace{-7mm}
\end{table}

\subsection{Influence of the $\hbar ^2/2m$ and subspace dimensionality $d$}
%\label{sec:hy_hbar_d}

The hyperparameter $\hbar ^2/2m$ or say $F_{\mbox{factor}}$ controls the frequency distribution across the basis vectors since the maximal frequency of a vector with energy $E$ at the local pixel value $V$ is $\sqrt{(E-V)/(\hbar ^2/2m)}$. Hence, the maximal oscillation frequency of the basis vector decreases with increasing $F_{\mbox{factor}}$. As a consequence, low-energy basis vectors become more prominent to distinguish low and high pixel regions using different levels of frequency. Thus, optimal subspace dimensionality $d$ decreases as $F_{\mbox{factor}}$ grows. These optimal choices vary with the image patch size and noise statistics. Table~\ref{tab:tab_planckdata} and Table~\ref{tab:tab_subspadim} show these optimal values that give the best output PSNR for the first seven sample images in Fig.~\ref{fig:samples} (in the primary manuscript) corrupted with image independent (\textit{e.g.}, Gaussian) and dependent (\textit{e.g.}, Poisson) noises with four different noise intensities.

%%%%%%%%%%%%%% Planck constant \hbar_factor value  %%%%%%%%%%%%%%%%%

\begin{table}[h!]
\begin{scriptsize}

\begin{center}

\caption{Optimal $F_{\textrm{factor}}$ values for De-QuIP}
%\resizebox{.8\hsize}{!}{
\label{tab:tab_planckdata}

\begin{tabular}{cc  ccc c ccc}
\thickhline

\multirow{3}{*}{Sample} & \multirow{3}{*}{Input} & \multicolumn{3}{c}{Gaussian case} && \multicolumn{3}{c}{Poisson case}\\
% \cline{3-5} \cline{7-9}
 & & \multicolumn{3}{c}{Patch size} && \multicolumn{3}{c}{Patch size}\\
		%	\cline{3-5}
			  & SNR(dB) & $5 \times 5$ & $7 \times 7$ & $11 \times 11$ && $5 \times 5$ & $7 \times 7$ & $11 \times 11$\\
\thickhline

\multirow{4}{*}{house}	&22& 1.4714  & 2.4250 & 2.2000 && 1.7000 & 2.2667 & 2.6000\\
						&16& 1.9000  & 1.6733 & 2.3000 && 1.6000 & 1.9167 & 1.8500\\
						&8 & 1.9800  & 1.5333 & 1.9833 && 1.9900 & 2.1000 & 1.0000\\  \vspace*{1mm}
						&2 & 1.5000  & 1.2000 & 1.7300 && 2.9000 & 1.9000 & 1.5800\\
						
	%	\cline{2-5}

\multirow{4}{*}{lake}	&22& 1.6000 & 2.1125 & 2.4000 && 1.4889 & 1.8000 & 1.6000\\
						&16& 1.5500 & 1.9633 & 2.8000 && 1.6333 & 1.9900 & 2.0000\\
						&8 & 2.2000 & 1.4083 & 2.0650 && 2.5000 & 1.8500 & 2.0000\\  \vspace*{1mm}
						&2 & 1.6632 & 2.1000 & 2.3000 && 2.8333 & 2.6000 & 1.7000\\
						
%		\cline{2-5}

\multirow{4}{*}{lena}	&22& 1.3850 & 1.9000  & 2.4000 && 1.6000 & 2.0400 & 2.5000\\
						&16& 1.8500 & 1.7800  & 2.3000 && 1.8000 & 1.8500 & 2.1000\\
						&8 & 2.0000 & 1.5500  & 2.0200 && 1.9200 & 2.2667 & 1.7000\\  \vspace*{1mm}
						&2 & 1.2400 & 0.8571  & 2.1000 && 2.6500 & 2.2333 & 2.0000\\
						
%		\cline{2-5}

\multirow{4}{*}{hill}	&22& 1.3570 & 1.5000 & 1.8000 && 1.6500 & 1.5182 & 1.9000\\
						&16& 2.4500 & 2.2500 & 2.0000 && 2.0000 & 2.2000 & 1.9400\\
						&8 & 2.6222 & 2.7400 & 2.5000 && 3.0000 & 3.0000 & 2.2800\\  \vspace*{1mm}
						&2 & 1.6167 & 2.1444 & 2.0000 && 2.5000 & 2.9000 & 3.2000\\
						
%		\cline{2-5}

\multirow{4}{*}{fingerprint}	&22 & 1.2500 & 1.3800 & 1.4500 && 1.6000 & 1.5000 & 1.5000\\
							&16 & 1.7000 & 1.7500 & 2.1000 && 1.5400 & 1.6000 & 2.3000\\
							&8  & 3.3000 & 3.0000 & 1.8000 && 3.8000 & 3.0000 & 1.9000\\  \vspace*{1mm}
							&2  & 2.7000 & 2.0000 & 2.0000 && 3.8000 & 3.0000 & 3.6000\\

	%	\cline{2-5}

\multirow{4}{*}{saturn}	&22& 1.3500 & 1.3900 & 1.9000 && 1.5500 & 1.8778 & 2.2600\\
						&16& 1.7818 & 1.7000 & 1.7000 && 1.5600 & 1.4889 & 1.8000\\
						&8 & 1.5909 & 1.8400 & 1.8000 && 2.1143 & 2.0000 & 1.9600\\  \vspace*{1mm}
						&2 & 1.6286 & 1.7100 & 1.7000 && 1.8333 & 1.6000 & 2.0429\\

	%	\cline{2-5}

\multirow{4}{*}{flintstones}	&22& 1.4000 & 1.5500 & 1.5000 && 1.8000 & 1.4000 & 1.5000\\
							&16& 1.9333 & 1.9000 & 1.9000 && 1.7000 & 1.8400 & 1.5000\\
							&8 & 1.8000 & 2.0000 & 1.9000 && 3.0000 & 1.9700 & 1.9000\\
							&2 & 3.0000 & 2.5750 & 1.8000 && 3.8000 & 2.9000 & 1.9000\\
						
\thickhline

\end{tabular}
%}
\end{center}

\end{scriptsize}
%\vspace*{-6mm}
\end{table}

%%%%%%%%%%%%% subspace dimensionality d %%%%%%%%%%%%%

\begin{table}[h!]

\begin{tiny}

\begin{center}
\caption{Optimal subspace dimensionality $d$ for De-QuIP}
%\resizebox{.8\hsize}{!}{
\label{tab:tab_subspadim}
\begin{tabular}{cc  ccc c ccc}
\thickhline

\multirow{3}{*}{Sample} & \multirow{3}{*}{Input} & \multicolumn{3}{c}{Gaussian case} && \multicolumn{3}{c}{Poisson case}\\
% \cline{3-5} \cline{7-9}
 & & \multicolumn{3}{c}{Patch size} && \multicolumn{3}{c}{Patch size}\\
		%	\cline{3-5}
			  & SNR(dB) & $5 \times 5$ & $7 \times 7$ & $11 \times 11$ && $5 \times 5$ & $7 \times 7$ & $11 \times 11$\\
\thickhline

\multirow{4}{*}{house}	&22& 16 & 39 & 120 && 15 & 33 & 85\\
						&16& 10 & 24 & 111 && 10 & 25 & 74\\
						&8 & 6  & 11 & 56 &&  3 & 11 & 24\\  \vspace*{1mm}
						&2 & 3  & 4  & 24 &&  2 &  6 & 18\\
	%	\cline{2-5}

\multirow{4}{*}{lake}	&22& 24 & 48 & 84 && 24 & 48 & 120\\
						&16& 21 & 40 & 64 && 22 & 36 & 101\\
						&8 & 7  & 15 & 25 && 8 & 14 & 36\\  \vspace*{1mm}
						&2 & 3  & 7  & 11 && 3 &  6 & 24\\
	%	\cline{2-5}

\multirow{4}{*}{lena}	&22& 19 & 35 & 100 && 18 & 33 & 98\\
						&16& 12 & 21 & 58 && 11 & 22 & 90\\
						&8 & 7  & 8  & 27 &&  5 &  8 & 28\\  \vspace*{1mm}
						&2 & 2  & 4  & 12 &&  2 &  4 & 17\\
%		\cline{2-5}

\multirow{4}{*}{hill}	&22& 25 & 48 & 120 && 24 & 48 & 120\\
						&16& 20 & 40 & 111 && 22 & 43 & 115\\
						&8 &  6 & 13 & 41 &&  6 & 13 & 41\\  \vspace*{1mm}
						&2 &  3 & 6  & 11 &&  2 &  6 & 11\\
	%	\cline{2-5}

\multirow{4}{*}{fingerprint}	&22& 13 & 28 & 86 && 13 & 26 & 86\\
							&16& 8  & 18 & 56 &&  8 & 17 & 47\\
							&8 & 4  & 9  & 28 &&  4 &  8 & 18\\  \vspace*{1mm}
							&2 & 3  & 7  & 19 &&  3 &  7 & 13\\

%		\cline{2-5}

\multirow{4}{*}{saturn}	&22& 8 & 17 & 51 && 8 & 17 & 47\\
						&16& 7 & 11 & 30 && 8 & 12 & 30\\
						&8 & 3 & 6  & 15 && 3 &  4 & 15\\  \vspace*{1mm}
						&2& 2 & 5   & 7 && 2 &  3 & 7\\
						
%		\cline{2-5}

\multirow{4}{*}{flintstones}	&22& 24 & 48 & 120 && 24 & 47 & 120\\
							&16& 13 & 41 & 118 && 14 & 37 & 118\\
							&8 &  8 & 13 & 41 &&  8 & 13 & 32\\
							& 2&  5 & 10 & 25 &&  4 &  8 & 24\\
							
\thickhline

\end{tabular}
%}
\end{center}

\vspace{-1mm}
\end{tiny}
\end{table}

%%%%%%%%%%%% plot d and curve fitting for d %%%%%%%%%%%%%

%\begin{figure}[t!]
%\centering
%\includegraphics[width=.84\textwidth]{figures/hyper/hy_d_P.pdf}
%\vspace{-3mm}
%\caption{Optimal subspace dimensionality $d$ value as a function of SNR for $5 \times 5$, $7 \times 7$ and $11 \times 11$ size patches respectively are shown in columns one, two and three for the Poisson noise models. Here the bars indicate the minimum and maximum values of the optimal $d$. The bottom and top edges of the blue boxes indicate the 25th and 75th percentiles and the central mark and green star indicate the median and mean values. The red line is the best linear curve fitted to the data points corresponding to the mean of the optimal $d$ values.}
%\label{fig:hy_d_fit_P}
%\vspace{-4mm}
%\end{figure}

%The optimal $d$ values are shown in Fig.~\ref{fig:hy_d_fit_P} as a function of SNR using box-plots for a fixed patch size, for the Poisson noise case.

The best-fitted curve parameters to the optimal $F_{\mbox{factor}}$ and $d$ are regrouped in Table~\ref{tab:tab_hbar_d_fit}. These rules give an efficient way to select the hyperparameters close to their optimality with a minimal performance loss depending on the size of the given patch and the intensity of the noise.

%%%%%%%%%%%% curve fitting for d and hbar_factor %%%%%%%%%%%%%

\begin{table}[t!]
\begin{scriptsize}

\begin{center}
\caption{Curve fitting parameters used in determining $d$ and $F_{\mbox{factor}}$ for various patch sizes for Gaussian and Poisson noise models. Also, the associative $\ell_2$ errors, PSNR (dB) loss and SSIM loss in curve fitting to the optimal $d$ and $F_{\mbox{factor}}$.}
%\vspace{-2mm}
%\resizebox{.7\hsize}{!}
%{
\label{tab:tab_hbar_d_fit}
\begin{tabular}{c l  r r r}
\thickhline

\multirow{2}{*}{~} & \multirow{2}{*}{~} & \multicolumn{3}{c}{Size of the patches}\\

			 && $5 \times 5$ & $7 \times 7$ & $11 \times 11$ \\
\thickhline

\multirow{9}{*}{\begin{turn}{90}Gaussian\end{turn}}

& Slope ($m_2$)		 & $0.7783$ & $1.7000$ & $4.2500$ \\

& Intercept ($c_2$) 	& $0.7315$ & $0.5345$ & $4.8210$\\

& $\ell_2$ error for $d$ fit & $21.1673$ & $43.3499$ & $112.4127$\\

& Parameter $l_1$ & $0.5287$ & $1.2630$ & $1.9161$ \\

& Parameter $l_2$ & $-4.4551$ & $-4.1915$ & $6.8223$ \\

& Parameter $l_3$ & $20.6204$ & $13.9698$ & $9.7995$ \\

& $\ell_2$ error for $F_{\mbox{factor}}$ fit & $2.3845$ & $2.6135$ & $1.9334$ \\

& PSNR(dB) loss	& 0.416 & 0.361 & 0.209\\ \vspace*{1mm}

& SSIM loss	& 0.0153 & 0.0129 & 0.0118\\

\multirow{9}{*}{\begin{turn}{90}Poisson\end{turn}}

& Slope ($m_2$)		 & $0.8202$ & $1.6030$ & $4.3990$ \\

& Intercept ($c_2$) 	& $0.8621$ & $0.5800$ & $2.8900$\\

& $\ell_2$ error for $d$ fit & $23.2670$ & $43.4135$ & $115.9894$\\

& Parameter $l_1$ & $0.8083$ & $1.5391$ & $1.8587$ \\

& Parameter $l_2$ & $-3.8975$ & $-4.4288$ & $11.6517$ \\

& Parameter $l_3$ & $16.8476$ & $10.1560$ & $3.9798$ \\

& $\ell_2$ error for $F_{\mbox{factor}}$ fit & $3.3802$ & $2.0652$  & $2.2571$ \\

& PSNR(dB) loss	& 0.487 & 0.594 & 0.485 \\

& SSIM loss	& 0.0175 & 0.0266 & 0.0386\\

\thickhline

\end{tabular}
%}
\end{center}

\end{scriptsize}
\vspace{-2mm}
\end{table}

\newpage ~~

\subsection{Denoising efficiency of the proposed scheme in comparison with standard methods}

For a comprehensive survey of the denoising ability of De-QuIP, rigorous comparisons with contemporary noise removal methods from the Table~\ref{tab:tab_denoi_diffpat}. Tables~\ref{tab:tab_psnrssimG} and~\ref{tab:tab_psnrssimP} compare quantitative performance in terms of PSNR and SSIM in the case of Gaussian and Poisson corrupted images, respectively.

%For a visual assessment, the zoomed segments of the denoised estimations of the Poisson corrupted (SNR $2$dB) \textit{House} image are shown in Fig.~\ref{fig:result_2P}.

%%%%%%%%%% denoising data for different patch size %%%%%%

\begin{table*}[t!]
\begin{footnotesize}

\setlength\tabcolsep{1pt}

\begin{center}
\caption{Comparison of denoising performance of De-QuIP with different patch sizes for different noise levels.}
\resizebox{.99\hsize}{!}{
\label{tab:tab_denoi_diffpat}
\begin{tabular}{cc cc c cc c cc  cc  cc c cc c cc}
\thickhline

\multirow{3}{*}{Sample} & \multirow{3}{*}{Input} & \multicolumn{8}{c}{Gaussian case} &&& \multicolumn{8}{c}{Poisson case}\\
%\multirow{3}{*}{Sample} & \multirow{3}{*}{Noise} & \multicolumn{8}{c}{Size of the patches}\\
		%	\cline{3-10}
			& & \multicolumn{2}{c}{$5 \times 5$} && \multicolumn{2}{c}{$7 \times 7$} && \multicolumn{2}{c}{$11 \times 11$} &&& \multicolumn{2}{c}{$5 \times 5$} && \multicolumn{2}{c}{$7 \times 7$} && \multicolumn{2}{c}{$11 \times 11$}\\
	%		\cline{3-4}  \cline{6-7}  \cline{9-10}
			& SNR(dB) & PSNR(dB) & SSIM && PSNR(dB) & SSIM && PSNR(dB) & SSIM &&~~~~& PSNR(dB) & SSIM && PSNR(dB) & SSIM && PSNR(dB) & SSIM\\
\thickhline

\multirow{4}{*}{house}	& 22 & 35.30 & 0.88 & & 35.45 & \bd{0.89} & & \bd{35.58} & \bd{0.89} &&& 34.94 & 0.87 && 35.10 & \bd{0.88} && \bd{35.14} & \bd{0.88} \\
						& 16 & 31.91 & \bd{0.83} & & 32.15 & \bd{0.83} & & \bd{32.29} & \bd{0.83} &&& 31.49 & \bd{0.82} && \bd{31.78} & \bd{0.82} && 31.73 & \bd{0.82} \\
						& 8  & 26.85 & 0.72 & & 27.45 & 0.75 & & \bd{27.91} & \bd{0.76} &&& 26.45 & 0.72 && 27.02 & 0.74 && \bd{27.27} & \bd{0.75} \\ \vspace*{1mm}
						& 2  & 23.05 & 0.60 & & 23.92 & 0.68 & & \bd{24.66} & \bd{0.72} &&& 22.65 & 0.59 && 23.48 & 0.66 && \bd{24.09} & \bd{0.70} \\
							
	%	\cline{2-10}

\multirow{4}{*}{lake}	& 22 & \bd{33.23} & \bd{0.92} & & 33.16 & 0.91 & & 32.80 & 0.90 &&& \bd{33.09} & \bd{0.91} && 33.04 & 0.90 && 32.72 & 0.90 \\
						& 16 & 28.81 & \bd{0.83} & & \bd{28.85} & 0.82 & & 28.63 & 0.81 &&& 28.54 & \bd{0.81} && \bd{28.60} & \bd{0.81} && 28.42 & \bd{0.81} \\
						& 8 & 24.05 & 0.69 & & 24.19 & \bd{0.71} & & \bd{24.37} & 0.69 &&& 23.75 & 0.66 && 23.98 & \bd{0.68} && \bd{24.11} & \bd{0.68} \\ \vspace*{1mm}
						& 2 & 21.09 & 0.57 & & 21.59 & 0.62 & & \bd{21.75} & \bd{0.63} &&& 20.90 & 0.56 && 21.33 & 0.61 && \bd{21.48} & \bd{0.63} \\
							
	%	\cline{2-10}

\multirow{4}{*}{lena}	& 22 & 35.05 & 0.89 & & 35.21 & 0.89 & & \bd{35.34} & \bd{0.90} &&& 34.86 & 0.88 && 35.05 & \bd{0.89} && \bd{35.16} & \bd{0.89} \\
						& 16 & 31.73 & 0.84 & & 32.00 & \bd{0.85} & & \bd{32.17} & \bd{0.85} &&& 31.49 & 0.83 && 31.78 & \bd{0.84} && \bd{32.34} & \bd{0.84} \\
						& 8  & 26.17 & 0.71 & & 27.67 & \bd{0.78} & & \bd{28.00} & \bd{0.78} &&& 26.93 & 0.74 && 27.40 & \bd{0.77} && \bd{27.61} & 0.76 \\ \vspace*{1mm}
						& 2  & 23.60 & 0.63 & & 24.53 & 0.71 & & \bd{25.04} & \bd{0.74} &&& 23.36 & 0.63 && 24.30 & \bd{0.71} && \bd{24.71} & \bd{0.71} \\
							
	%	\cline{2-10}

\multirow{4}{*}{hill}	& 22 & 31.54 & 0.82 & & \bd{31.58} & \bd{0.83} & & 31.55 & \bd{0.83} &&& 32.01 & 0.82 && \bd{32.16} & \bd{0.83} && 32.13 & \bd{0.83} \\
						& 16 & 27.95 & 0.69 & & 28.06 & \bd{0.70} & & \bd{28.10} & \bd{0.70} &&& 28.25 & \bd{0.70} && 28.37 & \bd{0.70} && \bd{28.39} & \bd{0.70} \\
						& 8  & 24.42 & \bd{0.55} & & 24.49 & \bd{0.55} & & \bd{24.61} & \bd{0.55} &&& 24.58 & \bd{0.55} && \bd{24.63} & \bd{0.55} && 23.58 & 0.54 \\ \vspace*{1mm}
						& 2  & 21.97 & 0.46 & & 22.41 & 0.48 & & \bd{22.61} & \bd{0.49} &&& 22.08 & 0.46 && 22.46 & 0.48 && \bd{22.54} & \bd{0.49} \\
						
%		\cline{2-10}

\multirow{4}{*}{fingerprint}	& 22 &	32.35 & 0.93 & & 32.50 & 0.93 & & \bd{32.54} & \bd{0.94} &&& 33.39 & 0.94 && 32.15 & 0.93 && \bd{33.49} & \bd{0.95} \\
							& 16 & 28.12 & 0.86 & & 28.46 & \bd{0.87} & & \bd{28.65} & \bd{0.87} &&& \bd{28.63} & \bd{0.87} && 28.16 & 0.86 && 28.24 & 0.86 \\
							& 8  & 23.36 & 0.72 & & 23.31 & 0.72 & & \bd{23.63} & \bd{0.73} &&& \bd{23.65} & \bd{0.74} && 23.07 & 0.72 && 23.40 & 0.73 \\ \vspace*{1mm}
							& 2  & \bd{20.03} & \bd{0.59} & & 19.80 & 0.57 & & 20.01 & 0.58 &&& \bd{19.90} & \bd{0.59} && 19.58 & 0.56 && 19.56 & 0.56 \\

	%	\cline{2-10}

\multirow{4}{*}{saturn}	& 22 & 38.94 & 0.89 & & 39.36 & 0.92 & & \bd{39.53} & \bd{0.94} &&& 40.64 & 0.97 && 40.85 & \bd{0.98} && \bd{40.87} & \bd{0.98} \\
						& 16 & 34.67 & 0.79 & & 35.27 & 0.83 & & \bd{35.63} & \bd{0.87} &&& 36.00 & 0.94 && 36.31 & \bd{0.95} && \bd{36.42} & 0.94 \\
						& 8  & 28.94 & 0.61 & & 29.87 & 0.67 & & \bd{30.60} & \bd{0.74} &&& 30.44 & 0.89 && 31.00 & \bd{0.90} && \bd{31.40} & 0.89 \\ \vspace*{1mm}
						& 2  & 24.45 & 0.46 & & 25.97 & 0.55 & & \bd{27.03} & \bd{0.62} &&& 26.40 & 0.82 && 27.26 & \bd{0.86} && \bd{27.46} & 0.85 \\
						
%		\cline{2-10}

\multirow{4}{*}{flintstones}	& 22 & \bd{32.20} & \bd{0.87} & & 32.16 & \bd{0.87} && 31.97 & 0.86 &&& 33.20 & \bd{0.88} && 33.08 & \bd{0.88} && \bd{32.99} & \bd{0.88} \\
							& 16 & 28.65 & \bd{0.80} & & \bd{28.69} & 0.79 && 28.47 & 0.78 &&& \bd{29.04} & \bd{0.80} && 29.00 & 0.78 && 28.77 & 0.78 \\
							& 8  & 23.48 & 0.67 & & \bd{23.78} & \bd{0.68} && 23.70 & 0.66 &&& 23.44 & 0.65 && \bd{23.84} & \bd{0.68} && 23.64 & 0.64 \\
							& 2  & 19.74 & 0.52 & & 19.87 & \bd{0.56} && \bd{20.03} & \bd{0.56} &&& 19.50 & 0.51 && \bd{19.69} & \bd{0.53} && 19.63 & \bd{0.53} \\
							
\thickhline

\end{tabular}
}
\end{center}

\end{footnotesize}
%\vspace*{-4mm}
\end{table*}

%%%%%%%% Quantitative denoising results using different methods for Gaussian noise%%%%%%%%%%%%%%%%%%%%%%%%%%%%

\begin{table*}[h!]

\begin{center}
%\begin{scriptsize}

\setlength\tabcolsep{4pt}
\caption{Quantitative denoising results for Gaussian corrupted images (average over 10 independent noise realizations). The best values are highlighted by color.}

\label{tab:tab_psnrssimG}
\begin{tabular}{cc  cccc cccc cccc cccc}
\thickhline

\multirow{3}{*}{Sample} & \multirow{3}{*}{Input} & \multicolumn{16}{c}{Methods}\\
%	\cline{3-18}
			
& & \multicolumn{2}{c}{PND} & \multicolumn{2}{c}{PGPCA} & \multicolumn{2}{c}{PLPCA} & \multicolumn{2}{c}{BM3D} & \multicolumn{2}{c}{DL} & \multicolumn{2}{c}{GSP} & \multicolumn{2}{c}{QAB} & \multicolumn{2}{c}{De-QuIP} \\

& SNR & PSNR & SSIM & PSNR & SSIM & PSNR & SSIM & PSNR & SSIM & PSNR & SSIM & PSNR & SSIM & PSNR & SSIM & PSNR & SSIM\\

\thickhline

\multirow{4}{*}{house}	& 22 &33.17&0.831 &34.28&0.863 &34.80&0.866 &\bd{35.89}&\bd{0.896} &32.86&0.827 &35.70&0.891 &33.77&0.856 &35.60&0.892 \\
						& 16 &30.78&0.800 &31.13&0.803 &31.47&0.802 &\bd{33.05}&\bd{0.848} &29.72&0.747 &32.91&0.839 &30.16&0.767 &32.86&0.843 \\
						& 8  &27.52&0.753 &26.83&0.672 &26.65&0.605 &\bd{28.73}&\bd{0.786} &25.31&0.587 &28.19&0.760 &25.71&0.601 &28.13&0.764 \\  \vspace*{1mm}
						& 2  &24.25&0.679 &23.45&0.491 &22.63&0.371 &\bd{24.96}&0.706 &21.20&0.405 &23.93&0.589 &22.92&0.464 &24.78&\bd{0.719} \\

% \cline{2-18}

\multirow{4}{*}{lake}	& 22 &30.29&0.855 &32.36&0.895 &32.56&0.895 &33.07&0.919 &30.90&0.863 &32.09&0.913 &32.22&0.878 &\bd{33.27}&\bd{0.926} \\
						& 16 &26.78&0.780 &28.21&0.815 &28.39&0.814 &28.92&\bd{0.857} &27.22&0.768 &28.45&0.836 &28.59&0.795 &\bd{29.11}&0.842 \\
						& 8  &23.61&0.697 &23.87&0.653 &23.77&0.606 &24.43&\bd{0.739} &23.25&0.588 &24.30&0.722 &24.24&0.599 &\bd{24.52}&0.705 \\ \vspace*{1mm}
						& 2  &21.42&0.623 &21.23&0.483 &20.73&0.388 &\bd{21.97}&\bd{0.652} &20.16&0.419 &21.27&0.548 &20.79&0.441 &21.80&0.631 \\
% \cline{2-18}

\multirow{4}{*}{lena}	& 22 &33.52&0.858 &34.78&0.881 &35.04&0.883 &35.50&0.898 &34.08&0.868 &35.09&0.893 &34.36&0.876 &\bd{35.81}&\bd{0.904} \\
						& 16 &31.10&0.822 &31.75&0.831 &31.95&0.829 &\bd{32.70}&\bd{0.861} &30.93&0.801 &32.43&0.852 &31.02&0.801 &32.41&0.852 \\
						& 8  &27.94&0.773 &27.65&0.716 &27.35&0.648 &\bd{28.75}&\bd{0.799} &26.34&0.638 &28.18&0.767 &26.61&0.657 &28.31&0.778 \\ \vspace*{1mm}
						& 2  &25.21&0.716 &24.27&0.536 &23.34&0.415 &\bd{25.57}&0.728 &22.27&0.446 &24.93&0.664 &22.76&0.476 &25.04&\bd{0.737} \\
% \cline{2-18}

\multirow{4}{*}{hill}	& 22 &29.35&0.742 &30.98&0.812 &31.29&0.819 &31.36&0.818 &29.80&0.761 &30.97&0.811 &30.56&0.810 &\bd{31.59}&\bd{0.826} \\
						& 16 &26.90&0.640 &27.86&0.690 &28.07&0.700 &28.32&\bd{0.710} &27.17&0.649 &28.15&0.708 &27.24&0.671 &\bd{28.34}&\bd{0.710} \\
						& 8  &24.34&0.534 &24.55&0.539 &24.42&0.515 &\bd{24.98}&\bd{0.575} &23.85&0.478 &24.86&0.571 &23.91&0.492 &24.78&0.563 \\ \vspace*{1mm}
						& 2  &22.49&0.475 &22.24&0.405 &21.75&0.345 &\bd{22.72}&\bd{0.493} &21.14&0.354 &22.33&0.447 &21.39&0.375 &22.61&0.491 \\
% \cline{2-18}

\multirow{4}{*}{fingerprint}	& 22 &28.40&0.845 &31.14&0.912 &31.30&0.914 &31.47&0.917 &29.79&0.882 &31.23&0.896 &30.32&0.894 &\bd{32.53}&\bd{0.934} \\
							& 16 &25.57&0.757 &27.27&0.830 &27.46&0.835 &27.92&0.850 &25.97&0.783 &26.99&0.824 &27.16&0.815 &\bd{28.61}&\bd{0.868} \\
							& 8  &22.58&0.666 &22.73&0.680 &23.02&0.676 &\bd{23.77}&\bd{0.734} &21.09&0.575 &23.22&0.698 &22.95&0.671 &23.65&0.733 \\ \vspace*{1mm}
							& 2  &20.05&0.563 &18.92&0.471 &19.50&0.474 &\bd{20.73}&\bd{0.613} &17.85&0.385 &20.21&0.546 &20.04&0.527 &20.56&0.605 \\
% \cline{2-18}

\multirow{4}{*}{saturn}	& 22 &40.95&0.955 &39.32&0.935 &39.63&0.929 &\bd{42.26}&\bd{0.970} &37.55&0.891 &41.20&0.943 &38.63&0.916 &39.70&0.937 \\
						& 16 &37.86&0.904 &35.80&0.907 &36.02&0.883 &\bd{38.64}&\bd{0.937} &33.74&0.776 &38.01&0.881 &34.44&0.850 &36.93&0.873 \\
						& 8  &32.23&0.775 &30.70&0.753 &30.09&0.647 &\bd{33.16}&\bd{0.861} &28.00&0.541 &32.58&0.757 &29.44&0.712 &31.25&0.735 \\ \vspace*{1mm}
						& 2  &28.13&0.640 &26.68&0.544 &25.31&0.398 &\bd{28.31}&\bd{0.747} &23.36&0.347 &27.64&0.608 &25.27&0.601 &27.18&0.615 \\
% \cline{2-18}

\multirow{4}{*}{flintstones}	& 22 &28.54&0.766 &30.62&0.827 &30.83&0.831 &31.31&0.847 &29.15&0.793 &30.91&0.841 &30.34&0.811 &\bd{32.20}&\bd{0.865} \\
							& 16 &25.63&0.702 &27.36&0.755 &27.61&0.758 &\bd{28.61}&\bd{0.802} &25.64&0.695 &28.33&0.776 &27.68&0.752 &28.46&0.781 \\
							& 8  &21.86&0.620 &22.15&0.586 &22.31&0.567 &\bd{23.96}&\bd{0.705} &21.12&0.520 &23.61&0.662 &22.54&0.561 &23.67&0.662 \\ \vspace*{1mm}
							& 2  &18.79&0.521 &18.67&0.415 &18.62&0.364 &\bd{20.12}&\bd{0.585} &17.86&0.361 &19.38&0.498 &19.17&0.435 &19.99&0.555 \\
% \cline{2-18}

\multirow{4}{*}{ridges}	& 22 &46.41&0.985 &47.52&0.980 &47.18&0.973 &\bd{50.58}&\bd{0.993} &44.49&0.968 &47.76&0.980 &46.47&0.956 &49.23&0.982 \\
						& 16 &42.32&0.961 &43.34&0.950 &42.97&0.938 &\bd{45.37}&\bd{0.978} &40.17&0.922 &42.60&0.931 &41.44&0.898 &43.56&0.940 \\
						& 8  &32.65&0.816 &38.03&0.856 &37.09&0.828 &\bd{38.60}&\bd{0.922} &33.60&0.779 &37.06&0.859 &34.68&0.754 &38.28&0.851 \\ \vspace*{1mm}
						& 2  &25.99&0.599 &33.07&0.720 &31.91&0.672 &\bd{33.56}&\bd{0.837} &28.97&0.618 &33.15&0.727 &29.86&0.643 &33.39&0.749 \\
% \cline{2-18}

\multirow{4}{*}{peppers}	& 22 &31.23&0.842 &32.83&0.871 &33.06&0.872 &\bd{33.79}&\bd{0.897} &31.32&0.840 &33.34&0.891 &32.08&0.849 &33.52&0.888 \\
						& 16 &28.46&0.801 &29.39&0.814 &29.70&0.811 &\bd{30.58}&\bd{0.853} &28.02&0.762 &30.25&0.839 &28.67&0.758 &30.43&0.847 \\
						& 8  &24.54&0.744 &24.93&0.674 &25.03&0.639 &\bd{26.24}&\bd{0.771} &23.96&0.608 &26.07&0.752 &23.94&0.612 &26.19&0.755 \\ \vspace*{1mm}
						& 2  &21.73&0.687 &21.67&0.519 &21.38&0.420 &\bd{22.83}&0.683 &20.43&0.434 &21.96&0.579 &20.49&0.457 &22.14&\bd{0.696} \\
% \cline{2-18}

\multirow{4}{*}{bridge}	& 22 &27.91&0.751 &29.90&0.839 &30.08&0.845 &30.11&0.846 &28.72&0.799 &29.43&0.824 &29.27&0.817 &\bd{30.72}&\bd{0.864} \\
						& 16 &25.16&0.608 &26.28&0.685 &26.46&0.697 &\bd{26.57}&0.698 &25.52&0.655 &26.33&0.693 &25.78&0.662 &26.49&\bd{0.699} \\
						& 8  &22.53&0.464 &22.82&0.499 &22.82&0.493 &\bd{23.12}&\bd{0.511} &22.27&0.467 &23.00&0.508 &22.31&0.472 &22.84&0.503 \\ \vspace*{1mm}
						& 2  &20.69&0.393 &20.61&0.372 &20.29&0.337 &\bd{21.00}&\bd{0.421} &19.71&0.333 &20.60&0.389 &20.04&0.362 &20.72&0.403 \\
% \cline{2-18}

\multirow{4}{*}{cameraman}	& 22 &30.25&0.812 &32.66&0.880 &32.97&0.881 &\bd{33.54}&\bd{0.907} &31.54&0.858 &32.71&0.882 &32.13&0.865 &32.91&0.892 \\
							& 16 &27.29&0.757 &28.78&0.793 &29.08&0.792 &\bd{29.99}&\bd{0.843} &27.98&0.758 &29.17&0.840 &28.18&0.751 &29.36&0.841 \\
							& 8  &24.28&0.695 &24.42&0.635 &24.45&0.579 &\bd{25.80}&\bd{0.755} &23.72&0.581 &25.54&0.751 &23.93&0.612 &25.57&0.751 \\
							& 2  &21.78&0.617 &21.45&0.434 &21.03&0.347 &\bd{22.77}&\bd{0.674} &20.28&0.393 &22.13&0.597 &20.51&0.402 &22.49&0.636 \\
							
\thickhline

\end{tabular}
%\end{scriptsize}

\vspace{6mm}
\begin{tabular}{lcl c lcl}
PND & $\rightarrow$ & Principal Neighborhood Dictionaries \cite{tasdizen2009principal} && DL & $\rightarrow$ &  Dictionary Learning \cite{Elad2006image}\\
PGPCA & $\rightarrow$ & Patch-based Global Principal Component Analysis \cite{deledalle2011image} && GSP & $\rightarrow$ & Graph Signal Processing \cite{Pang2017graph}\\
PLPCA & $\rightarrow$ & Patch-based Local Principal Component Analysis \cite{deledalle2011image} && QAB & $\rightarrow$ & Quantum Adaptive Basis \cite{dutta2021quantum}\\
BM3D & $\rightarrow$ &  Block-Matching and 3D Filtering \cite{Dabov2007Image} && De-QuIP & $\rightarrow$ & Denoising by Quantum Interactive Patches (proposed)\\
\end{tabular}

\end{center}
%\vspace*{-5mm}
\end{table*}

%\vspace{-4mm}

%%%%%%%% Quantitative denoising results using different methods for Poisson noise%%%%%%%%%%%%%%%%%%%%%%%%%%%%

\begin{table}[h!]

\begin{center}
%\begin{scriptsize}

\setlength\tabcolsep{3pt}
\caption{Quantitative denoising results for Poisson corrupted images (average over 10 independent noise realizations). The best values are highlighted by color.}
\vspace{-3mm}

\label{tab:tab_psnrssimP}
\begin{tabular}{cc  cccc cccc}
\thickhline

\multirow{3}{*}{Sample} & \multirow{3}{*}{Input} & \multicolumn{8}{c}{Methods}\\
%	\cline{3-18}
			
& & \multicolumn{2}{c}{PNLPCA} & \multicolumn{2}{c}{ATBM3D} & \multicolumn{2}{c}{QAB} & \multicolumn{2}{c}{De-QuIP} \\

& SNR & PSNR & SSIM & PSNR & SSIM & PSNR & SSIM & PSNR & SSIM\\

\thickhline

\multirow{4}{*}{house}	& 22 &33.19&0.848 &\bd{35.36}&\bd{0.879} &32.85&0.833 &35.27&\bd{0.879} \\
						& 16 &30.64&0.815 &\bd{33.13}&\bd{0.851} &29.18&0.738 &32.79&0.839 \\
						& 8  &26.72&0.706 &\bd{28.48}&0.731 &25.63&0.619 &27.72&\bd{0.757} \\  \vspace*{1mm}
						& 2  &20.93&0.519 &\bd{23.96}&0.549 &21.02&0.413 &23.94&\bd{0.651} \\

% \cline{2-18}

\multirow{4}{*}{lake}	& 22 &30.23&0.884 &32.13&\bd{0.913} &31.21&0.845 &\bd{32.43}&0.892 \\
						& 16 &27.45&0.832 &\bd{28.69}&\bd{0.854} &27.76&0.789 &28.62&0.839 \\
						& 8  &22.42&0.675 &\bd{24.50}&0.698 &23.29&0.567 &24.28&\bd{0.707} \\ \vspace*{1mm}
						& 2  &19.38&0.494 &\bd{21.65}&0.534 &20.70&0.418 &21.51&\bd{0.637} \\
% \cline{2-18}

\multirow{4}{*}{lena}	& 22 &32.68&0.852 &34.92&0.886 &34.29&0.872 &\bd{35.15}&\bd{0.889} \\
						& 16 &30.74&0.846 &\bd{32.39}&\bd{0.856} &30.65&0.791 &31.87&0.827 \\
						& 8  &26.19&0.686 &\bd{28.08}&0.751 &25.83&0.617 &27.78&\bd{0.768} \\ \vspace*{1mm}
						& 2  &21.76&0.541 &24.23&0.599 &21.88&0.436 &\bd{24.63}&\bd{0.707} \\
% \cline{2-18}

\multirow{4}{*}{hill}	& 22 &30.88&0.792 &\bd{31.81}&0.813 &30.89&0.824 &31.69&\bd{0.824} \\
						& 16 &26.24&0.673 &\bd{28.14}&\bd{0.696} &27.44&0.682 &27.89&\bd{0.696} \\
						& 8  &22.95&0.535 &\bd{24.98}&\bd{0.555} &24.45&0.494 &24.71&0.548 \\ \vspace*{1mm}
						& 2  &19.59&0.413 &22.37&0.445 &21.80&0.397 &\bd{22.48}&\bd{0.480} \\
% \cline{2-18}

\multirow{4}{*}{fingerprint}	& 22 &27.95&0.846 &31.40&0.913 &30.74&0.905 &\bd{32.39}&\bd{0.924} \\
						& 16 &26.81&0.820 &\bd{28.41}&0.860 &27.62&0.829 &28.36&\bd{0.862} \\
						& 8  &21.94&0.686 &\bd{23.63}&0.719 &22.91&0.679 &23.56&\bd{0.732} \\ \vspace*{1mm}
						& 2  &19.42&0.538 &\bd{20.16}&\bd{0.569} &20.07&0.549 &20.01&0.565 \\
% \cline{2-18}

\multirow{4}{*}{saturn}	& 22 &39.36&0.942 &\bd{41.15}&0.976 &38.94&0.906 &40.96&\bd{0.978} \\
						& 16 &35.85&0.909 &\bd{37.39}&\bd{0.957} &34.42&0.849 &36.71&0.942 \\
						& 8  &28.55&0.851 &\bd{30.82}&0.874 &28.75&0.714 &30.31&\bd{0.890} \\ \vspace*{1mm}
						& 2  &24.87&0.740 &26.20&0.765 &24.58&0.591 &\bd{27.12}&\bd{0.845} \\
% \cline{2-18}

\multirow{4}{*}{flintstones}	& 22 &30.08&0.820 &31.72&0.843 &30.29&0.789 &\bd{32.28}&\bd{0.864} \\
						& 16 &28.12&0.796 &\bd{29.46}&\bd{0.814} &27.23&0.678 &29.31&0.806 \\
						& 8  &22.51&0.631 &\bd{23.93}&0.650 &23.67&0.529 &23.76&\bd{0.668} \\ \vspace*{1mm}
						& 2  &18.23&0.435 &\bd{19.97}&0.479 &21.08&0.389 &19.72&\bd{0.532} \\
% \cline{2-18}

\multirow{4}{*}{ridges}	& 22 &42.67&0.942 &\bd{48.98}&\bd{0.991} &45.59&0.947 &47.93&0.986 \\
						& 16 &39.95&0.894 &\bd{43.88}&\bd{0.988} &42.23&0.902 &42.97&0.965 \\
						& 8  &31.92&0.630 &\bd{38.33}&\bd{0.898} &34.64&0.734 &37.09&0.865 \\ \vspace*{1mm}
						& 2  &26.42&0.491 &\bd{33.00}&0.815 &29.83&0.602 &31.79&\bd{0.816} \\
% \cline{2-18}

\multirow{4}{*}{peppers}	& 22 &31.85&0.860 &\bd{33.35}&\bd{0.893} &31.42&0.816 &33.32&0.889 \\
						& 16 &30.02&0.847 &\bd{30.49}&\bd{0.855} &27.78&0.737 &29.97&0.830 \\
						& 8  &25.08&0.693 &\bd{25.97}&\bd{0.743} &23.02&0.587 &25.45&0.737 \\ \vspace*{1mm}
						& 2  &20.61&0.532 &\bd{22.19}&0.591 &19.74&0.434 &21.69&\bd{0.676} \\
% \cline{2-18}

\multirow{4}{*}{bridge}	& 22 &28.59&0.734 &29.22&0.800 &30.49&0.847 &\bd{30.58}&\bd{0.860} \\
						& 16 &25.99&0.660 &\bd{26.61}&0.690 &26.44&0.689 &26.50&\bd{0.694} \\
						& 8  &22.37&0.471 &\bd{23.16}&\bd{0.512} &23.13&0.472 &22.65&0.496 \\ \vspace*{1mm}
						& 2  &19.46&0.351 &\bd{20.94}&\bd{0.413} &20.17&0.352 &20.56&0.398 \\
% \cline{2-18}

\multirow{4}{*}{cameraman}	& 22 &29.48&0.855 &\bd{32.45}&0.881 &31.77&0.837 &32.41&\bd{0.890} \\
						& 16 &28.03&0.821 &\bd{29.86}&\bd{0.839} &27.78&0.720 &28.76&0.818 \\
						& 8  &23.87&0.645 &\bd{25.83}&\bd{0.734} &23.12&0.524 &24.87&0.716 \\ \vspace*{1mm}
						& 2  &19.39&0.470 &\bd{22.37}&0.546 &19.41&0.359 &22.08&\bd{0.612} \\
					
\thickhline

\end{tabular}
%\end{scriptsize}

\vspace{2mm}
\begin{tabular}{lcl c lcl}
PNLPCA & $\rightarrow$ & Poisson Non-Local Principal Component Analysis \cite{salmon2014poisson} && ATBM3D & $\rightarrow$ & Anscombe Transform with Block-Matching and 3D Filtering \cite{Makitalo2011optimal}\\
QAB & $\rightarrow$ & Quantum Adaptive Basis \cite{dutta2021quantum} && De-QuIP & $\rightarrow$ & Denoising by Quantum Interactive Patches (proposed)\\
\\
\end{tabular}

\end{center}
\vspace{-3mm}
\end{table}

\newpage ~~

%\subsection{Application to ultrasound (US) image despeckling}

%The despeckling performance of De-QuIP is investigated through a phantom as well as four real cancer and two non-cancer thyroid US images acquired with a 7.5 MHz linear probe. For the quantitative analysis, the contrast-to-noise-ratio (CNR) and resolution loss (RL) are regrouped in Table~\ref{tab:tab_usdespeckling} and in Fig~\ref{fig:result_US_supple} for a visual demonstrations.

%\vspace*{20cm}
\subsection{Application to ultrasound (US) image despeckling}
\label{sec:usdesp}
%\vspace*{-1mm}

\begin{figure}[b!]
\centering
\includegraphics[width=.6\textwidth]{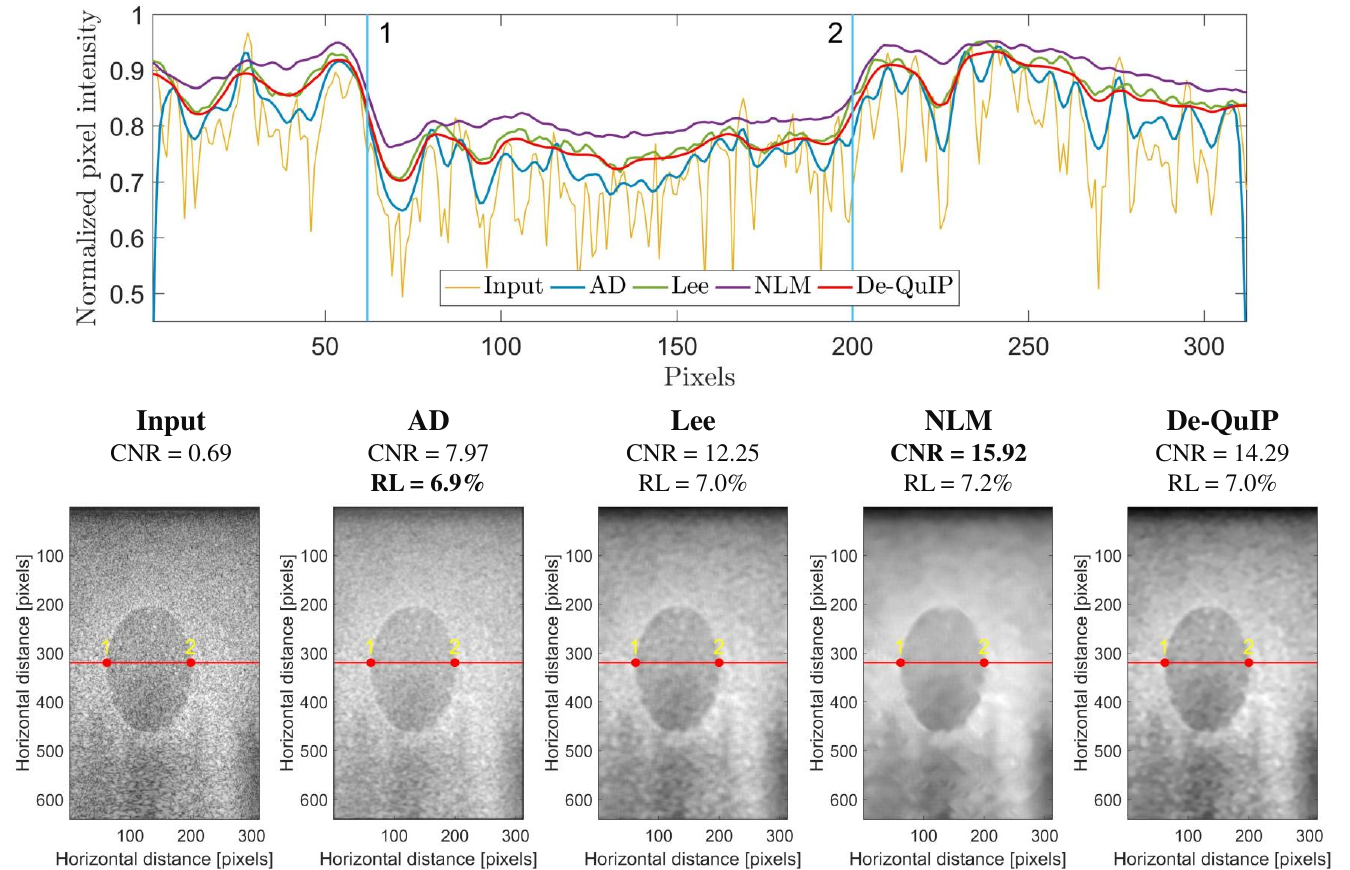}
%\vspace*{6mm}

\includegraphics[width=.6\textwidth]{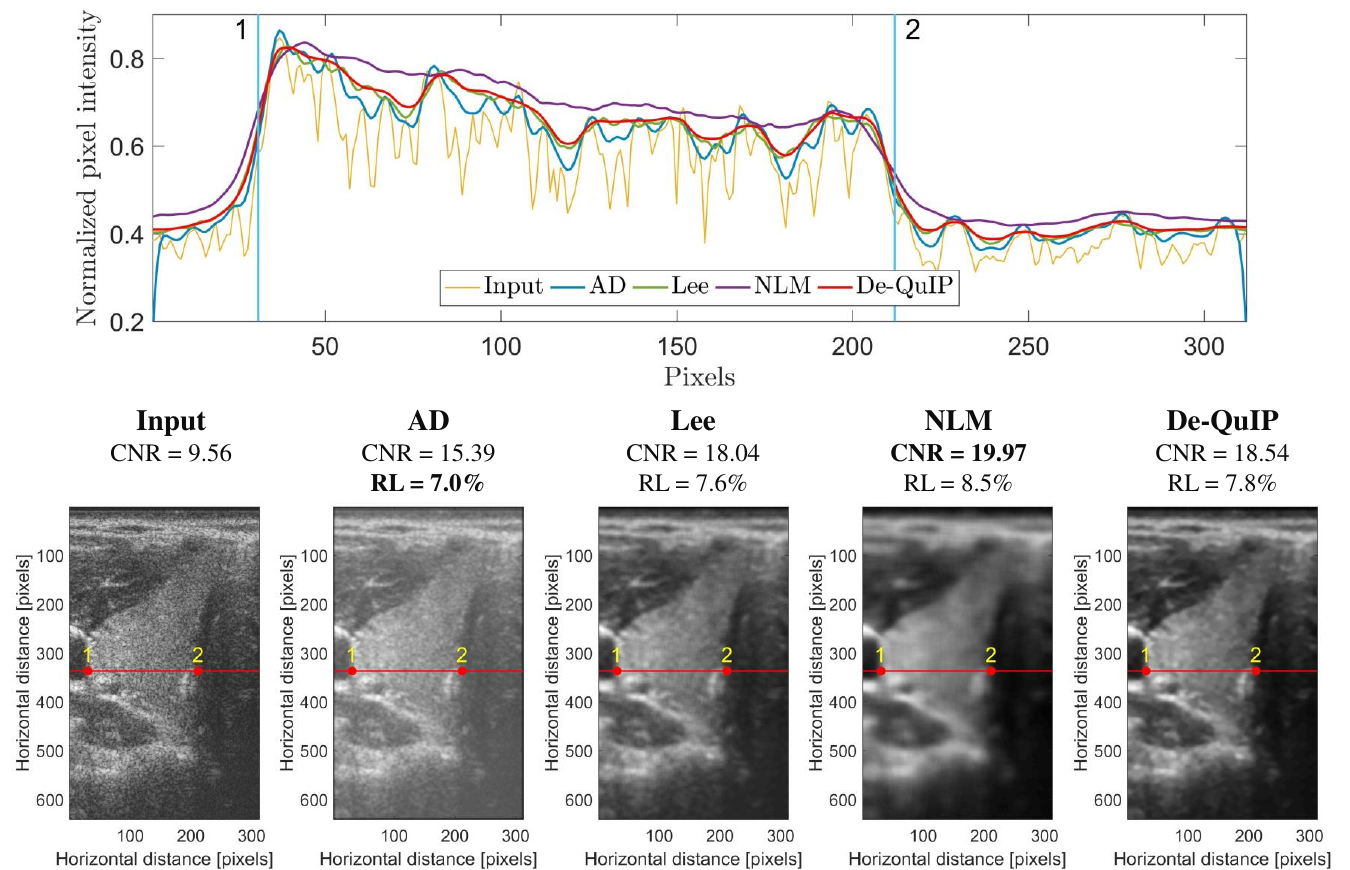}

\includegraphics[width=.6\textwidth]{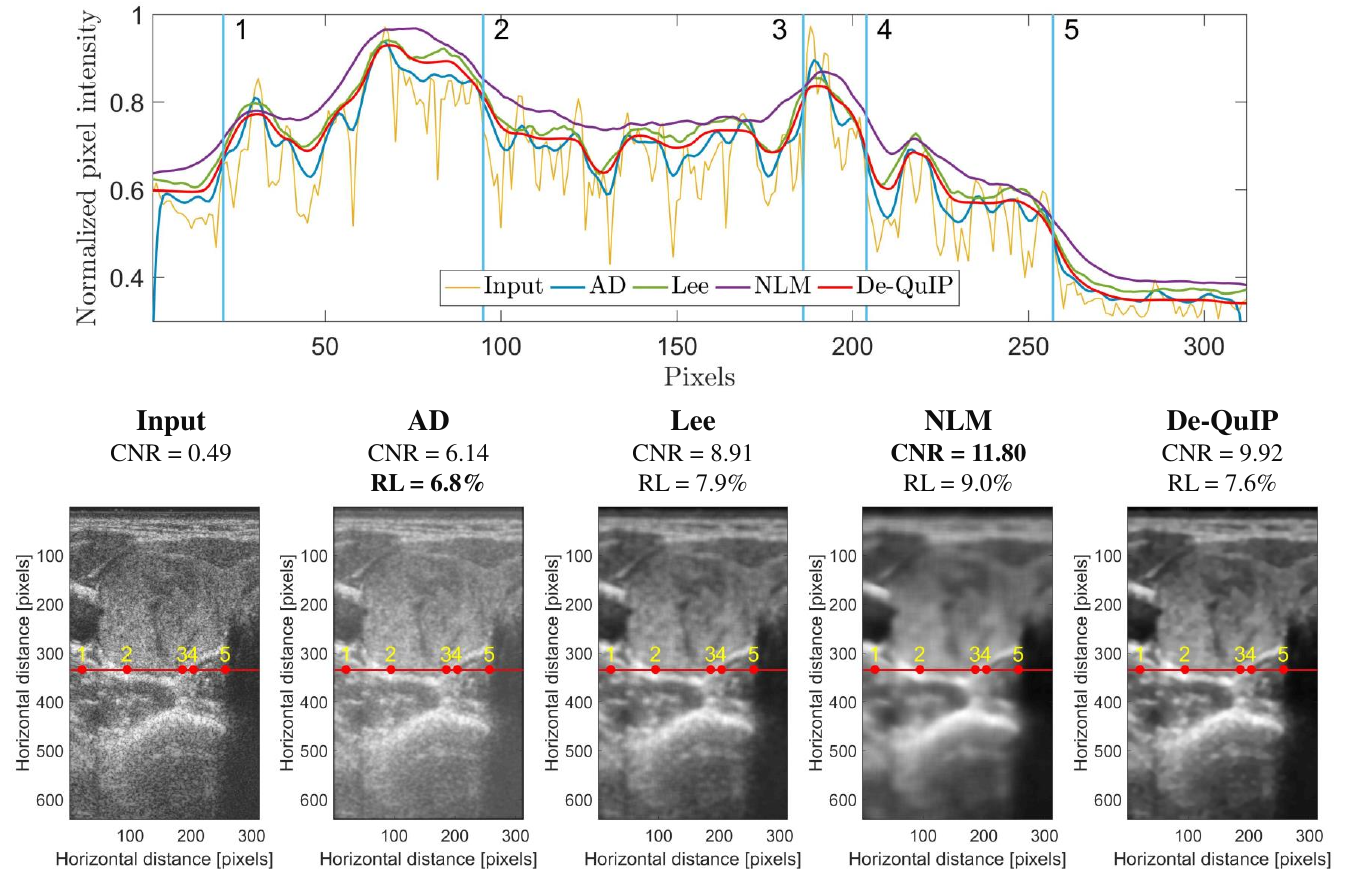}
\vspace*{3mm}
\caption{US image despeckling results using different methods. The top one is the phantom image, the second one is the non-cancer image, and the bottom one is associated with cancer images. The normalized pixel intensities of the extracted red lines from speckled and despeckled US images are shown.}
\label{fig:result_US_supple}

\end{figure}

The despeckling performance of De-QuIP is investigated through a phantom as well as four real cancer and two non-cancer thyroid US images acquired with a 7.5 MHz linear probe. For the quantitative analysis, the contrast-to-noise-ratio (CNR) and resolution loss (RL) are regrouped in Table~\ref{tab:tab_usdespeckling} and in Fig.~\ref{fig:result_US_supple} for a visual demonstrations. Observation shows that De-QuIP offers a better image contrast (higher CNR than AD, Lee and slightly lower than NLM, which over-smooths the images and yields poor resolution) while having less spatial resolution loss (De-QuIP has less spatial resolution loss compared to the native US image). Note that these images are chosen arbitrarily, that is, the quality of the results should not depend on the data tested.

%%%%%%%% Quantitative US despeckling results using different methods %%%%%%%%%%%%%%%%

%\vspace*{10mm}

\begin{table}[h!]

\begin{center}
%\begin{scriptsize}

% \setlength\tabcolsep{3pt}
\caption{Quantitative despeckling results of real medical US images using different methods. The best values are highlighted by color.}
\vspace*{-3mm}

\label{tab:tab_usdespeckling}
\begin{tabular}{c cc  cccc cccc}
\thickhline

\multirow{3}{*}{Sample} && & \multicolumn{8}{c}{Methods}\\
%	\cline{3-18}
			
& Input && \multicolumn{2}{c}{AD} & \multicolumn{2}{c}{Lee} & \multicolumn{2}{c}{NLM} & \multicolumn{2}{c}{De-QuIP} \\

& CNR &&  CNR & RL (\%) &  CNR & RL (\%) &  CNR & RL (\%) & CNR & RL (\%)\\

\thickhline
%~~\vspace*{1mm}

phantom		& 0.69 &&	7.97&\bd{6.9}& 12.25&7.0  &\bd{15.92}&7.2  &14.29&7.0 \\

non-cancer 1		&9.56 &&	15.39&\bd{7.0}& 18.04&7.6  &\bd{19.97}&8.5  &18.54&7.8 \\

non-cancer 2		&1.86 &&	7.05&\bd{6.0}& 9.77&6.7  &\bd{11.89}&8.4  &10.54&7.3 \\

cancer 1		& 1.41 &&	4.59&\bd{7.5}& 5.41&8.0  &8.43&8.8  &\bd{9.10}&8.1 \\

cancer 2	& 0.49 &&	6.14&\bd{6.8}& 8.91&7.9  &\bd{11.80}&9.0  &9.92&7.6 \\

cancer 3	& 0.96 &&	4.12&\bd{6.9}& 6.12&8.4  &\bd{8.20}&9.4  &6.40&8.5 \\

cancer 4	& 1.22 &&	5.35&\bd{7.3}& 6.90&9.2  &\bd{9.04}&11.6  &7.24&9.6 \\

% \cline{2-18}

\thickhline

\end{tabular}
%\end{scriptsize}

\vspace{2mm}
\begin{tabular}{lcl c lcl}
AD & $\rightarrow$ &  Anisotropic Diffusion \cite{Yu2002speckle} && NLM & $\rightarrow$ & Non-Local Means \cite{tasdizen2009principal}\\
Lee & $\rightarrow$ & Lee Filtering \cite{Lee1980digital} && De-QuIP & $\rightarrow$ & Denoising by Quantum Interactive Patches (proposed)\\
\\
\\
\end{tabular}

\end{center}

\end{table}

%\vspace{-4mm}

%%%%%%%%%%%% US despeckling %%%%%%%%%%%%%

%\bibliographystyle{IEEEbib}
%\section*{References}
%\bibliographystyle{elsarticle-num}

%\bibliography{bibDeQuIP2022}

\end{document}